\documentclass[a4paper,11pt]{article}
\pdfoutput=1 
\usepackage{jcappub}
\usepackage[T1]{fontenc} 
\usepackage[utf8]{inputenc}
\usepackage{graphicx}
\usepackage[normalem]{ulem}

\usepackage{color}

\title{\boldmath On the Measurement of the Helicity of Intergalactic Magnetic Fields Using Ultra-High-Energy Cosmic Rays}

\author[a]{Rafael {Alves~Batista}}
\author[b,c]{Andrey Saveliev}

\affiliation[a]{Universidade de S{\~a}o Paulo, Instituto de Astronomia, Geof{\'i}sica e Ci{\^e}ncias Atmosf{\'e}ricas; Rua do Mat{\~a}o, 1226, 05508-090, S{\~a}o Paulo-SP, Brazil}
\affiliation[b]{Immanuel Kant Baltic Federal University, Institute of Physics, Mathematics and Information Technology; Ul. Aleksandra Nevskogo 14, 236041 Kaliningrad, Russia}
\affiliation[c]{Lomonosov Moscow State University, Faculty of Computational Mathematics and Cybernetics; GSP-1, Leninskiye Gory 1-52, 119991 Moscow, Russia}

\emailAdd{rafael.ab@usp.br}
\emailAdd{andrey.saveliev@desy.de}

\abstract{
The origin of the first magnetic fields in the Universe is a standing problem in cosmology. Intergalactic magnetic fields (IGMFs) may be an untapped window to the primeval Universe, providing further constrains on magnetogenesis. We demonstrate the feasibility of using ultra-high-energy cosmic rays (UHECRs) to constrain the helicity of IGMFs by performing simulations of cosmic-ray propagation in simple magnetic field configurations. We show that the first harmonic moments of the arrival distribution of UHECRs may be used to measure the absolute value of the helicity and its sign.
}

\begin{document}
\maketitle
\flushbottom

\section{Introduction}

Intergalactic magnetic fields (IGMFs) may be fossil records of some cosmological process taking place in early phases of the Universe, thereby carrying imprints of the processes from whence they originated. For instance, phase transitions such as the quantum chromodynamics (QCD)~\cite{PhysRevLett.51.1488,1989ApJ...344L..49Q,PhysRevD.55.4582,Tevzadze:2012kk} and the electroweak (EW)~\cite{Vachaspati:1991nm,Enqvist:1993np,PhysRevD.53.662,Grasso:1997nx,Fujita:2016igl} one, as well as inflation~\cite{PhysRevD.37.2743,Ratra:1991bn,Byrnes:2011aa,Ferreira:2014hma}, have been suggested as mechanisms for magnetogenesis. Alternative explanations postulate their origin much later in time, for example during structure formation~\cite{Kulsrud:1996km}. These fields are believed to have served as seeds for structures to acquire their current magnetisation.

We define IGMFs as pervasive fields filling the whole Universe, not bound to any particular structure. The strength of IGMFs is believed to be $B \lesssim 1 \; \text{nG}$~\cite{PhysRevD.80.123012,Ade:2015cva}, thereby not being measurable inside any structures such as filaments and galaxy clusters. Furthermore, they are prone to contamination by feedback and magnetohydrodynamical processes in the immediate vicinity of structures. For this reason, the measurement of IGMFs should ideally be carried out in cosmic voids, the low-density regions of the cosmic web that fill most of the volume of the Universe. This is, however, rather difficult, and requires indirect measurement techniques.

Upper limits on the strength of IGMFs have been available for some time and were obtained using a variety of methods including Faraday Rotation~\cite{PhysRevD.80.123012,han2017a} and anisotropies of the cosmic microwave background (CMB) (see \cite{Jedamzik:2018itu} and the references therein). Lower limits, on the other hand, are much harder to derive. A promising method proposed over two decades ago consists in the observation of gamma-ray-induced electromagnetic cascades in the intergalactic medium~\cite{Aharonian:1993vz,Plaga:1995ins,Coppi:1996ze}. Nevertheless, it was not until recently, with the advent of imaging air Cherenkov telescopes such as H.E.S.S., VERITAS, and MAGIC, combined with space telescopes such as Fermi, that we were able to study individual sources of high-energy gamma rays with high enough precision to attempt to constrain IGMFs with gamma rays. A number of such works has been done in the past decade~\cite{JETPLett.85.10.473,PhysRevD.80.123012,2010ApJ...722L..39A,2010MNRAS.406L..70T,Taylor:2011bn,Takahashi:2013lba,2014arXiv1410.7717C}. The fundamental idea is to observe electromagnetic cascades triggered by TeV gamma rays from extreme blazars. The charged leptonic component of these cascades is deflected away from the line of sight, resulting in a suppression of the observed signal in the GeV range. This signal would be clearly visible in blazar spectra under ideal conditions. A similar effect could also be seen in the arrival directions by observing the so-called blazar pair haloes. Moreover, the distribution of arrival times of gamma rays from flaring sources could provide us hints of the strength and coherence length of the intervening field (see~\cite{PhysRevD.80.123012} for further details).

Electromagnetic cascades provide lower limits for the strength of IGMFs of the order of $B  \gtrsim 10^{-17}\, \text{G}$. These results are, however, still disputed, due to claims~\cite{0004-637X-752-1-22,0004-637X-758-2-102,Schlickeiser:2013eca,SavelievIGM,Chang:2014cta,Broderick:2018nqf} that plasma instabilities could provide similar signatures even in the absence of IGMFs, thus rendering the inferred limits invalid.

Another astroparticle approach to obtain information on IGMFs is to use ultra-high-energy cosmic rays (UHECRs), since they may carry imprints of the intervening fields~\cite{Ryu:2009pf}, including anisotropies in their angular distributions~\cite{Takami:2012uw,Mollerach:2016mko,Hackstein:2016pwa}, or, if individual UHECR sources can be identified, specific morphological features in their arrival directions~\cite{Harari:2015mal}. Another possibility, similar to the case of electromagnetic cascades, is to look at the UHECR flux suppression~\cite{Mollerach:2013dza,Batista:2014xza}.
 Finally, it is also possible to constrain IGMFs using spectra of secondary particles produced by UHECR such as gamma rays~\cite{Essey:2009ju,Essey:2010nd}. One should bear in mind that, in the case of UHECRs, disentangling the IGMF signal from those due to, for example, fields in filaments, clusters, and galaxies would be extremely difficult.

Besides the strength and coherence length of IGMFs, their topological properties are equally important. A proxy to describe the overall geometry of the field is the magnetic helicity $\mathcal{H}$,  defined as
\begin{equation} \label{HelDef}
\mathcal{H} = \int \mathbf{A} \cdot \mathbf{B} \,  {\rm d}^{3}r \,,
\end{equation}
where $\mathbf{A}$ is the magnetic vector potential and $\mathbf{B}=\nabla \times \mathbf{A}$ is the magnetic field. 

Magnetic helicity is a crucial quantity for understanding the (intergalactic) magnetic fields as it connects their geometrical structures and time evolutions in a unique way. As mentioned above, it describes the topological properties of the field, which follows from the fact that Eq.~\ref{HelDef} contains both the vector field $\mathbf{A}$ and its curl, which describes rotations. Thus, the scalar product between these two quantities gives a measure of how strongly the vector potential follows a helical structure. Formally, one can establish this connection by proving that magnetic helicity is strongly related to the linking number of infinitisimally thin magnetic flux tubes -- a formal description of magnetic field lines~\cite{Moffatt1969,0741-3335-41-12B-312}.

In general, helicity is defined for magnetic fields for which no field lines are crossing the boundary\footnote{Note, however, that there are suggestions on how to drop this condition~\cite{JFluidMech_147.133}).} as this ensures invariance under electromagnetic gauge transformations. In addition, for ideal magnetohydrodynamics, i.e.~with electrical conductivity tending to infinity ($\sigma \rightarrow \infty$), one can show that the overall helicity is conserved, which is also known as the First Woltjer Theorem \cite{1958PNAS...44..489W}.

The fact that helicity is virtually conserved\footnote{Magnetic helicity is conserved for infinite conductivities. In reality, conductivity is very large but finite.} is also responsible for different regimes of time evolution for the IGMFs depending whether it has a zero value or not. In particular, it might be responsible for the so-called inverse cascade of the magnetic spectrum~\cite{Sigl:2002kt,0004-637X-640-1-335,Saveliev:2013uva}, i.e.~an efficient transport of energy from small to large scales, even though a non-helical inverse cascade may also be possible~\cite{Brandenburg:2014mwa}.  Thus, one possible approach for measuring magnetic helicity would be to compare theoretical models for the IGMF evolution with actual measurements of the spectral features. Another possibility, which has gained some interest recently, is again to use electromagnetic cascades, since depending on its magnitude, helicity might leave some specific imprints on blazar pair haloes~\cite{PhysRevD.87.123527,Tashiro:2013ita,Tashiro:2014gfa,Long:2015bda,Chen11072015,AlvesBatista:2016urk,Duplessis:2017rde}.

Following the considerations above, it is reasonable to assume that UHECRs, too, may be used to determine the helicity content of IGMFs. However, up to the present day, to the best of our knowledge, there has been only one work~\cite{Kahniashvili:2005yp} addressing this possibility. The authors of the aforementioned work claim that for specific configurations of sources imprints of the helicity of the intervening field may be found, and propose a method to obtain this information. 

Conversely, helical magnetic fields may significantly affect the propagation of UHECRs leaving specific signatures on their spectrum, composition, or arrival distributions. A detailed understanding of magnetic fields is essential when building phenomenological models to interpret the measurements and, most importantly, when attempting to use UHECRs for particle astronomy. A number of works~\cite{Hackstein:2016pwa,AlvesBatista:2017vob,Hackstein:2017pex} has discussed the effects of extragalactic magnetic fields on the propagation of UHECRs, though with conflicting conclusions due to the different assumptions and owing to our lack of knowledge about magnetic field distributions in the Universe. Nevertheless, even in extreme scenarios in which voids are highly magnetised ($B \sim 1 \; \text{nG}$), UHE proton astronomy may be possible in most of the sky for typical magnetic field configurations~\cite{AlvesBatista:2017vob}.    

In this paper we study the effect of helical magnetic fields on the propagation of UHECRs, and show that such particles may be used to constrain the helicity of IGMFs.
The paper is structured as follows: in Sec.~\ref{sec:HelSources} we present an analytical treatment of the influence of magnetic helicity on the propagation of UHECRs; in Sec.~\ref{sec:Simulations} we describe the Monte Carlo simulations of UHECR propagation, and derive constraints on the magnetic helicity in Sec.~\ref{sec:Constraints}, before discussing the results in Sec.~\ref{sec:discussion}; finally, in Sec.~\ref{sec:CO} we draw our conclusions and present the prospects for detecting IGMFs using the presented method or a derivation thereof.
\section{Analytical Treatment} \label{sec:HelSources}

We study a simple case that can be treated analytically. To this end, we follow Ref.~\cite{Long:2015bda} for the magnetic field parametrisation. We consider a single mode of the IGMF that can be written as:
\begin{equation} \label{helB}
\mathbf{B} = B_{0} \sigma \sin\left(\frac{2 \pi}{\lambda} z + \psi \right) \mathbf{\hat{x}} + B_0 \cos\left(\frac{2 \pi}{\lambda} z + \psi \right) \mathbf{\hat{y}} \, ,
\end{equation}
where $B_{0}$ is the average magnetic field, $\psi$ is a arbitrary phase, $\lambda$ is the coherence length and $\sigma = -1,0,1$ corresponds to negative, zero and positive helicity, respectively. It should be noted that by defining the magnetic field in this way, we ensure that for the case $\psi = 0$, used in the following, the magnetic field points in the same direction (along the $y$-axis) at the position of the observer, i.e.~at the origin $(0,0,0)$, for all three cases, hence guaranteeing their comparability. 

The magnetic field in Eq.~\ref{helB} can be obtained by taking the curl of the magnetic vector potential ($\mathbf{A}$), which is therefore given by:
\begin{equation}
\mathbf{A} = B_{0} \frac{\lambda}{2 \pi} \sin\left(\frac{2 \pi}{\lambda} z + \psi \right) \mathbf{\hat{x}} + B_{0} \frac{\lambda}{2 \pi} \sigma \cos \left(\frac{2 \pi}{\lambda} z + \psi \right) \mathbf{\hat{y}} \,.
\end{equation}
Plugging it into the integrand of Eq.~\ref{HelDef}, we obtain
\begin{equation}
\mathbf{A} \cdot \mathbf{B} = \sigma \frac{\lambda}{2 \pi} B_{0}^{2}\,,
\end{equation}
which confirms that the sign of $\sigma$ corresponds to the sign of helicity.

\begin{figure}
\center
\includegraphics[width=0.495\columnwidth]{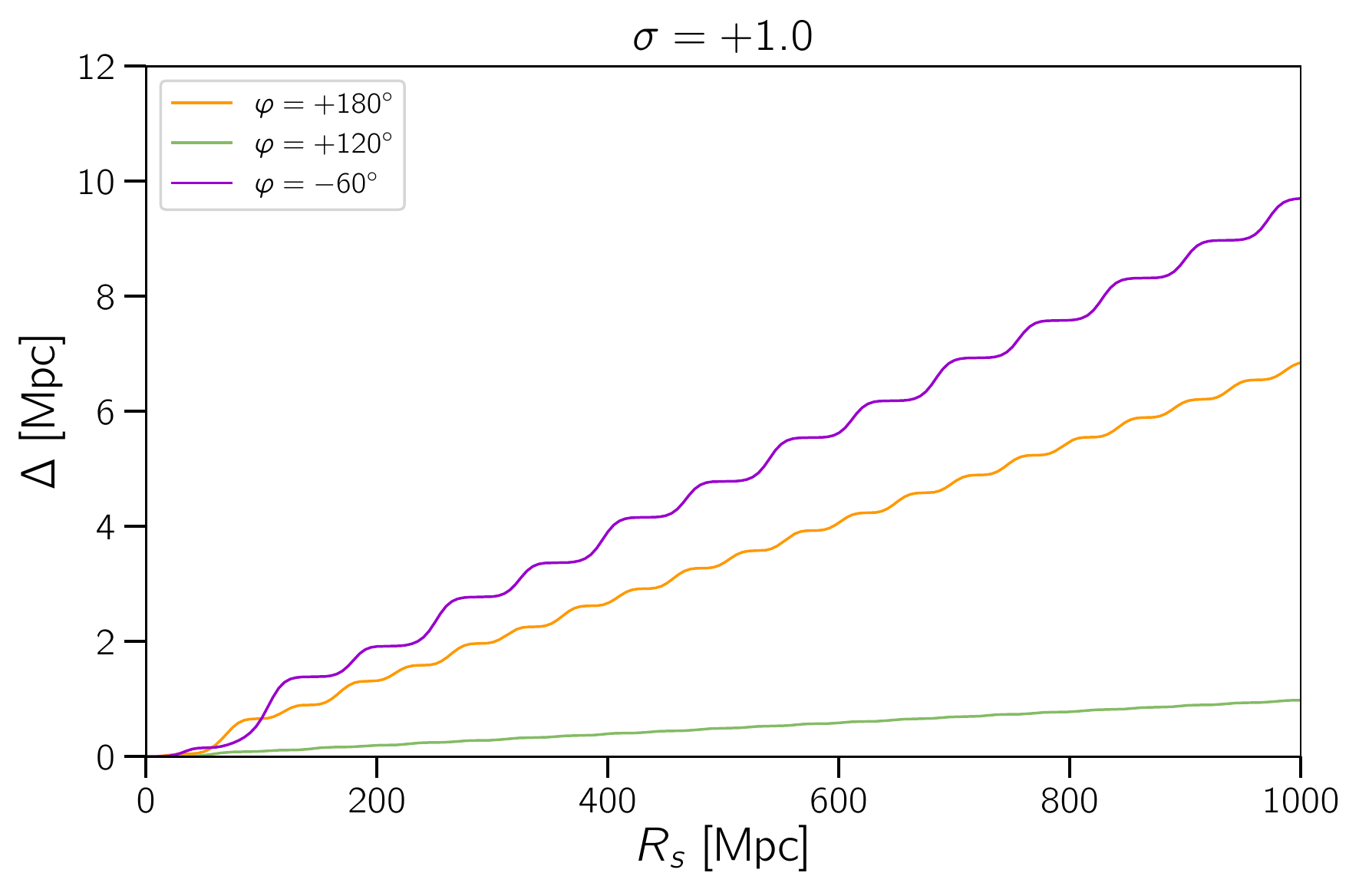}
\includegraphics[width=0.495\columnwidth]{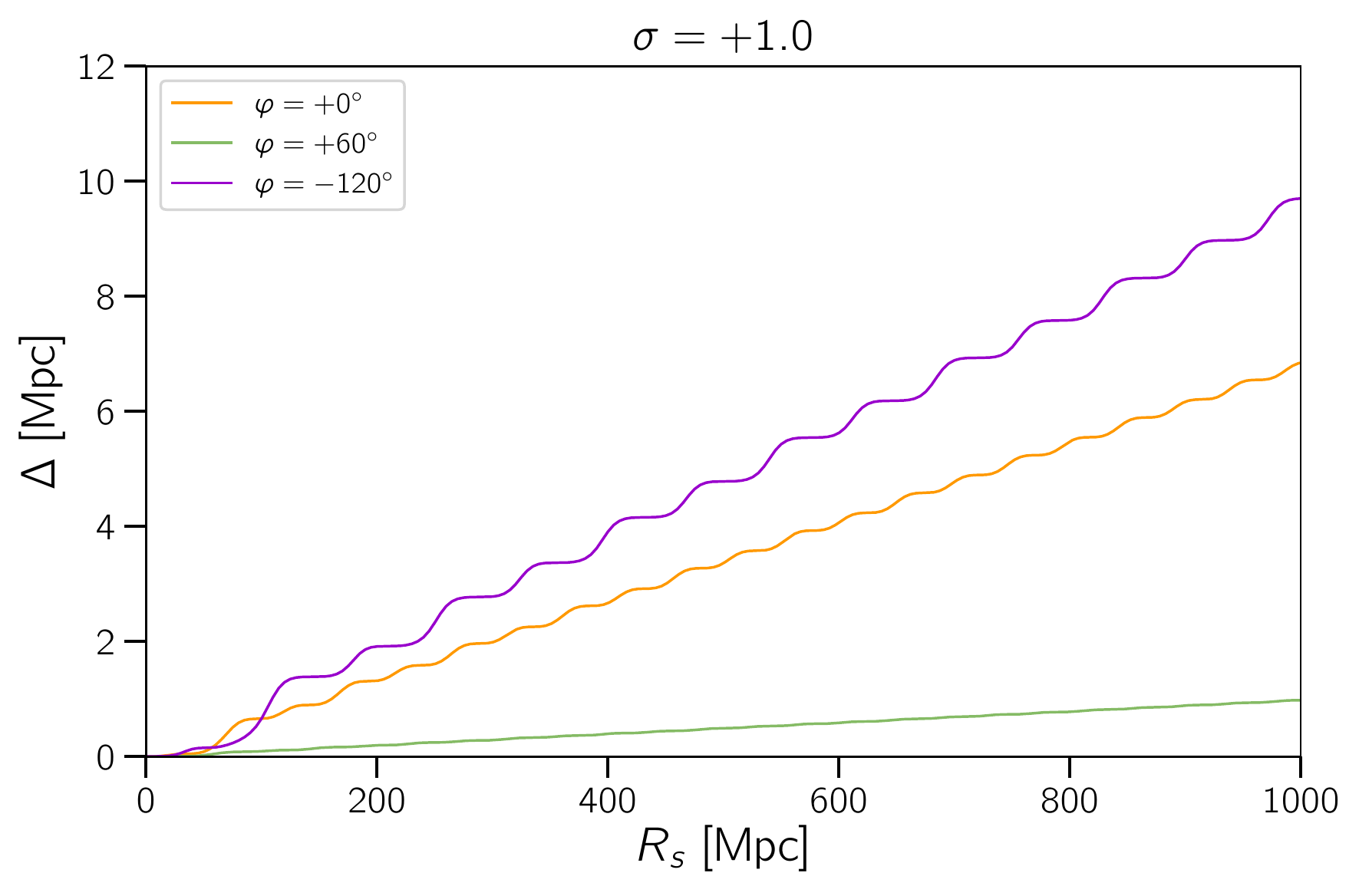}
\includegraphics[width=0.495\columnwidth]{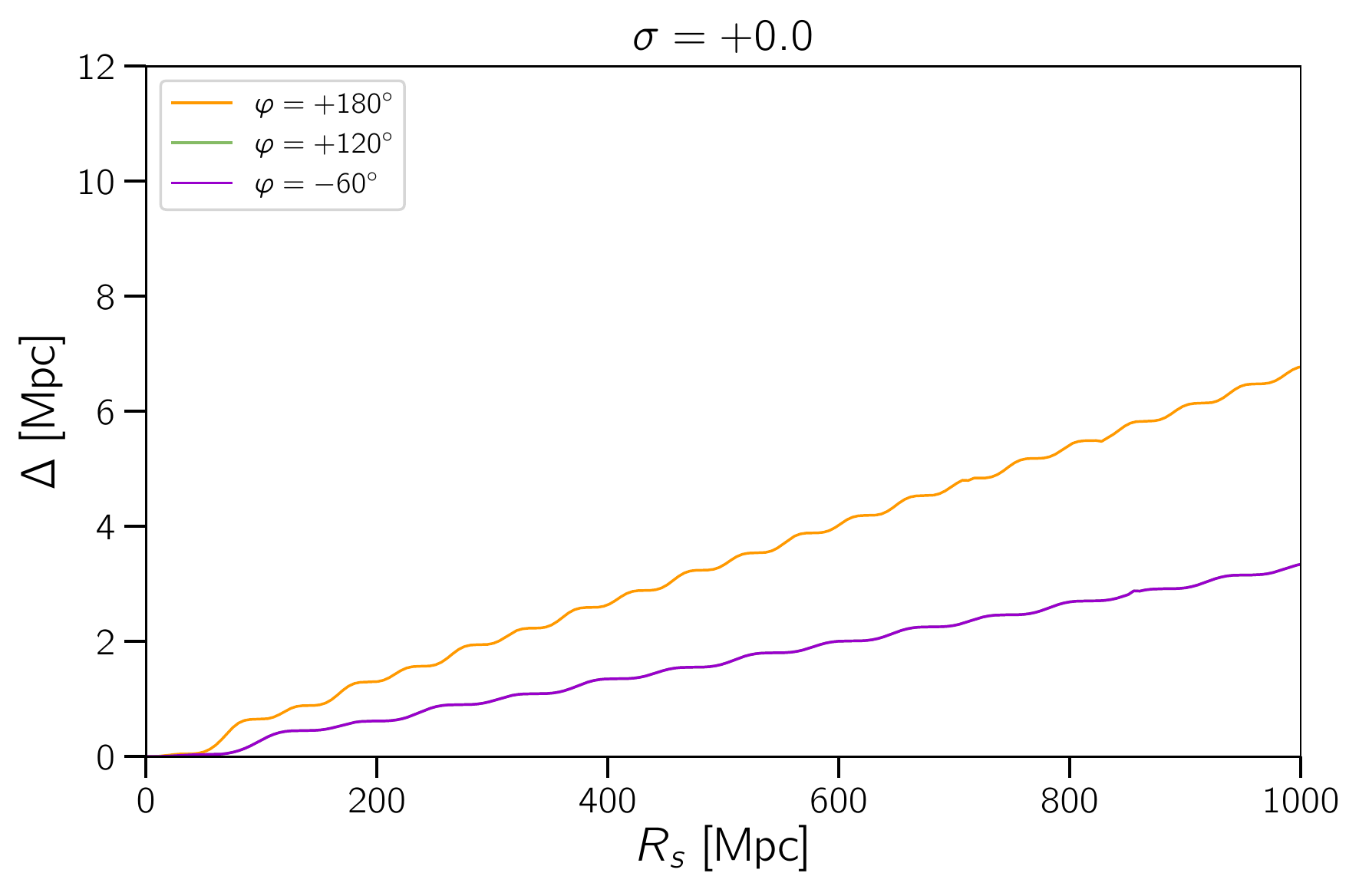}
\includegraphics[width=0.495\columnwidth]{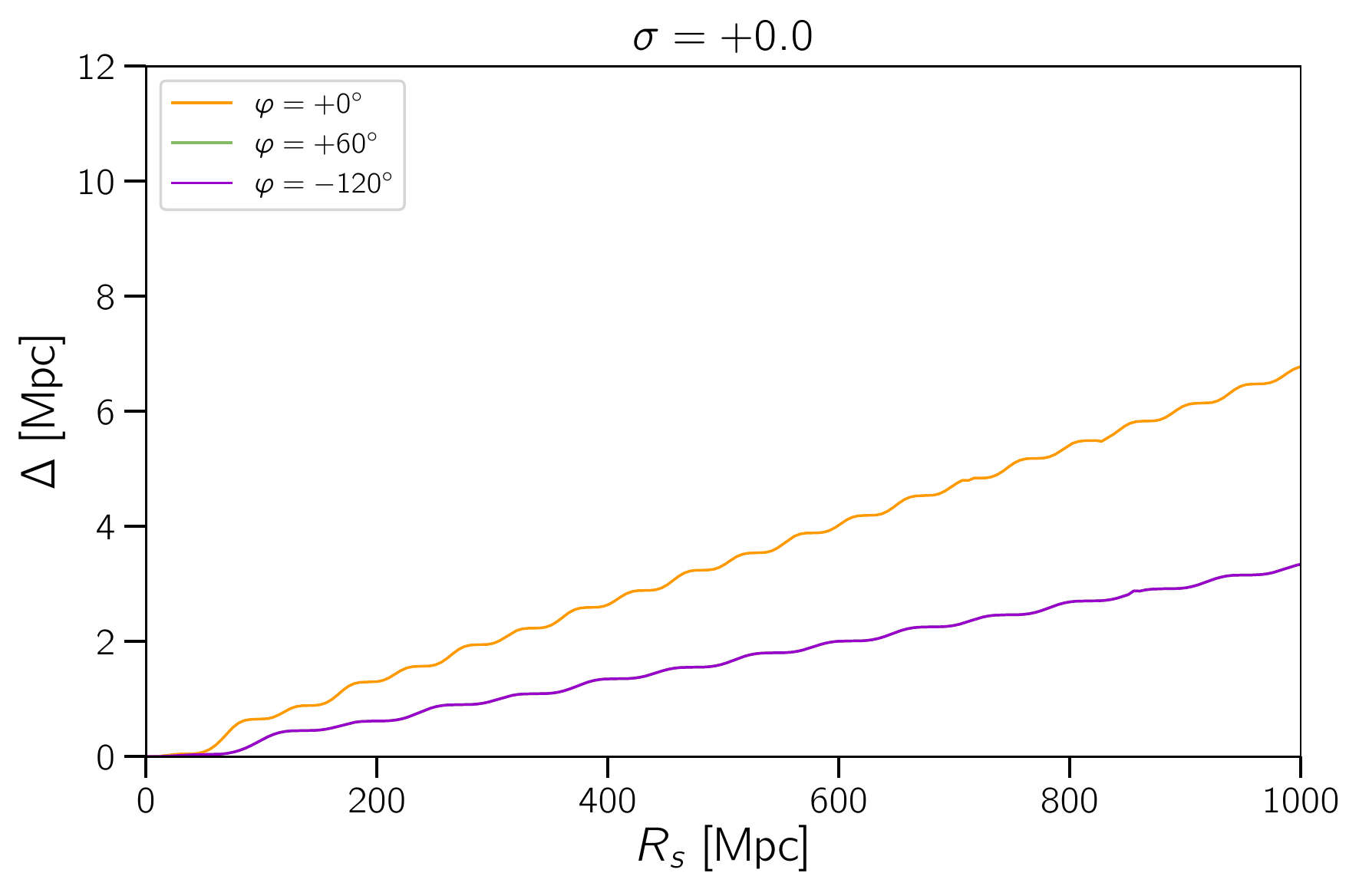}
\includegraphics[width=0.495\columnwidth]{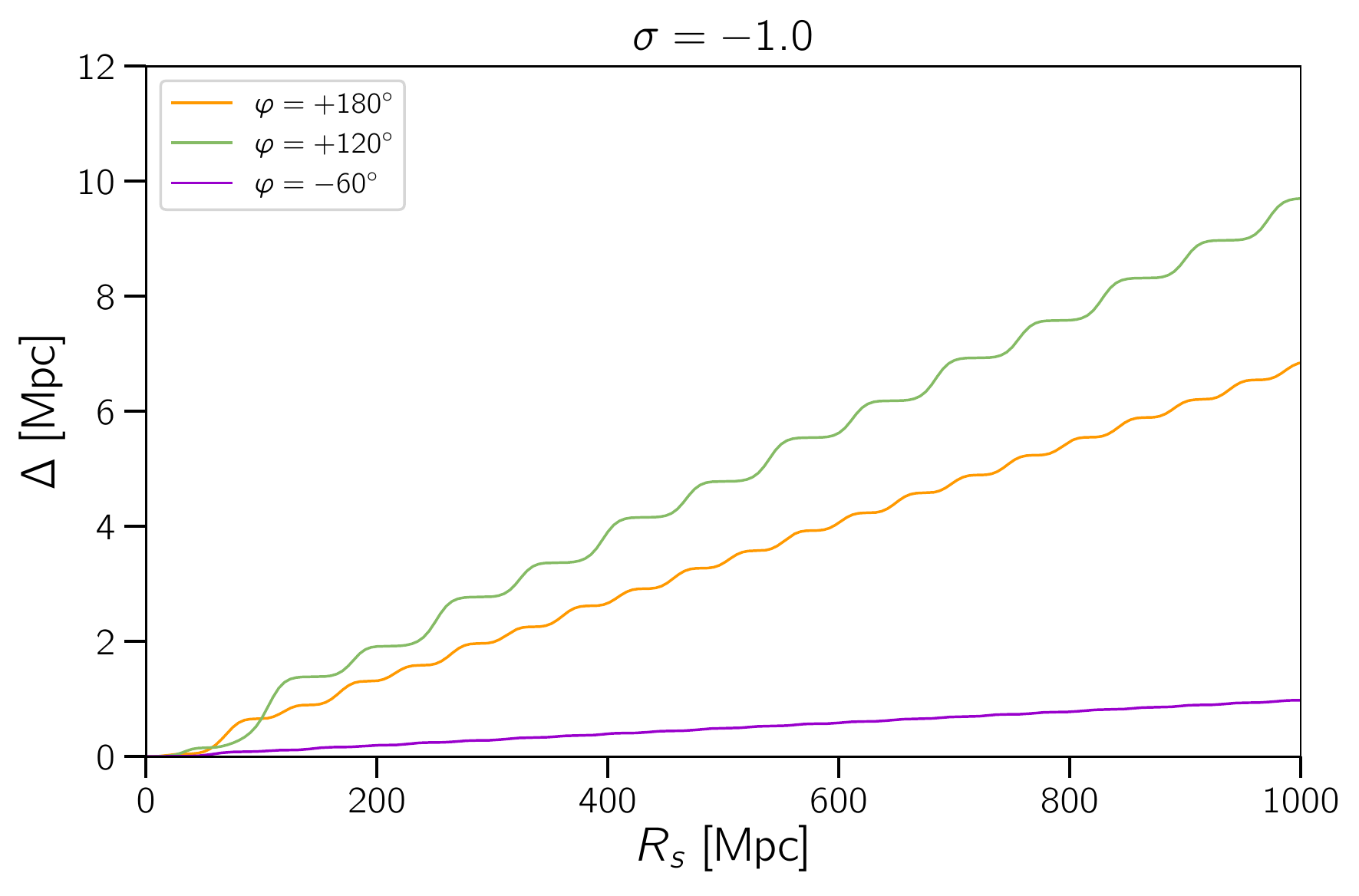}
\includegraphics[width=0.495\columnwidth]{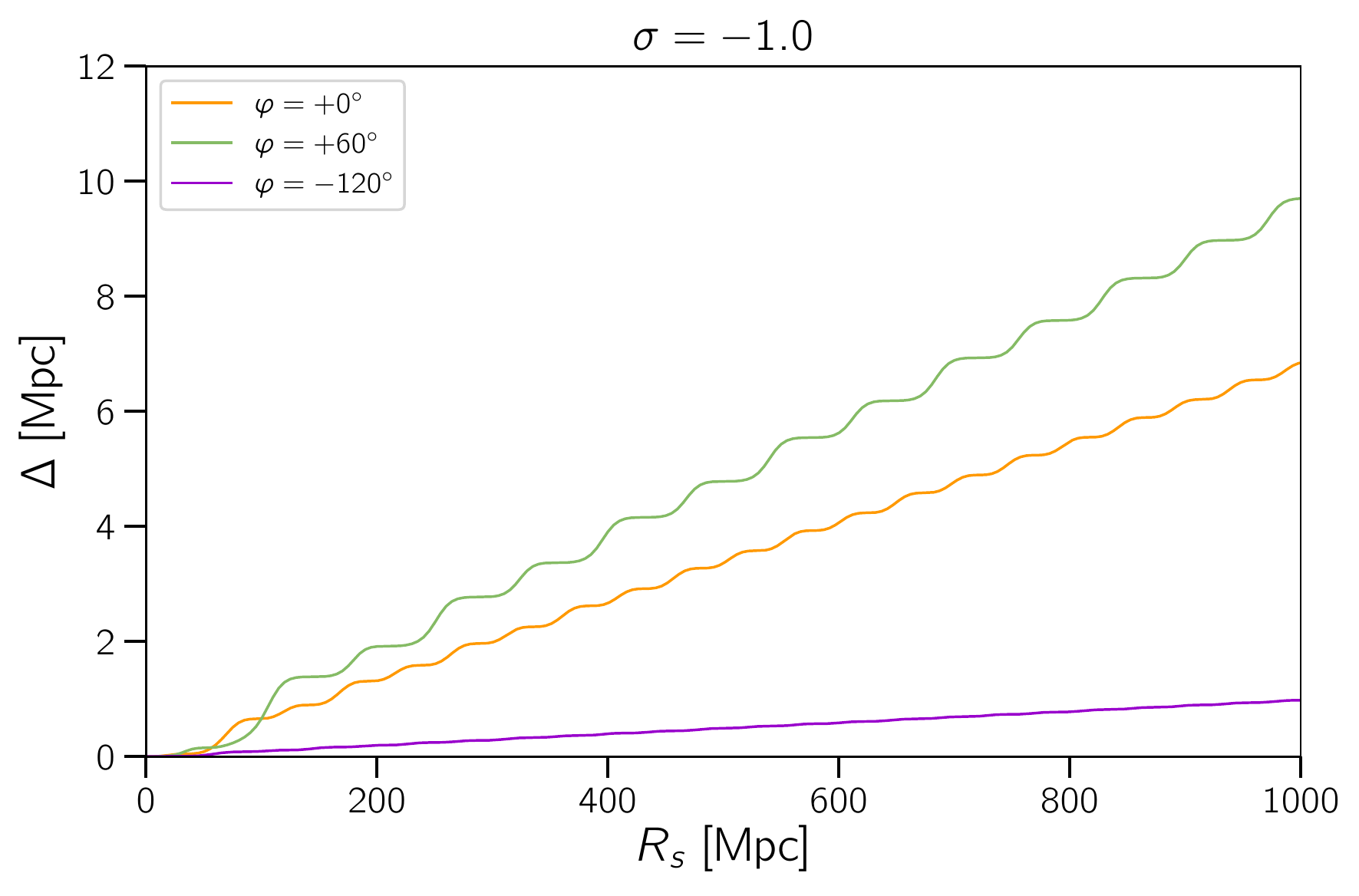}
\caption{Elongation of the trajectory ($\Delta$) of protons as a function of the source distance ($R_{\rm s}$) for $\sigma = -1$ (upper), $\sigma = 0$ (middle) and $\sigma = +1$ (lower panels). Note that for $\sigma = 0$ (middle panels) the cases $\varphi = -60^{\circ}$ and $\varphi = 120^{\circ}$ (left) are the same, which is also true for $\varphi = 60^{\circ}$ and $\varphi = -120^{\circ}$ (right). Hence, only one line for each equal pair is visible due to a complete overlap. In addition, also for $\sigma = 0$, there is a symmetry $\varphi \rightarrow -\varphi$, thus explaining why the curves for angles of the same magnitude, but different signs, are the same, respectively. The parameters used here are $\lambda = 10 \; {\rm Mpc}$, $E = 10^{20}\;{\rm eV}$, $B_{0} = 10^{-9}\,{\rm G}$ and $\mathbf{v_{0}} = v_{0} (\cos\varphi \, \mathbf{\hat{x}} + \sin\varphi \, \mathbf{\hat{y}})$.}
\label{fig:RsDelta}
\end{figure}

Now consider an UHECR with energy $E$, rest mass $m_{0}$ and charge $Z e$, where $Z$ is its atomic number. Its Lorentz factor is
\begin{equation}
\gamma = \frac{E}{m_{0} c^{2}}.
\end{equation} 
Hence the velocity of the cosmic ray can be written as
\begin{equation}
v_{0} = \sqrt{1 - \left(\frac{m_{0} c^{2}}{E} \right)^{2}} \, c\,.
\end{equation}

The equation of motion of a charged particle in an arbitrary magnetic field is given by the Lorentz force:
\begin{equation} \label{LorentzForce}
\mathbf{F} = \frac{{\rm d} \mathbf{p}}{{\rm d} t} = \frac{{\rm d} ( \gamma m_{0} \mathbf{v})}{{\rm d} t} = q \mathbf{v} \times \mathbf{B}.
\end{equation}
Assuming that the only interaction of the particle is with the magnetic field, its energy and hence Lorentz factor are both constant, such that it can be pulled out of the time derivative. Using $\mathbf{v} = \partial_{t} \mathbf{x}$, $\partial_{t}\mathbf{v} = \partial_{t}^{2}\mathbf{x}$ and Eq.~\ref{helB}, the Lorentz force equation (\ref{LorentzForce}) becomes
\begin{align}
\gamma m_{0} \partial_{t}^{2}x &= - q B_{0} \cos\left(\frac{2 \pi}{\lambda} z + \psi \right) \partial_{t}z\,, \label{DiffEq1} \\
\gamma m_{0} \partial_{t}^{2}y &= \sigma q B_{0} \sin\left(\frac{2 \pi}{\lambda} z + \psi \right) \partial_{t}z\,, \label{DiffEq2} \\
\gamma m_{0} \partial_{t}^{2}z &= q \left[ B_{0} \cos\left(\frac{2 \pi}{\lambda} z + \psi \right) \partial_{t}x - \sigma B_{0} \sin\left(\frac{2 \pi}{\lambda} z + \psi \right) \partial_{t}y \right]\,. \label{DiffEq3}
\end{align}
With the initial conditions $x(0) = x_{0}$, $y(0) = y_{0}$, $z(0) = z_{0}$, $\partial_{t}x(0) = v_{x0}$, $\partial_{t}y(0) = v_{y0}$, $\partial_{t}z(0) = v_{z0}$, Eqs. \ref{DiffEq1} and \ref{DiffEq2} can be written as
\begin{align}
\partial_{t}x(t) = v_{x0} + \frac{Z e B_{0} c^{2}}{E} \frac{\lambda}{2 \pi} \left[ \sin\left(\frac{2 \pi}{\lambda} z_{0} + \psi \right) - \sin\left(\frac{2 \pi}{\lambda} z(t) + \psi \right) \right]\,, \label{DiffEq1a} \\
\partial_{t}y(t) = v_{y0} + \sigma \frac{Z e B_{0} c^{2}}{E} \frac{\lambda}{2 \pi} \left[ \cos\left(\frac{2 \pi}{\lambda} z_{0} + \psi \right) - \cos\left(\frac{2 \pi}{\lambda} z(t) + \psi \right) \right]\,,
\end{align}
using which we obtain
\begin{equation} \label{DiffEq3a}
\begin{split}
\partial_{t}^{2}z(t) = \frac{Z e B_{0} c^{2}}{E} \Bigg\{ v_{x0} \cos\left(\frac{2 \pi}{\lambda} z(t) + \psi \right) - \sigma v_{y0}\sin\left(\frac{2 \pi}{\lambda} z(t) + \psi \right) \\
+ \frac{Z e B_{0} c^{2}}{E} \frac{\lambda}{2 \pi} \sin\left[ \frac{2 \pi}{\lambda}(z_{0} - z(t)) \right]\Bigg\}
\end{split}
\end{equation}
for Eq.~\ref{DiffEq3}. Eq.~\ref{DiffEq3a} is a differential equation that depends solely on $z(t)$, with the explicit dependence on $x(t)$ and $y(t)$ eliminated.

We define $t=0$ as the time when the particle with the velocity vector $\mathbf{v_{0}} = v_{x0} \, \mathbf{\hat{x}} + v_{y0} \, \mathbf{\hat{y}} + v_{z0} \, \mathbf{\hat{z}}$ arrives at an observer located at $(x_{0}, y_{0}, z_{0}) = (0, 0, 0)$. We also set $\psi = 0$, such that Eqs.~\ref{DiffEq1a}-\ref{DiffEq3a} reduce to
\begin{align}
\partial_{t}x(t) &= v_{x0} - \frac{Z e B_{0} c^{2}}{E} \frac{\lambda}{2 \pi} \sin\left[\frac{2 \pi}{\lambda} z(t) \right]\,, \label{DiffEq1b}\\
\partial_{t}y(t) &= v_{y0} + \sigma \frac{Z e B_{0} c^{2}}{E} \frac{\lambda}{2 \pi} \left\{ 1 - \cos\left[\frac{2 \pi}{\lambda} z(t) \right] \right\}\,, \label{DiffEq2b} \\
\partial_{t}^{2}z(t) &= \frac{Z e B_{0} c^{2}}{E} \Bigg\{ v_{x0} \cos\left[\frac{2 \pi}{\lambda} z(t) \right] - \sigma v_{y0}\sin\left[\frac{2 \pi}{\lambda} z(t) \right] - \frac{Z e B_{0} c^{2}}{E} \frac{\lambda}{2 \pi} \sin\left[ \frac{2 \pi}{\lambda}z(t) \right]\Bigg\}\,. \label{DiffEq3b}
\end{align}

We now assume that the sources are uniformly distributed on the surface of a sphere with radius $R_{\rm s}$, centred around the origin (i.e. around the observer). We compute the trajectories of the particles by numerically solving Eq.~\ref{DiffEq3b} for $z(t)$, which is then plugged into \ref{DiffEq1b} and \ref{DiffEq2b} to obtain $x(t)$ and $y(t)$ through a simple integration. Now, for a given arrival velocity $\mathbf{v_{0}}$, we calculate a value of $t$ ($t<0$), for which $x^{2}(t) + y^{2}(t) + z^{2}(t) = R_{\rm s}^{2}$, thus obtaining the position whence the cosmic ray was emitted.

The cosmic-ray trajectory depends upon $\mathbf{v_{0}}$ and the source distance $R_{\rm s}$, which are related to the elongation ($\Delta$) of the trajectory length of the particle, given by
\begin{equation}
\Delta(R_{\rm s}) = \left[v_{0} t - \sqrt{x^{2}(t) + y^{2}(t) + z^{2}(t)})\right]_{t=\max\{t|x^{2}(t) + y^{2}(t) + z^{2}(t) = R_{\rm s} \wedge t<0\}}\,,
\label{eq:Delta_Rs}
\end{equation}
which, for a given source distance $R_{\rm s}$, gives the difference between the actual length of the trajectory described by the particle and the corresponding distance in the absence of intervening magnetic fields.

This quantity, in the case of protons, is shown in Fig.~\ref{fig:RsDelta} for $\lambda = 10\,{\rm Mpc}$, $E = 10^{20}\,{\rm eV}$, and $B_{0} = 10^{-9}\,{\rm G}$ for different directions of $\mathbf{v_{0}}$ with $v_{z0} = 0$, parametrised by the angle $\varphi$: $\mathbf{v_{0}} = v_{0}(\cos\varphi, \sin\varphi,0)^{\rm T}$.
As one can see, while for the case with zero helicity ($\sigma = 0$) $\Delta$ is symmetric with respect to both the $x$ and the $y$ axes, this is not true for the $\sigma = -1,1$ cases, where only a symmetry with respect to the $y$-axis is present. Therefore, this may be used to distinguish the zero and the non-zero helicity cases. In addition, if the orientation of the magnetic field at the position of the observer is known, it is even possible to break the degeneracy and distinguish the $\sigma = -1$ and $\sigma = +1$ cases. In the same figure one can also see that for some values of $\varphi$ the corresponding curves show a oscillatory behaviour, e.g.~for $\sigma = +1$ and $\varphi = -60^{\circ}$. This is due to the fact that the corresponding particle trajectory is sinusoidal-like and therefore close to being two-dimensional, while the non-oscillatory curves correspond to helical trajectories.

The most important question is how UHECR measurements may be used to obtain information regarding $\sigma$. The treatment we have presented in this section is only approximate, insofar as it does not take energy losses into account. Since ${\rm d}E/{\rm d}x < 0$, i.e.~the energy of the particle decreases monotonically with the propagated distance, for two particles starting off with the same energy, the observed energy decreases with $\Delta$. Moreover, because for a fixed $R_{\rm s}$ the value of $\Delta$ depends on $\varphi$, the arrival energy also depends on $\varphi$. As a consequence, the spatial distribution of observed energies, $E_{\rm arrival}(\varphi)$, can be used to constrain the helicity of both IGMFs and magnetic fields in structures.

\section{Numerical Treatment} \label{sec:Simulations}

\subsection{Simulation Setup}

The simulations of UHECR propagation in the intergalactic space were performed using CRPropa 3~\cite{Batista:2016yrx}. We have implemented homogeneous helical magnetic fields in the code following Eq.~\ref{helB}. The equations of motion are obtained by solving Eq.~\ref{LorentzForce}, taking into account the main interactions of UHECRs with the cosmic microwave background (CMB) and the extragalactic background light (EBL), namely: the production of electron-positron pairs (Bethe-Heitler process); photopion production; photodisintegration and nuclear decay in the case of cosmic-ray nuclei. Adiabatic losses due to the expansion of the Universe are also taken into account. We have adopted the EBL model by Gilmore {\it et al.}~\cite{Gilmore:2011ks}; note that this choice should not significantly affect the conclusions drawn here.

We have  simulated $N_\text{evt} = 10^{7}$ events for equally luminous sources assumed to be isotropically distributed on the surface of 100 concentric spheres of radii $D$, up to $D_\text{max}$ (i.e.~$10^{5}$ events per sphere, cf.~App.~\ref{sec:AppIso}).  
This represents the distance below which the vast majority of particles arrive with energies $E_\text{obs} > 10^{19} \; \text{eV}$, corresponding to the minimal energy considered in this work for particles arriving at the observer. The value of $D_\text{max}$ depends on the composition of the cosmic rays as well as their typical interaction horizon, which is a function of the energy. We have used $D_\text{max} = 1900 \; \text{Mpc}$ ($z_\text{max} \approx 0.65$). This choice exceeds the energy loss length of the cosmic rays at the energies of interest for all scenarios considered.
With this choice of $N_\text{evt}$, the source distribution on each sphere is approximately isotropic and the anisotropy signal is dominated by the magnetic field rather than the realisation of the source distribution. Moreover, the effective source density is $n_{\rm s} \approx 10^{-3} \; \text{Mpc}^{-3}$, which is consistent with estimates by Auger~\cite{Abreu:2013kif}.

We have assumed that the cosmic rays emitted by the sources have energies $E = E_0$. A cosmic ray is emitted  at $(\mathbf{r_0}, t_0)$, wherein $\mathbf{r_0} \equiv (x_0, y_0, z_0)$ is the initial position vector and $t_0$ the time of emission. Detection occurs when the particle reaches the observer, modelled as a sphere. Additionally, we add a fourth dimension to this problem to account for the cosmological evolution of the Universe. In this case, the observer is no longer a three-dimensional sphere, but a four-dimensional volume which includes time (or conversely redshift). Therefore, only particles arriving at the observer within a given time window are accepted. The use of this four-dimensional setup is justified because adiabatic energy losses are not negligible, and because photon background fields, namely the CMB and EBL, evolve with redshift. For this reason we adopt a redshift window of width $\Delta z = 0.005$ around present time ($z=0$), which corresponds to a time window of $\Delta t \approx 22 \; \text{Mpc}/c$.

The deflection ($\delta$) of a cosmic ray of energy $E$ in a magnetic field of strength $B$ and coherence length $\lambda$ can be approximated by~\cite{DuNe}
\begin{equation}
	\delta \approx 0.5^\circ Z \,\left( \frac{R_{\rm s}}{\text{Mpc}} \right) \left( \frac{B}{\text{nG}} \right)  \left( \frac{E}{\text{EeV}} \right) .
	\label{eq:deflection}
\end{equation}
Note that this expression is approximately valid if energy losses are neglected and if the field is roughly homogeneous over distances $\sim R_s$. Larger deflections correspond to larger elongations of the trajectories ($\Delta$), such that $\delta$ would, to first order, increase with $\Delta$, hence ${\rm d} \delta \propto {\rm d} \Delta$. 

\begin{figure}
\center
\includegraphics[width=0.325\columnwidth]{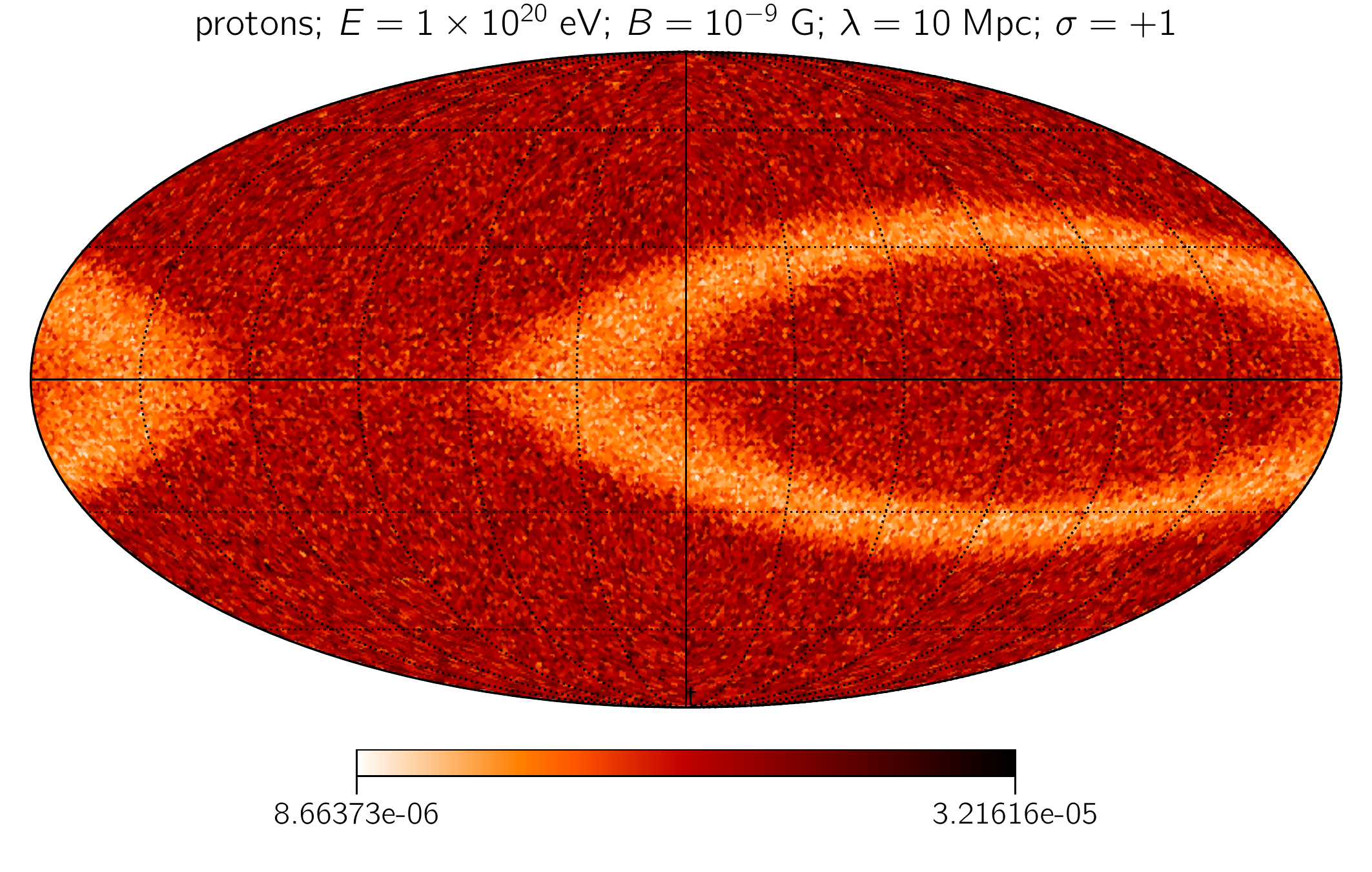}
\includegraphics[width=0.325\columnwidth]{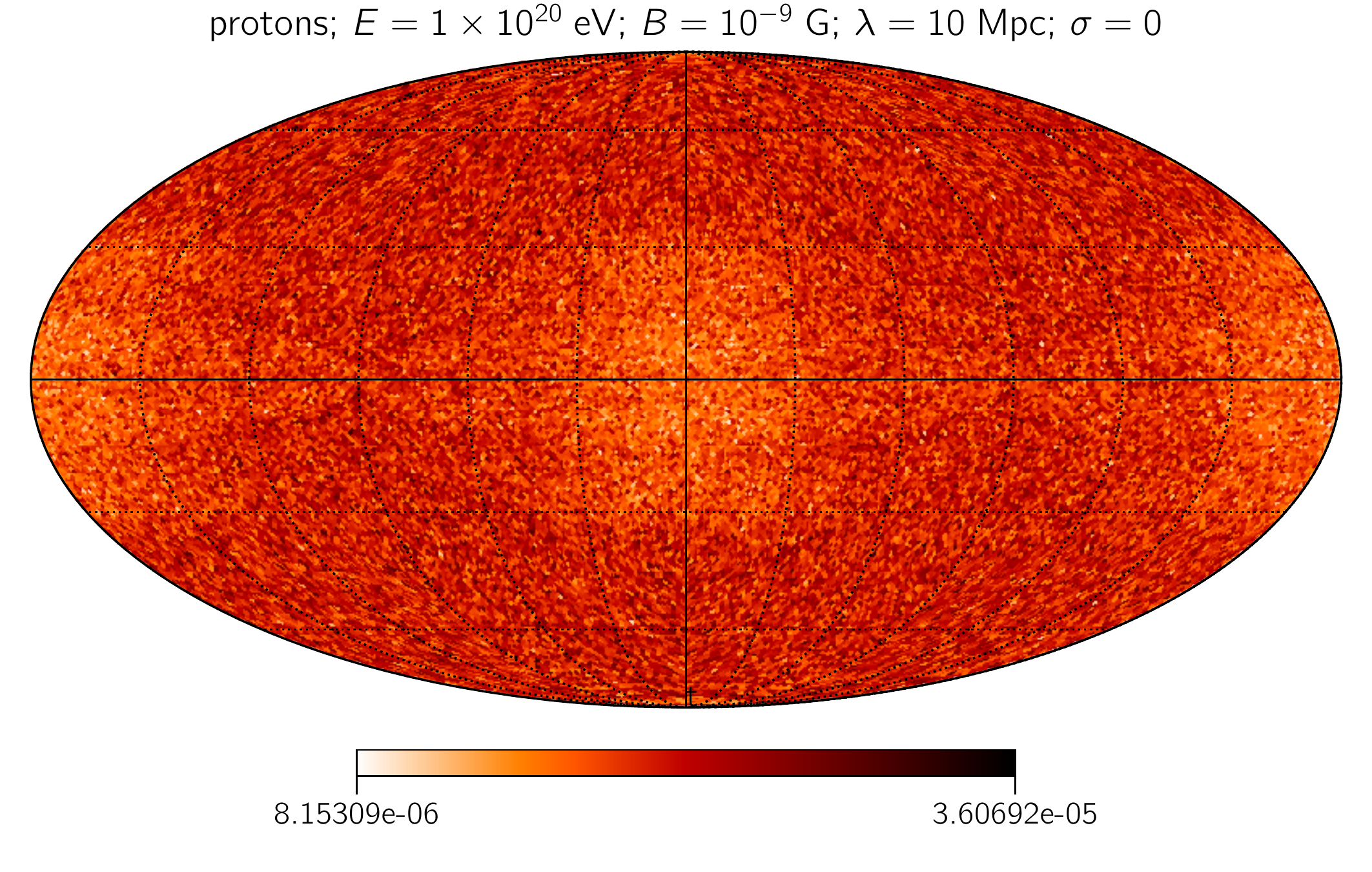}
\includegraphics[width=0.325\columnwidth]{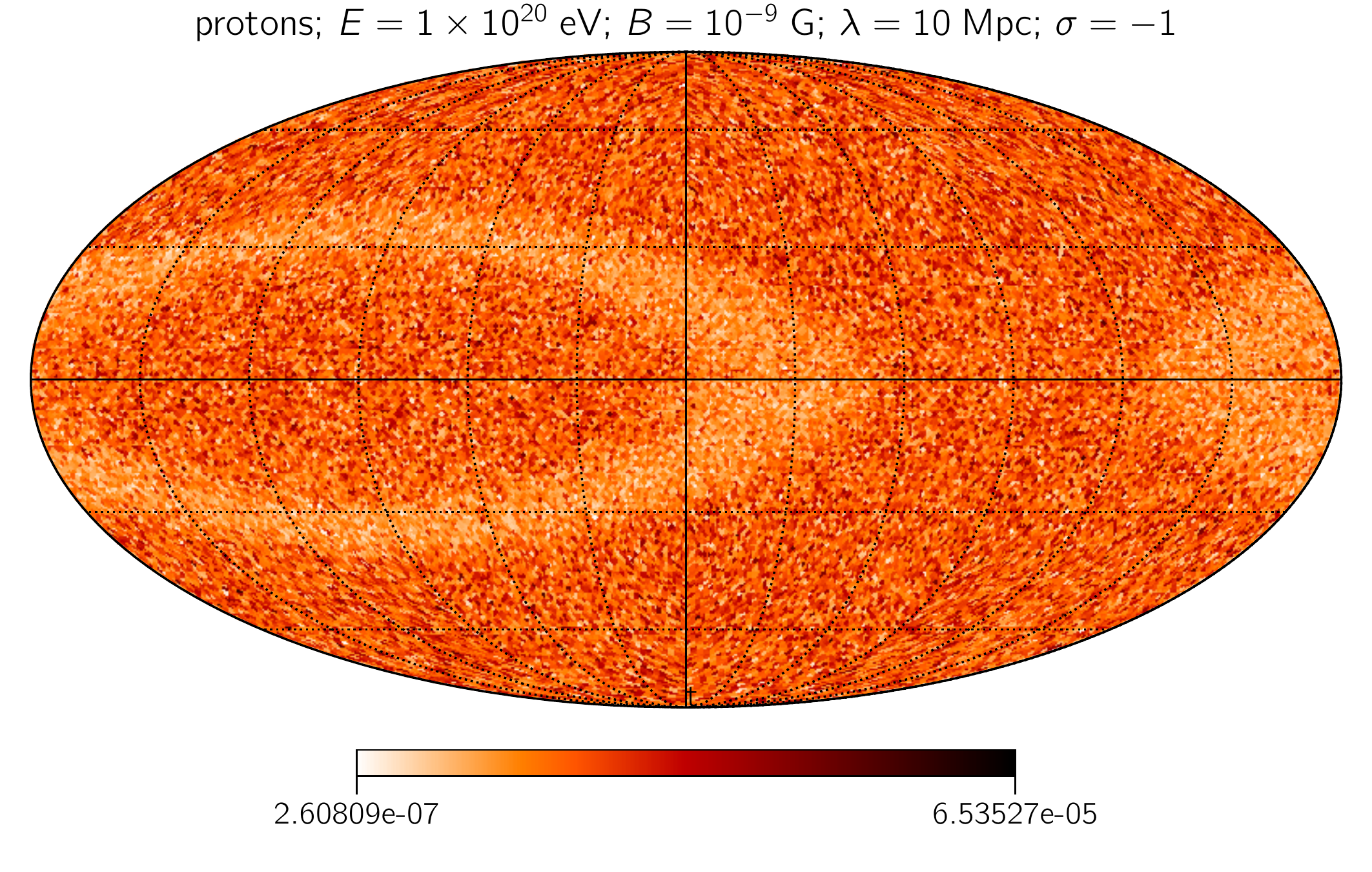}
\includegraphics[width=0.325\columnwidth]{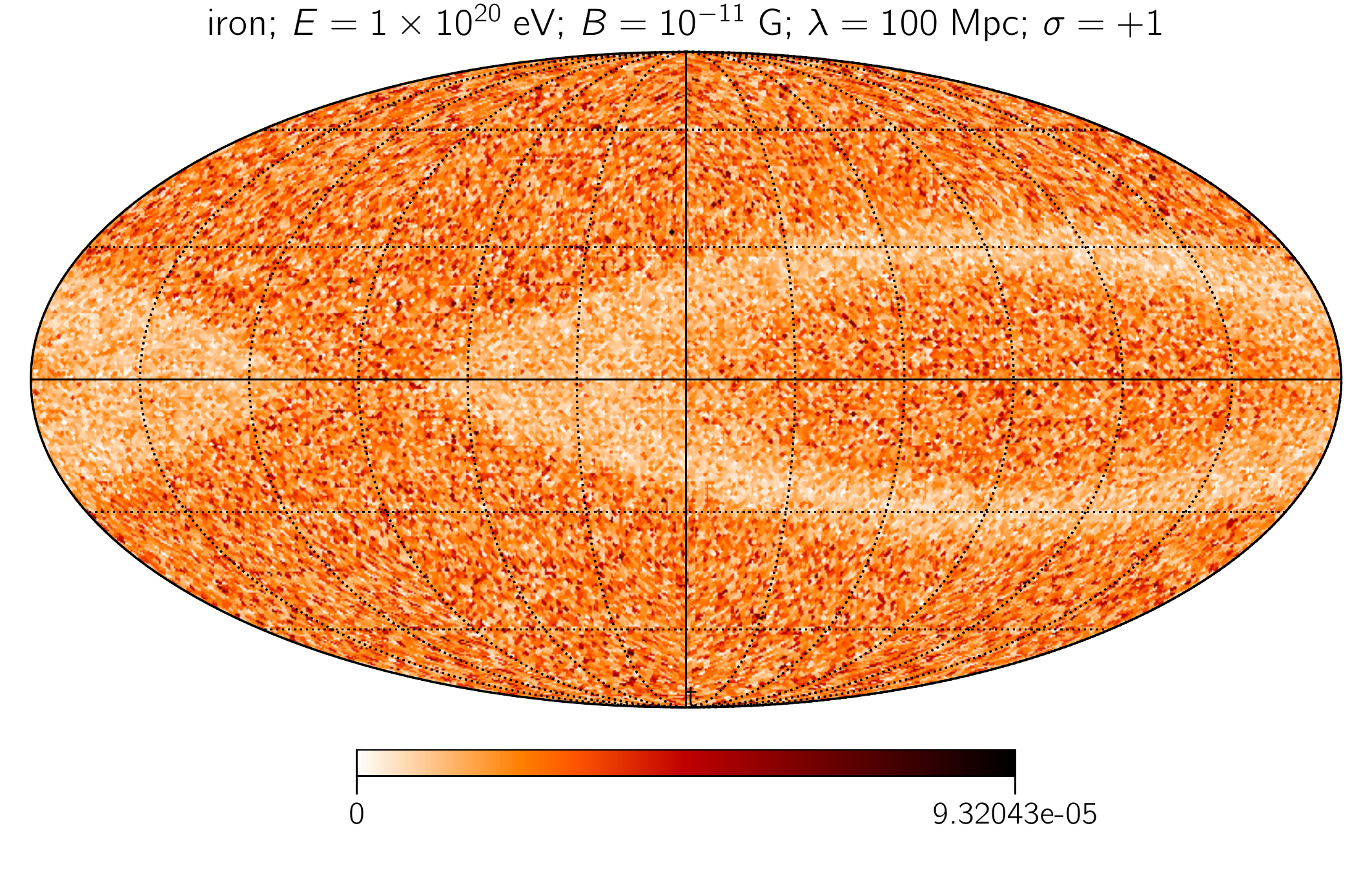}
\includegraphics[width=0.325\columnwidth]{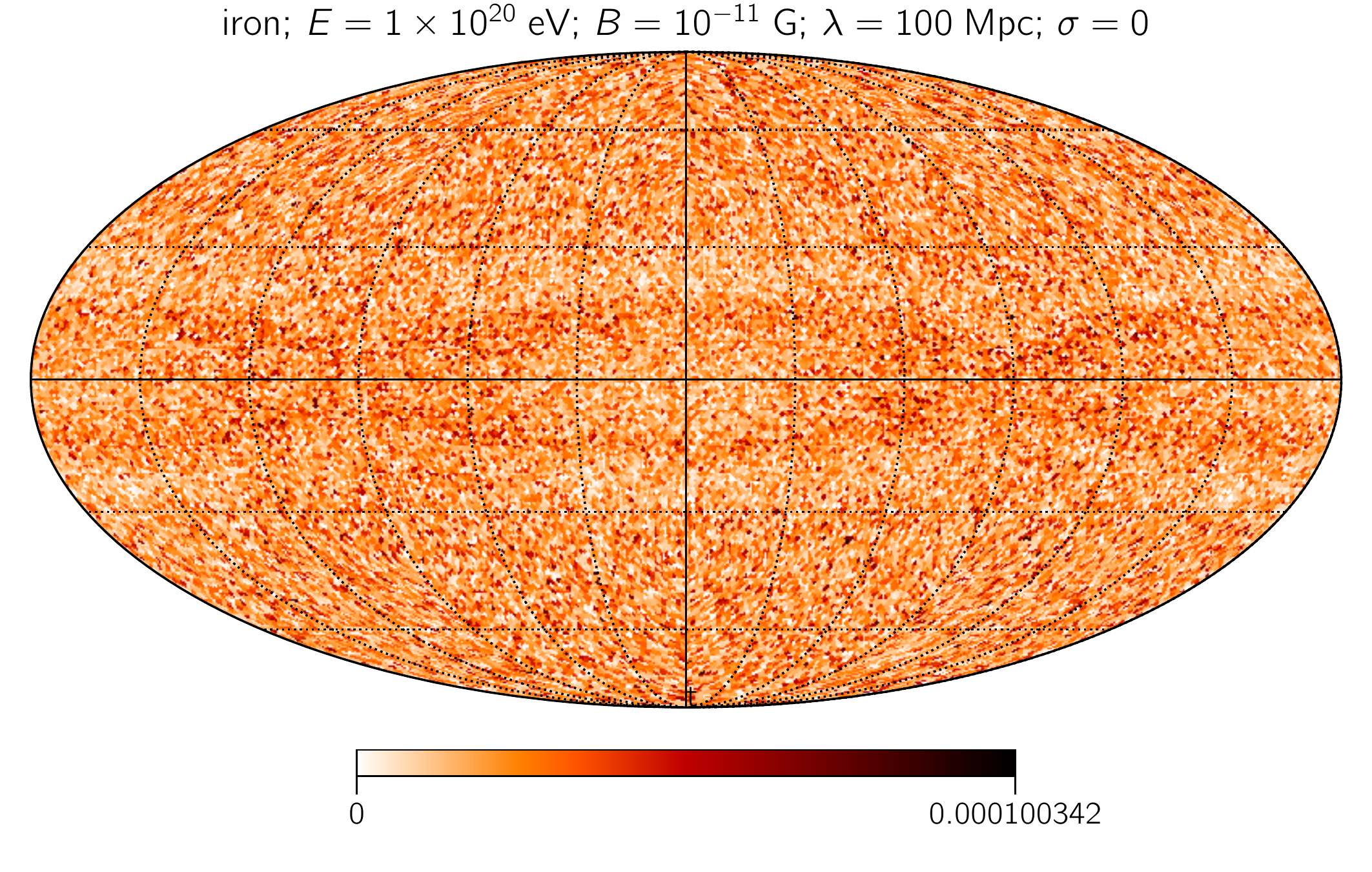}
\includegraphics[width=0.325\columnwidth]{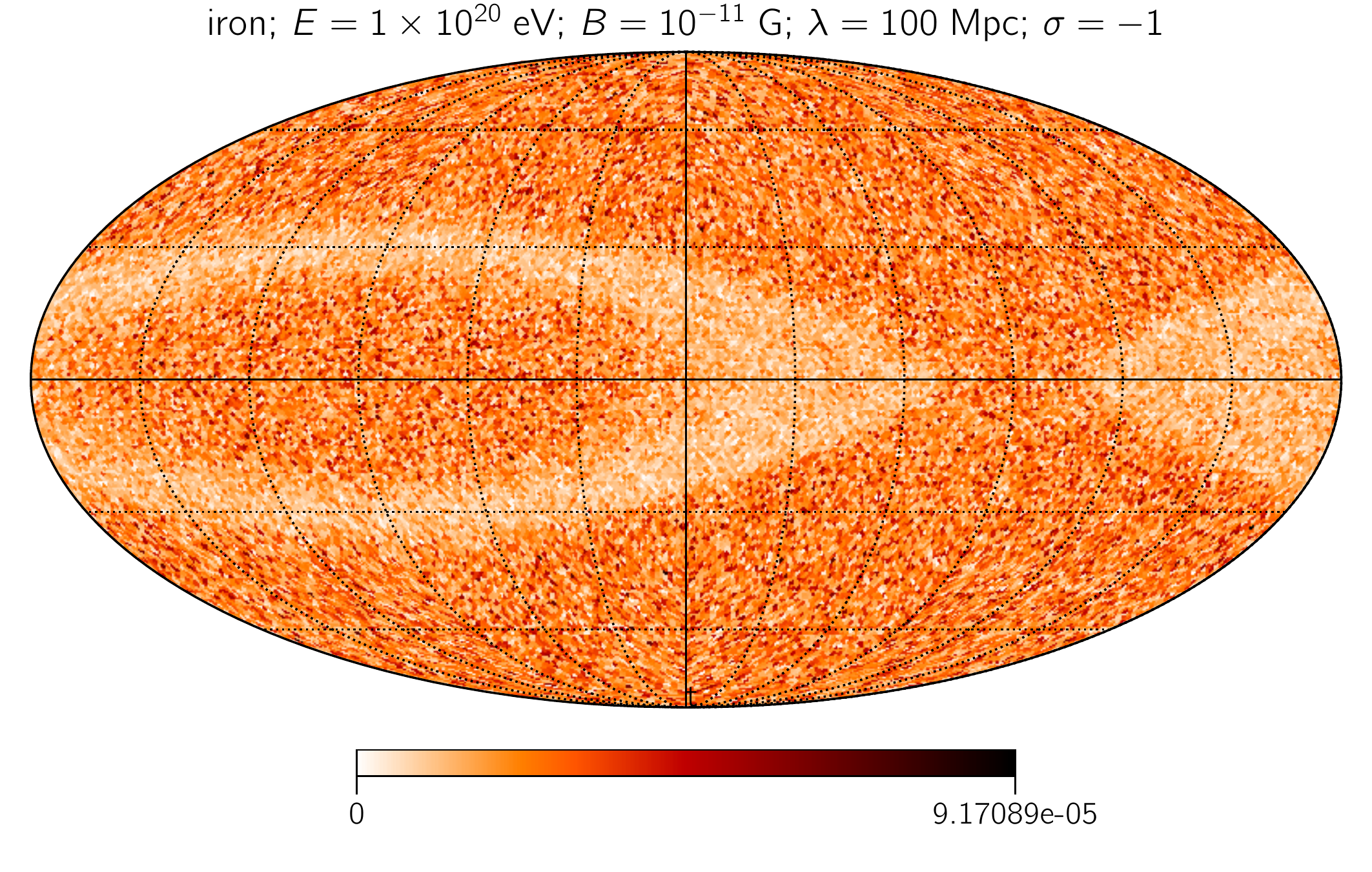}
\includegraphics[width=0.325\columnwidth]{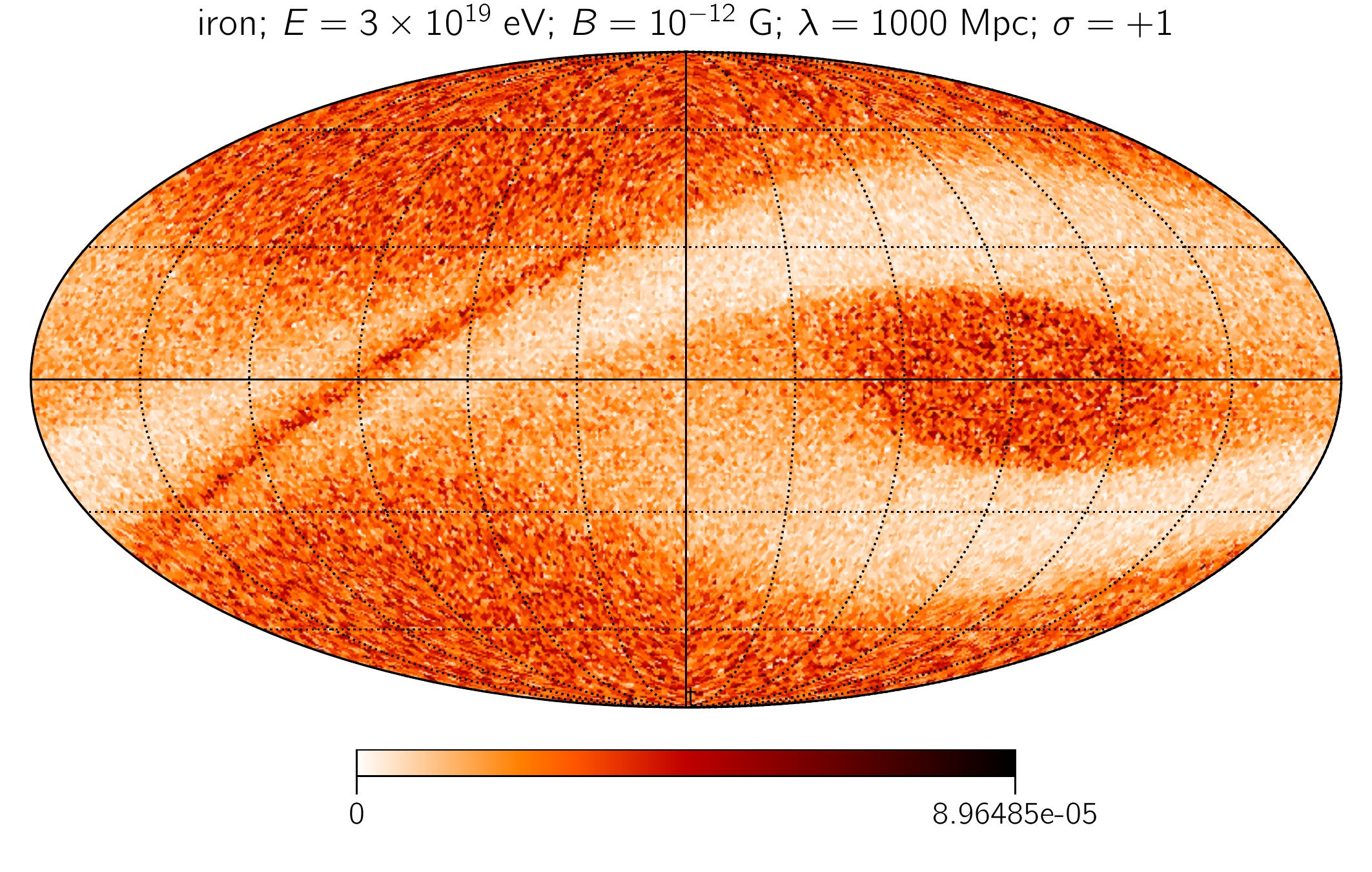}
\includegraphics[width=0.325\columnwidth]{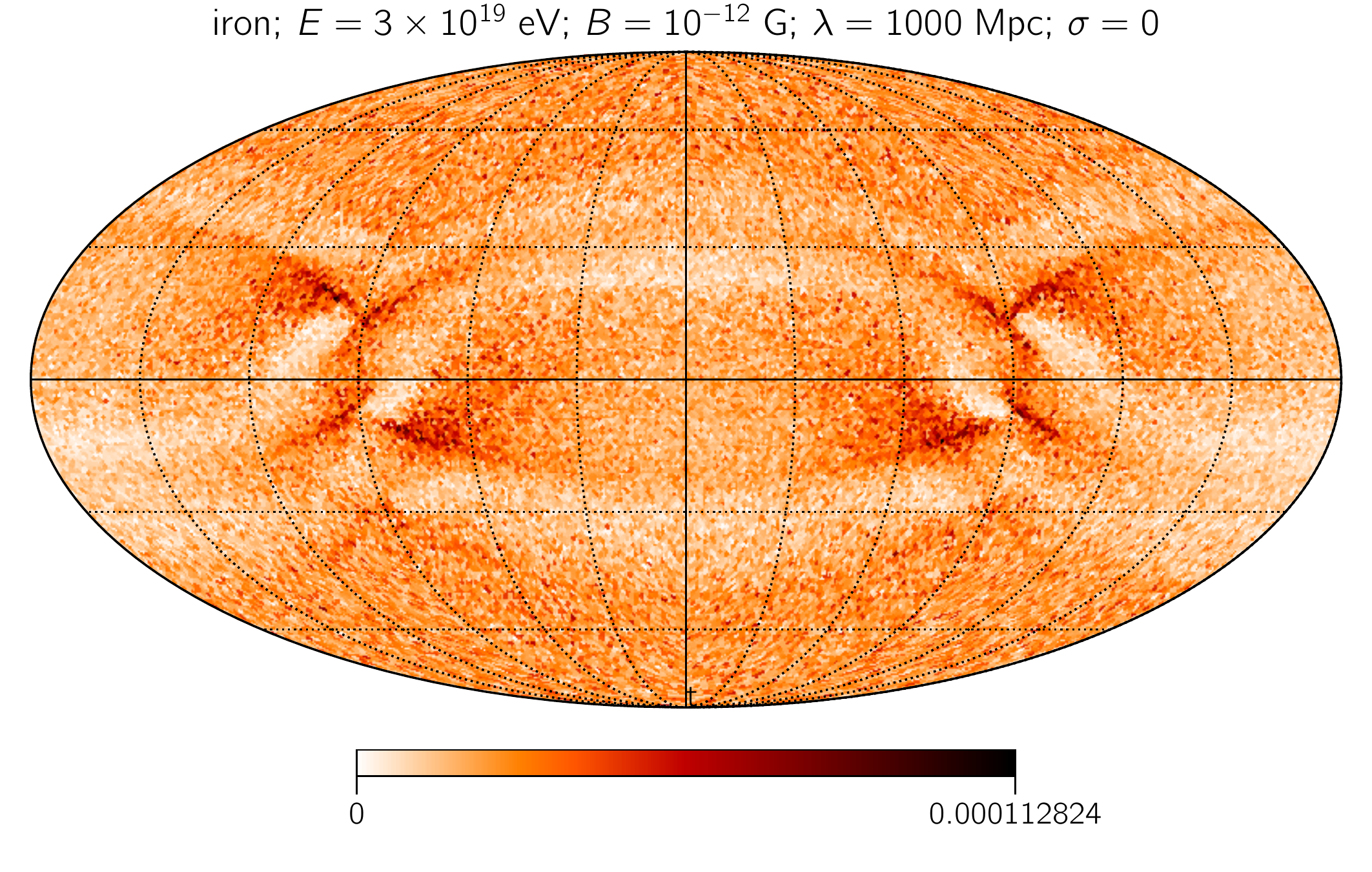}
\includegraphics[width=0.325\columnwidth]{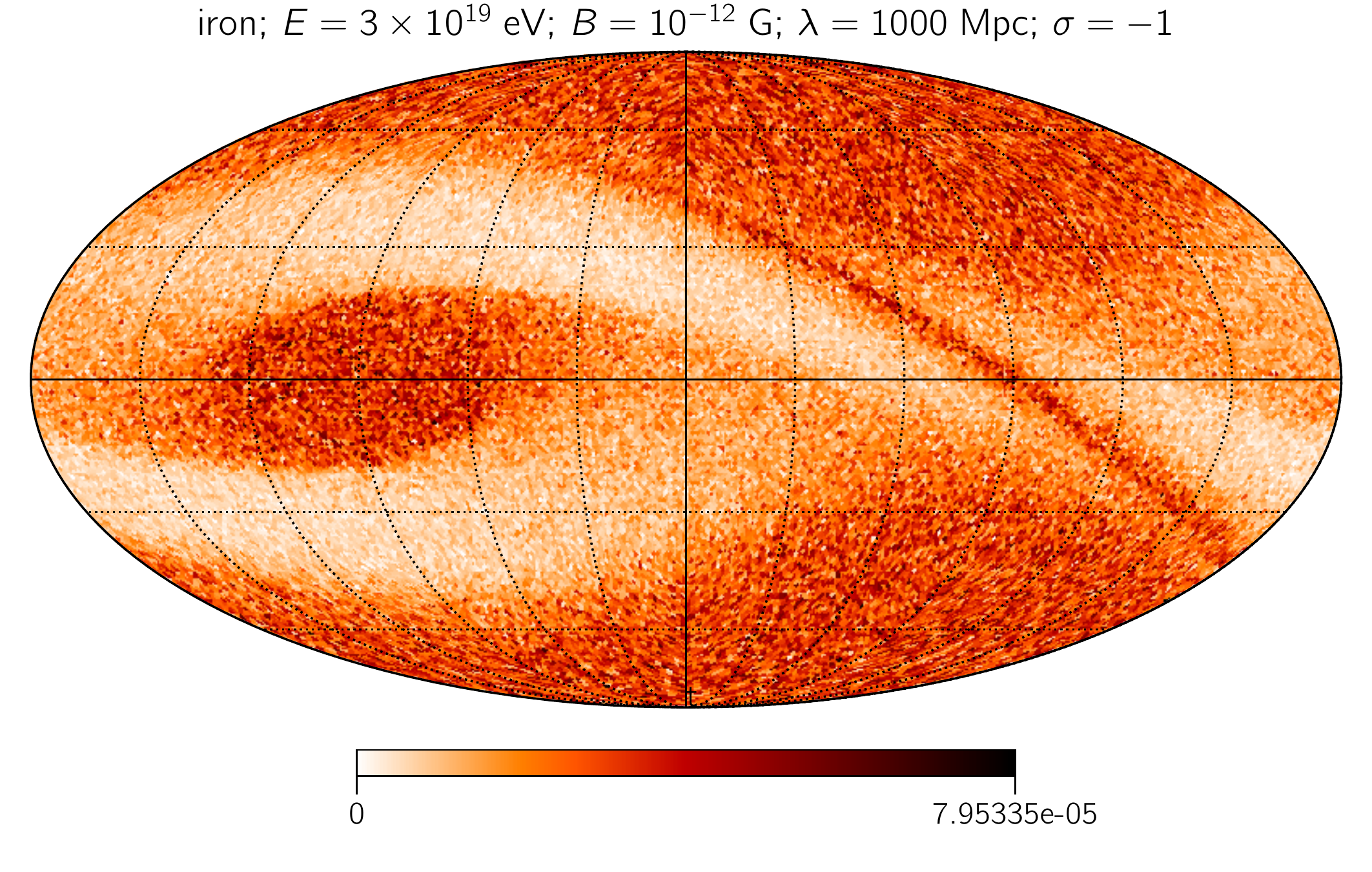}
\caption{Skymaps containing the arrival directions of UHECRs for $\sigma=+1$ (left), $\sigma=0$ (centre), $\sigma=-1$ (right). 
At the top the skymaps for $10^{20} \; \text{eV}$ protons are shown, for $B = 10^{-9} \; \text{G}$ and $\lambda = 10 \; \text{Mpc}$. The middle row represents the injection of $10^{20} \; \text{eV}$ iron, for $B = 10^{-11} \; \text{G}$ and $\lambda = 100 \; \text{Mpc}$. The lower row corresponds to the case of iron nuclei with $E = 3 \times 10^{19} \; \text{eV}$ for $B = 10^{-12} \; \text{G}$ and $\lambda = 1000 \; \text{Mpc}$. The colour bar shows the normalised number of events per pixel.}
\label{fig:skymap}
\end{figure}

We have chosen a convenient coordinate system such that the helical magnetic field is orthogonal to the $\mathbf{\hat{z}}$ direction, without loss of generality. This is justified because the skymaps can be arbitrarily rotated and rewritten in any basis for the purposes of calculations, and subsequently converted back into the original coordinate system.

\subsection{Harmonic Analysis of Simulated Data Sets} \label{subsec:harmonicAnalysis}

We study the impact of helicity on the propagation of UHECRs by performing a harmonic analysis of the simulated data. We expand the skymap ($\mathbf{\Phi}$) into spherical harmonics $Y_{lm}$ using HEALPix~\cite{Gorski:2004by}:
\begin{equation}
	\mathbf{\Phi}(\mathbf{\hat{n}}) = \sum\limits_{l=0}^{l_{\rm max}} \sum\limits_{m=-l}^{l} a_{lm} Y_{lm}(\mathbf{\hat{n}}),
\end{equation}
with $a_{lm}$ being the spherical harmonic coefficients, defined as
\begin{equation}
	a_{lm} = \int d\Omega \mathbf{\Phi}(\mathbf{\hat{n}}) Y_{lm} (\mathbf{\hat{n}}).
\end{equation}
Now we expand the skymap up to the largest scale of interest, $l=2$, following Ref.~\cite{Auger:2012an}:
\begin{equation}
	\mathbf{\Phi} (\mathbf{\hat{n}}) \approx \frac{\mathbf{\Phi_0}}{4 \pi} \left[ 1 + w_{\rm d} \mathbf{\hat{d}} \cdot \mathbf{\hat{n}} + \frac{1}{2} \sum\limits_{i,j} Q_{ij} n_i n_j \right], 
	\label{eq:harmonicExpansion}
\end{equation}
where $\mathbf{\hat{n}}$ is arbitrary direction, $\mathbf{\hat{d}}$ is the dipole unit vector, $Q_{ij}$ is the traceless quadrupole tensor, and $w_{\rm d}$ is the dipole amplitude that can be written as
\begin{equation}
	w_{\rm d} = \frac{\sqrt{3}}{a_{00}} \sqrt{a_{10}^2 + a_{11}^2 + a_{1-1}^2}.
\end{equation}

The dipole points to
\begin{equation}
	(\theta_{\rm d}, \varphi_{\rm d}) =  \left( \arcsin\left( \frac{\sqrt{3} a_{10}}{a_{00}} w_{\rm d} \right), \arctan\left( \frac{a_{1-1}}{a_{11}} \right) \right),
	\label{eq:dirDip}
\end{equation}
with $\theta_{\rm d}$ being the zenithal coordinate of the dipole, and $\varphi_{\rm d}$ the azimuthal coordinate.

The quadrupole tensor, $\mathbf{Q}$, can be diagonalised. The corresponding eigenvectors, $\lambda_\text{max}$, $\lambda_0$, and $\lambda_\text{min}$, can then be used to define a quadrupole amplitude ($w_q$), which reads~\cite{Aab:2014ila}:
\begin{equation}
	w_\text{q} = \frac{\lambda_\text{max} - \lambda_\text{min}}{2 + \lambda_\text{max} + \lambda_\text{min}}.
	\label{eq:quadAmp}
\end{equation}
Following Ref.~\cite{Aab:2014ila}, the other amplitude can be defined from the highest eigenvalue, $\lambda_\text{max}$. 
In the absence of a dipolar pattern, i.e., for $w_\text{d} = 0$ in Eq.~\ref{eq:harmonicExpansion}, $w_\text{d}$ as defined in Eq.~\ref{eq:quadAmp} is the maximum anisotropy contrast. For simplicity, we ignore the second dipole amplitude and refer to $w_\text{q}$ simply as \emph{the} quadrupole amplitude. This is merely a choice of an observable and does not imply that $w_\text{d} \neq 0$.

We analyse the effects of helicity on the dipole amplitude. We expect $\sigma = +1$ and $\sigma = -1$ to have similar behaviours with respect to the axis of symmetry ($\varphi=0$ meridian, by construction). For instance, if for $\sigma = +1$ the dipole points to $(\theta_{\rm d}, \varphi_{\rm d})$, then in the case of $\sigma=-1$ it would point to $(\theta_{\rm d}, -\varphi_{\rm d})$, as can be seen in Fig.~\ref{fig:skymap}. Therefore, the symmetry with respect to the transformation $\varphi \rightarrow -\varphi$ implies that the sign of $\varphi_{\rm d}$ is a suitable observable to constrain the sign of the helicity. 

Model uncertainties are computed assuming that the events have a Poisson distribution with event rate equal to the number of events in the corresponding pixels. These uncertainties propagate to all observables and for the angular quantities they are combined with the angular resolution (size of the pixel). 

\begin{figure}
	\includegraphics[width=0.495\columnwidth]{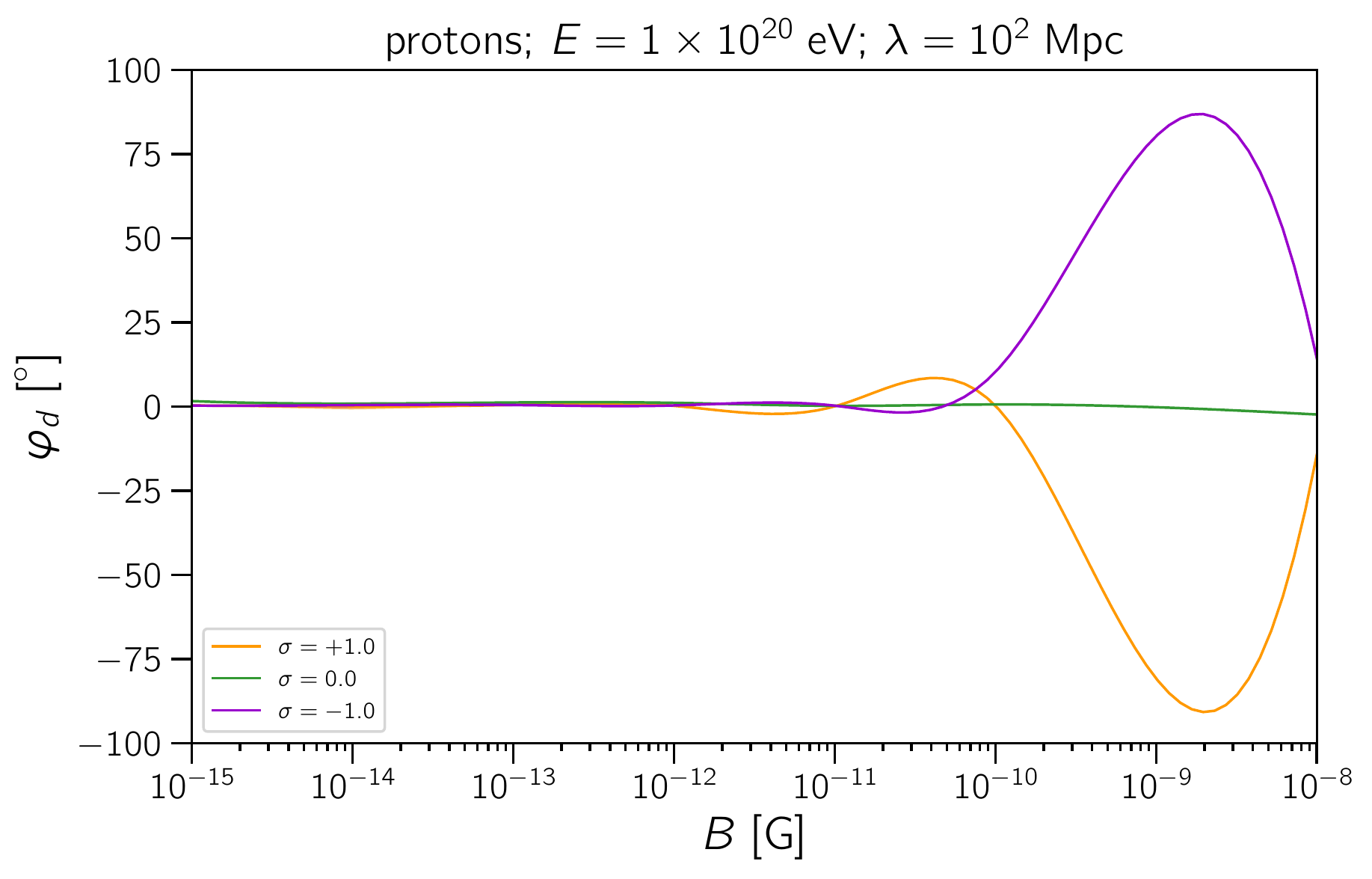}
	\includegraphics[width=0.495\columnwidth]{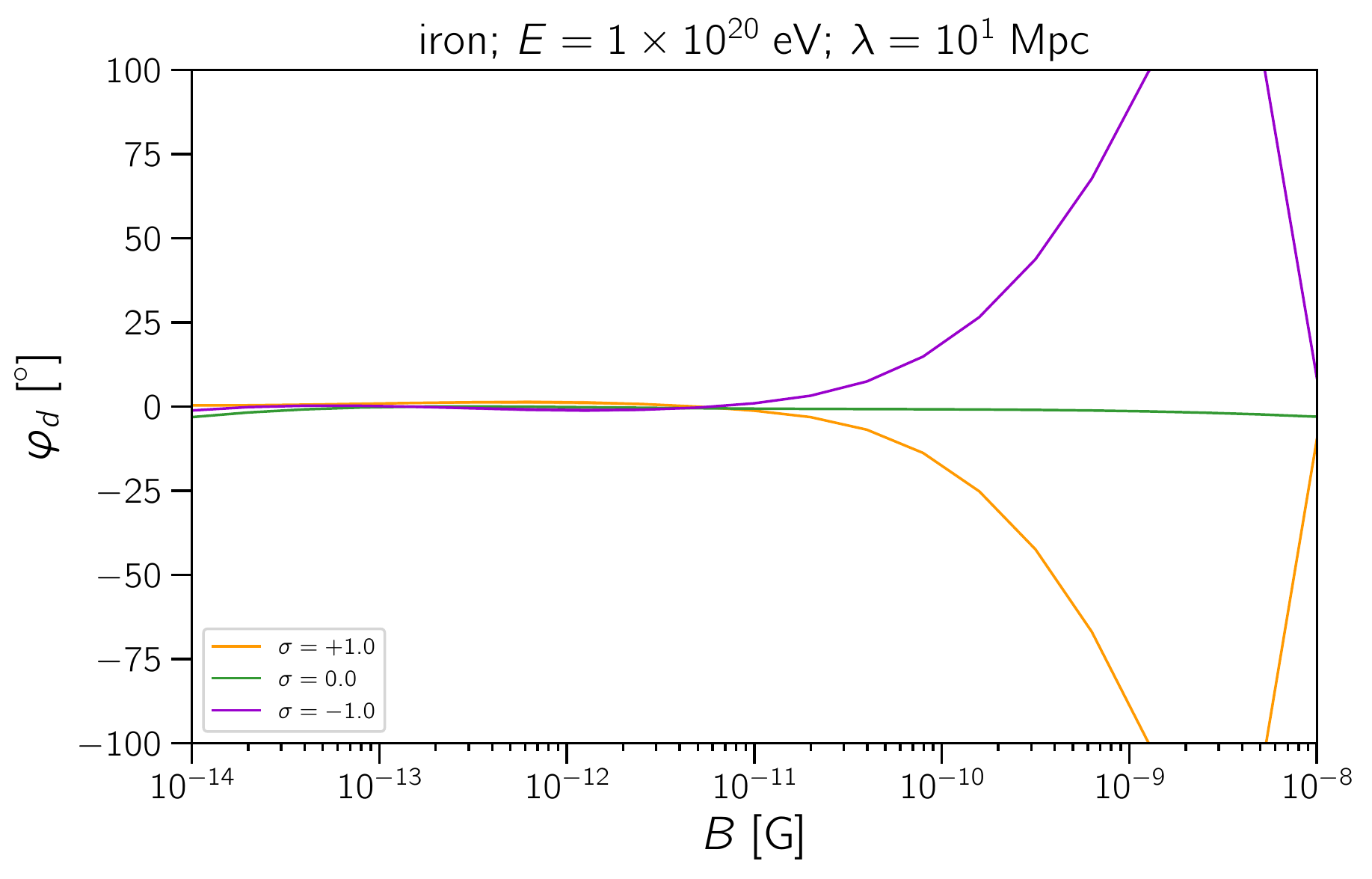}
	\includegraphics[width=0.495\columnwidth]{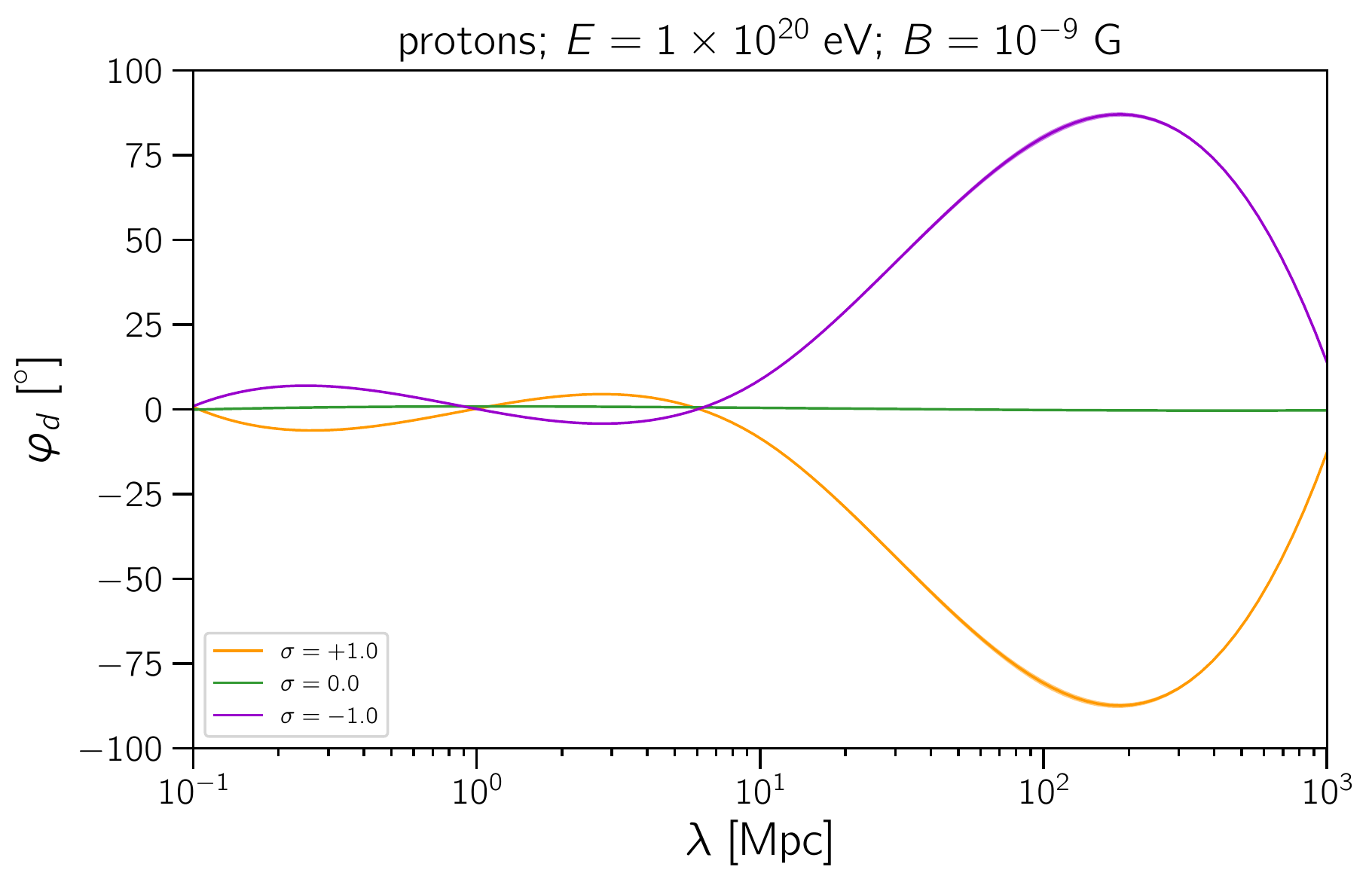}
	\includegraphics[width=0.495\columnwidth]{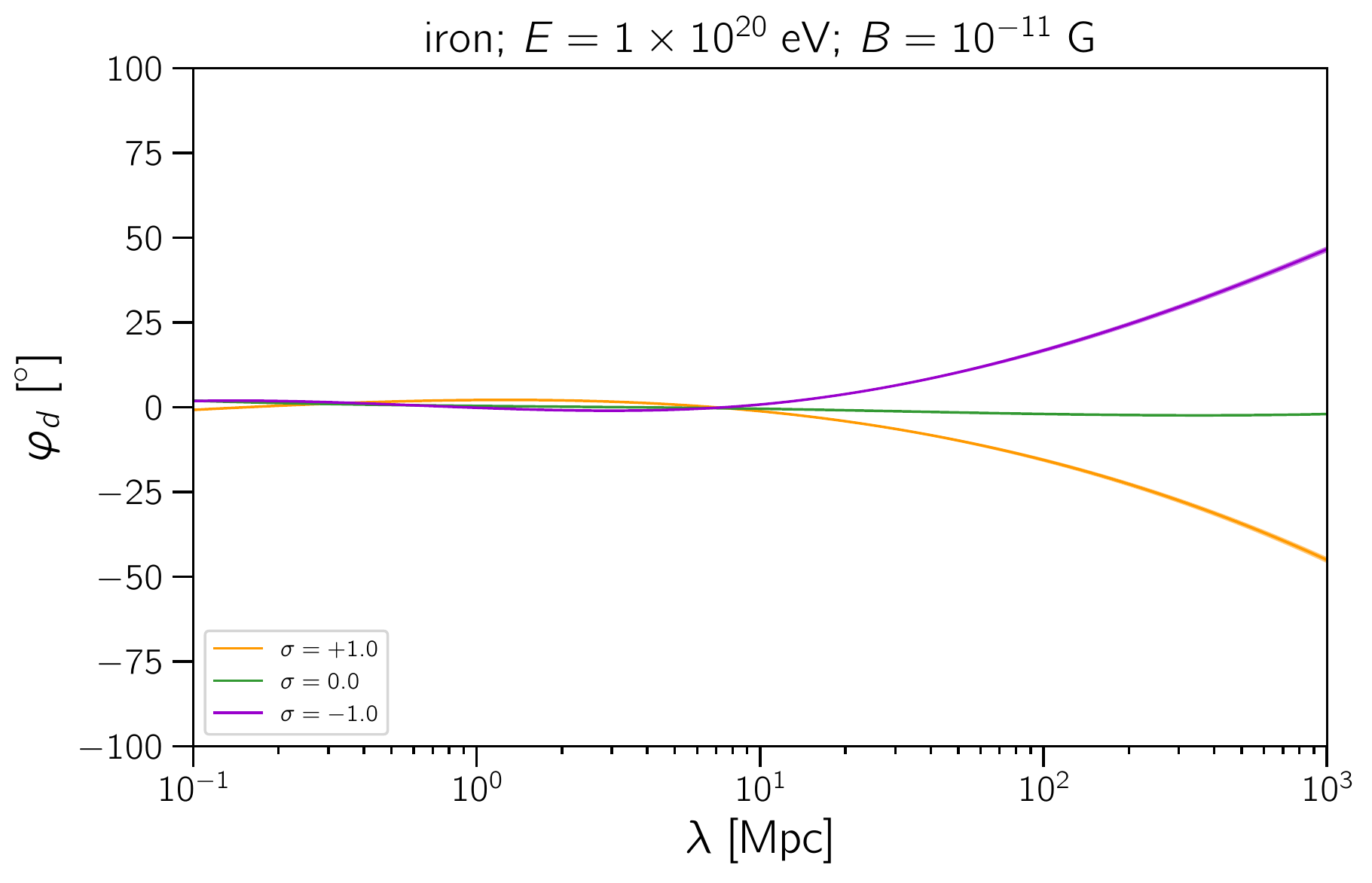}
	\caption{The azimuthal angle of the dipole ($\varphi_{\rm d}$) as a function of: the magnetic field for $\lambda = 100 \; \text{Mpc}$ (upper left) and $\lambda = 10 \; \text{Mpc}$ (upper right); the coherence length for $B = 10^{-9} \; \text{G}$ (lower left) and $B = 10^{-11} \; \text{G}$ (lower right panel). The panels displayed in the left column correspond to the pure proton scenario, whereas the ones on the right are for iron nuclei, both for an injected energy of $10^{20} \; \text{eV}$.}
	\label{fig:dipolePhi}
\end{figure}

In Fig.~\ref{fig:dipolePhi} we present the behaviour of $\varphi_{\rm d}$ as a function of the magnetic field strength, for iron primaries. The symmetry  $\varphi \rightarrow -\varphi$ for $\sigma = \pm 1$  becomes evident as the magnetic field increases. For $\sigma = 0$, the value of $\varphi_{\rm d}$ cannot be readily constrained. Similar behaviour is observed by changing the composition to protons, and if the energy is lowered to $E = 3 \times 10^{19} \; \text{eV}$.

By analysing Fig.~\ref{fig:dipolePhi} one reaches the somewhat trivial conclusion that opposite magnetic helicities lead to opposing behaviours of the arrival distribution of cosmic rays. Nevertheless, these figures serve an important purpose: they indicate for which set of parameters this conclusion is reliable. In particular, there are specific configurations for which the behaviours of $\sigma = \pm 1$ are maximal, providing an optimal window for constraining the average helicity of magnetic fields. A simple analysis of the positions of the peaks in Fig.~\ref{fig:dipolePhi} suggests that the maxima are located at $\mathcal{B} \equiv B \lambda \sim 10^{-7} \; \text{Mpc} \, \text{G}$ for $\sigma = \pm 1$, in the proton case.

\begin{figure}
	\includegraphics[width=0.495\columnwidth]{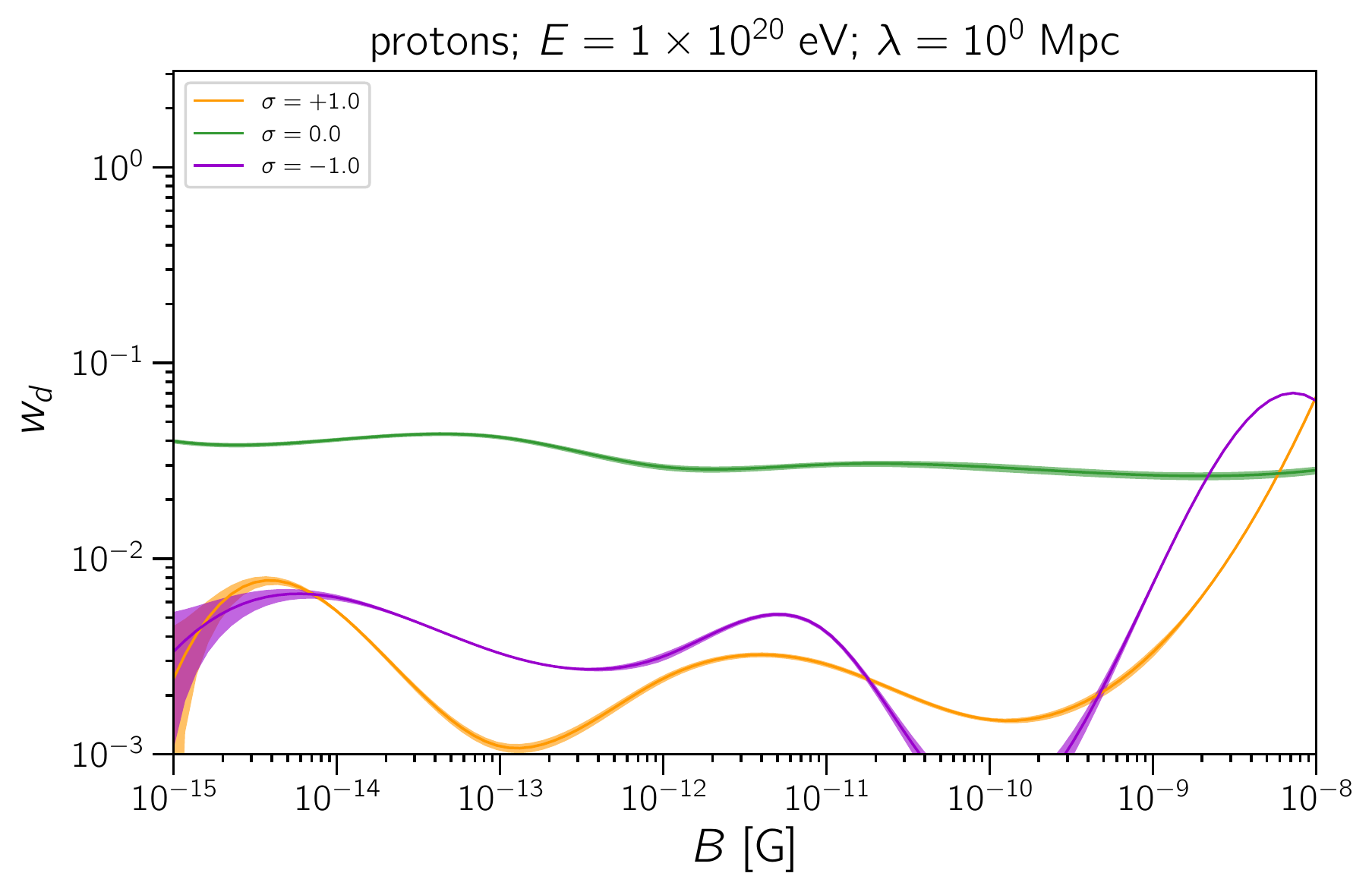}
	\includegraphics[width=0.495\columnwidth]{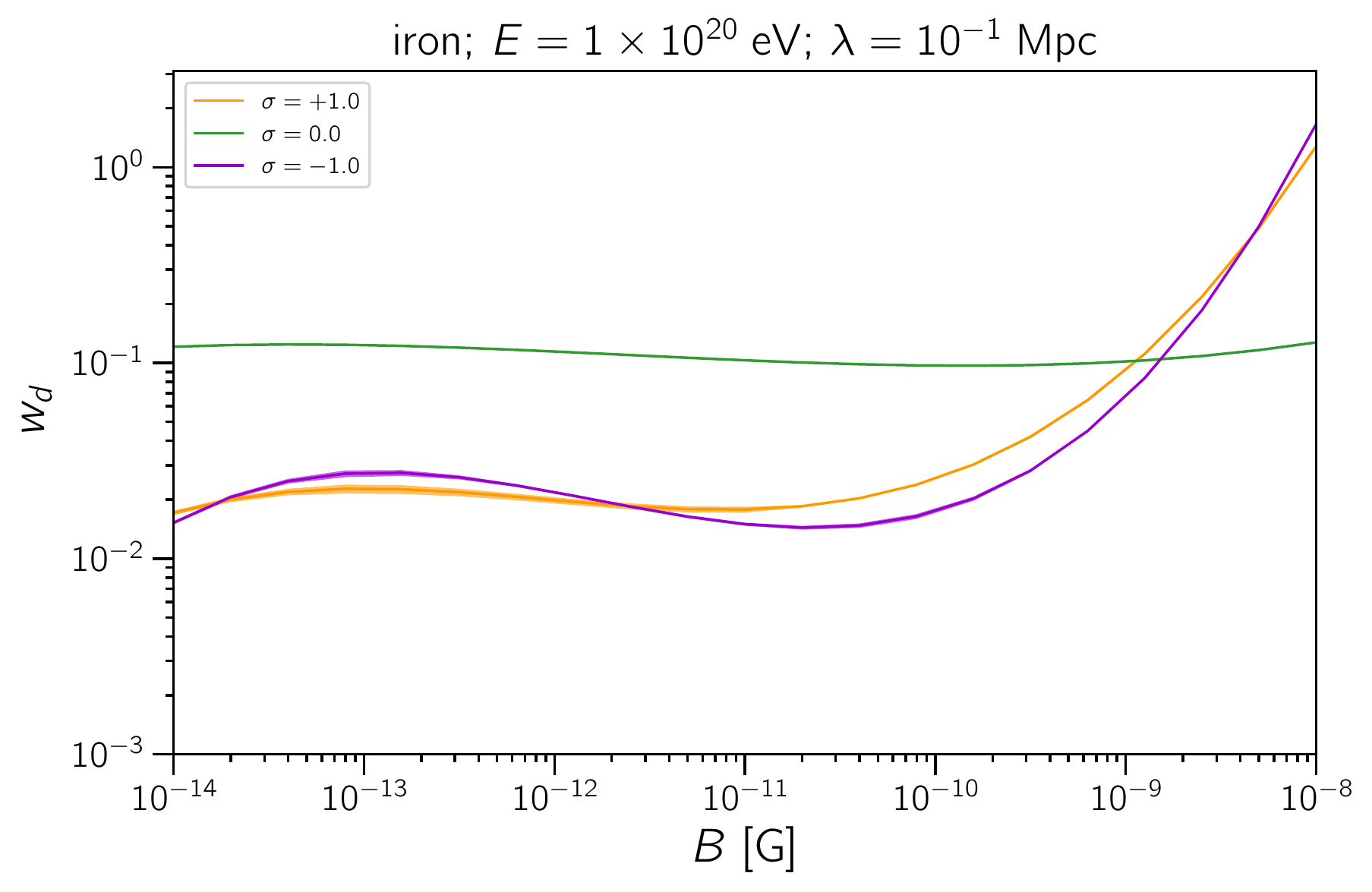}
	\includegraphics[width=0.495\columnwidth]{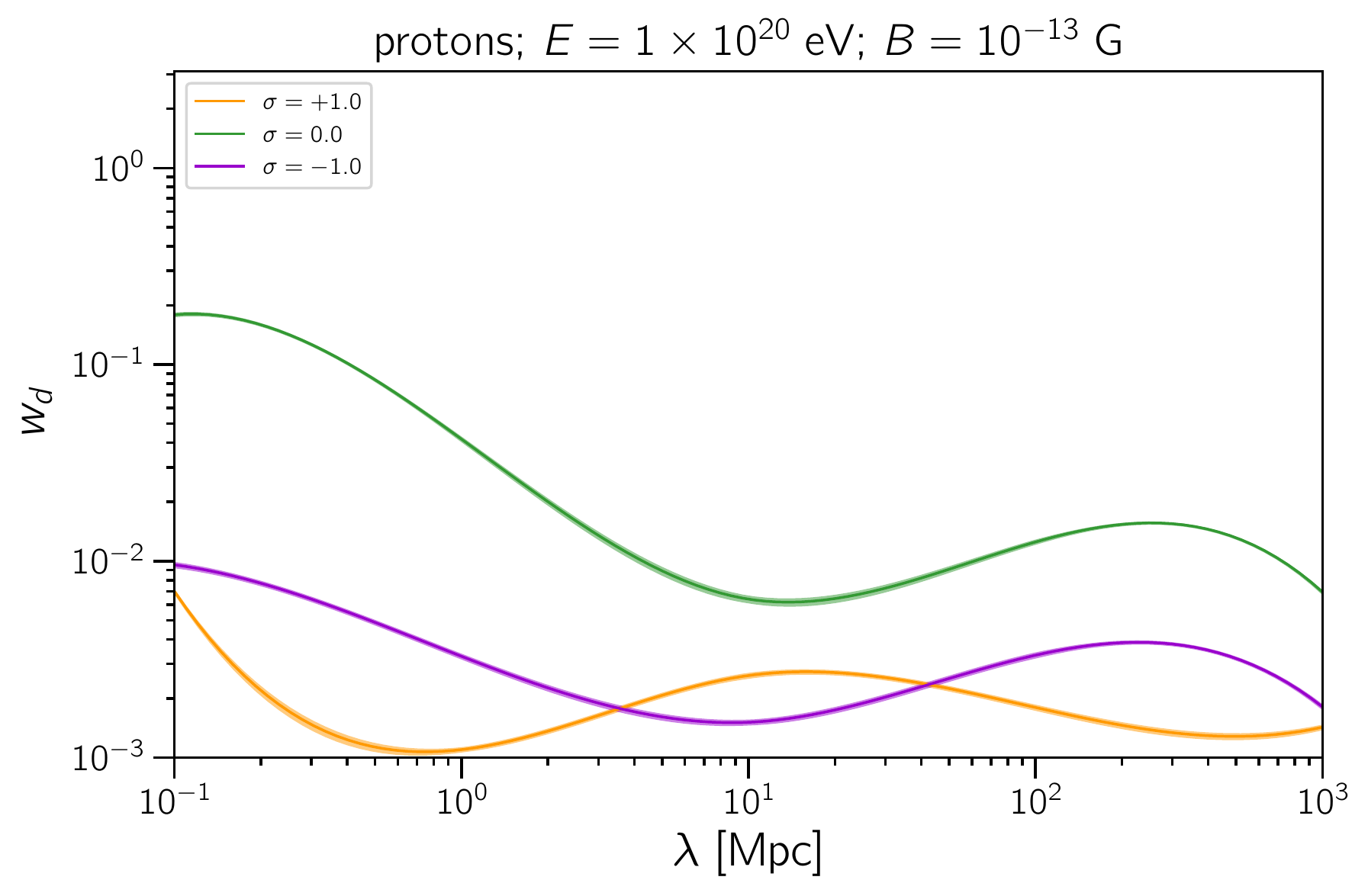}
	\includegraphics[width=0.495\columnwidth]{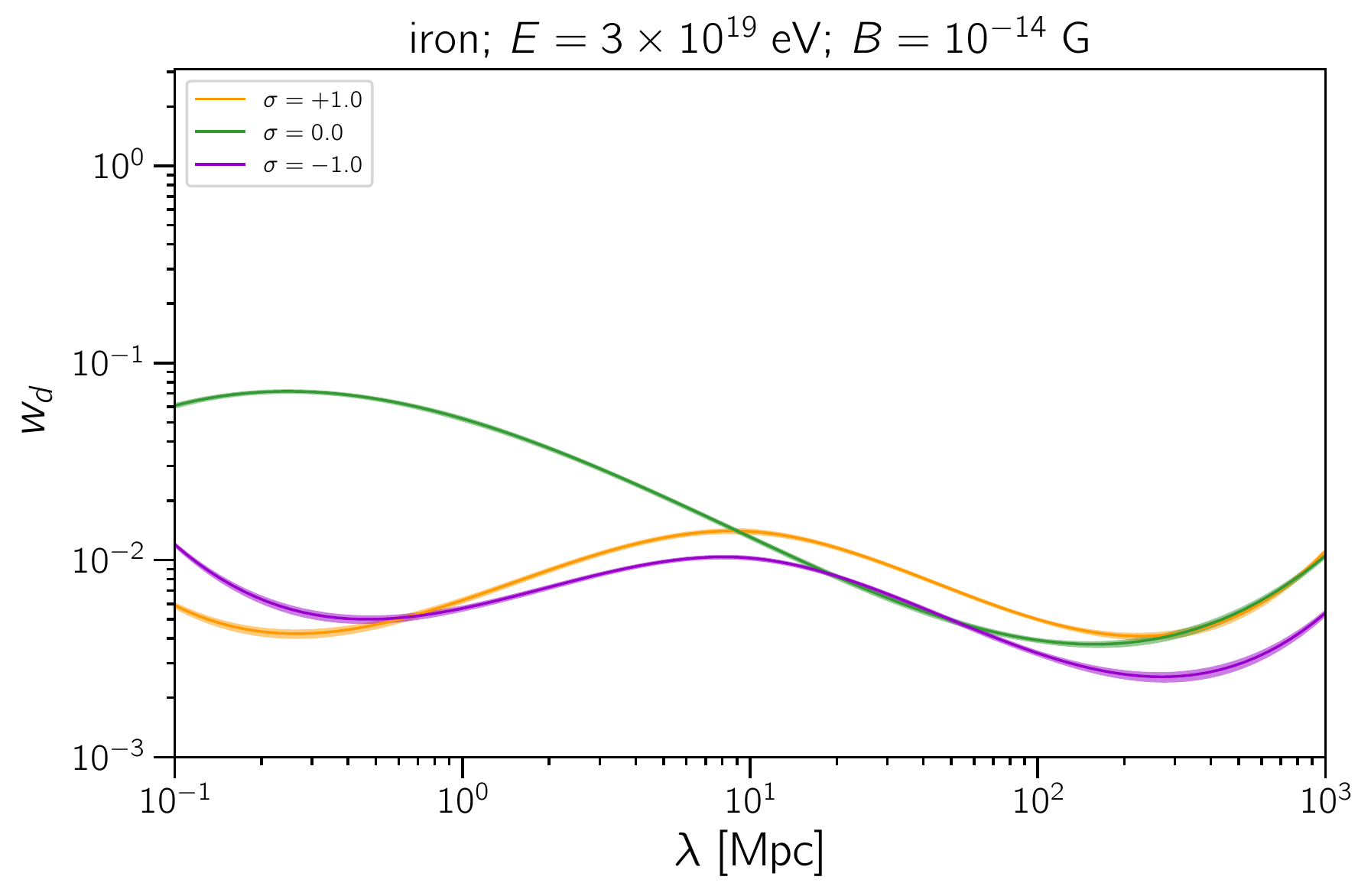}
	\caption{The dipole amplitude ($w_{\rm d}$) as a function of: the magnetic field (upper row), and coherence length (lower row). The upper panels are for the case of $10^{20} \; \text{eV}$ primaries with $\lambda = 1 \; \text{Mpc}$ (upper left) $\lambda = 100 \; \text{kpc}$ (upper right), the former corresponding to proton injection and the latter to iron. The case of $10^{20} \; \text{eV}$ protons and $B = 10^{-13} \; \text{G}$ is shown in the lower left panel, and the lower right panel corresponds to the case of $3 \times 10^{19} \; \text{eV}$ iron propagating through a magnetic field $B = 10^{-14} \; \text{G}$.}
	\label{fig:dipoleAmplitude}
\end{figure}

In Fig.~\ref{fig:dipoleAmplitude} the dipole amplitude is presented. Based on the skymaps shown in Fig.~\ref{fig:skymap}, it is reasonable to suppose that the absolute value of the helicity is related to the amplitude of the dipole. One expects that for $\sigma = \pm 1$ the dipole amplitude will increase with the magnetic field, up to the point where all cosmic rays would be completely isotropised. This is illustrated in in Fig.~\ref{fig:dipoleAmplitude}. 

Note in Fig.~\ref{fig:dipoleAmplitude} that, as expected, opposite helicities behave approximately in the same way for some combinations of $B$ and $\lambda$. One also notices that $w_{\rm d}(\sigma = 0) > w_{\rm d}(\sigma = \pm 1)$ for  $B \lesssim 10^{-11} \; \text{G}$ and fixed coherence lengths. 

The energy of a cosmic ray, evidently, plays an important role in their propagation, being crucial for arrival directions. One could argue that, in principle, all results derived here scale with the rigidity ($R \equiv E / Z$, where $Z$ is the atomic number of a nucleus). Thus, a $10^{20} \; \text{eV}$ proton would behave as a $26 \times 10^{20} \; \text{eV}$ iron nucleus. This argument is only approximately true as the energy of a cosmic ray can be degraded due to interactions with the CMB and EBL; moreover, photodisintegration could break down a nucleus into smaller constituents thereby affecting arrival directions. Ultimately, the validity of the aforementioned argument depends on a delicate interplay between the gyroradius of the cosmic ray, its energy loss length, and the coherence length of the magnetic field.

It is interesting to study the diffusive regime as a limiting case. Let us first define the critical energy, $E_{\rm c}$, which is the energy at which the Larmor radius of a particle equals the coherence length of the magnetic field. It can be written as~\cite{Batista:2014xza}:
\begin{equation}
	E_{\rm c} \simeq 0.9 Z \left( \dfrac{B}{\text{nG}} \right) \left( \dfrac{L_{\rm c}}{\text{Mpc}} \right) \; \text{EeV}.
\end{equation}
In the non-resonant regime we have $E > E_{\rm c}$, whereas for $E < E_{\rm c}$ diffusion is resonant. Another factor that plays a role in the former regime is the distance to the dominant sources ($D_\text{min}$) compared to the coherence length. If $D_\text{min} \gg \lambda$, diffusion is non-resonant, whereas for $D_\text{min} \ll \lambda$ it is quasi-rectilinear.

From Eq.~\ref{helB} one can see that the magnetic field has no component in the $\hat{z}$-direction. In particular, for $\sigma = 0$ it is oriented along the y-axis, i.e., $\mathbf{B} = |\mathbf{B}| \mathbf{\hat{y}}$, while for $\sigma = \pm 1$ the helicity is controlling the contribution and direction of the $\hat{x}$ component. As a consequence, for scenarios with $\mathcal{B} \lesssim \text{nG}\,\text{Mpc}$, we expect to see an increase in the value of $\varphi_{\rm d}$ as $E_{\rm c}$ approaches $E$; thereafter, the propagation of UHECRs becomes diffusive. This is confirmed by analysing Fig.~\ref{fig:dipolePhi}, in particular the position of the peaks. The same argument applies to Fig.~\ref{fig:dipoleAmplitude}; for a fixed coherence length (upper panels), the dipole amplitude starts to increase noticeably for $E_{\rm c} \sim E$, i.e. as $B$ increases. Therefore, in the diffusive regime ($E \lesssim Z \frac{\mathcal{B}}{\text{Mpc} \, \text{nG}} \, \text{eV}$) our results confirm the theoretical predictions for this limit.

\subsection{Comparison with the Analytical Predictions}

\begin{figure}
\centering
\includegraphics[scale=0.5]{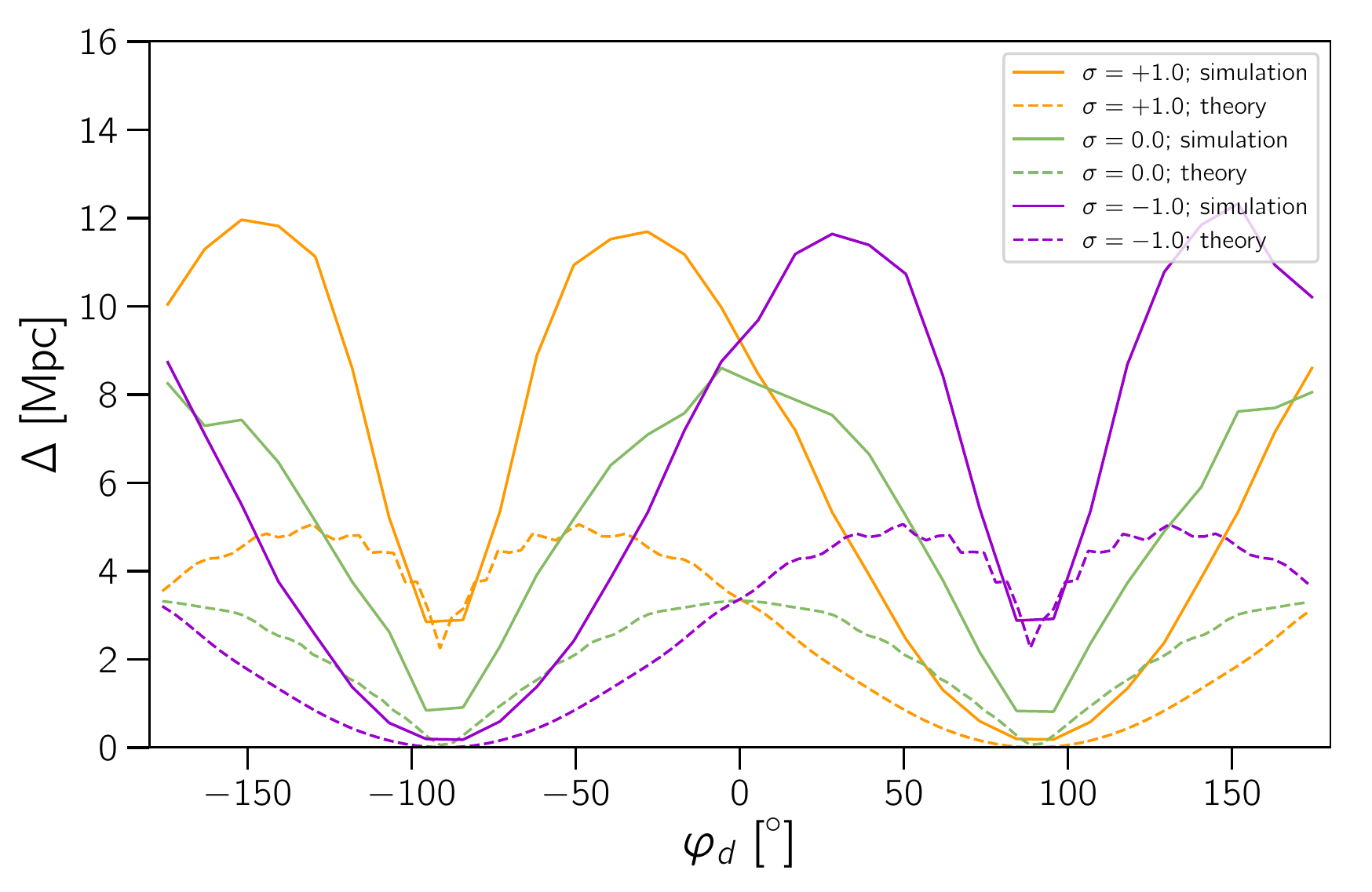}
\caption{Dependence of the additional travelled distance $\Delta$ on the arrival velocity angle $\varphi$ for $\sigma = -1$ (purple), $\sigma = 0$ (green) and $\sigma = +1$ (orange), for different arrival velocity directions with $v_{z0} = 0$ (i.e.~$\theta = 0$). The parameters used here are $\lambda = 10\,{\rm Mpc}$, $E = 10^{20}\,{\rm eV}$, $B_{0} = 10^{-9}\,{\rm G}$ and $R_{\rm s} = 500\,{\rm Mpc}$. The solid line depicts the results of the simulations while the dotted line shows the analytical predictions based on Eqs. \ref{DiffEq1b}-\ref{DiffEq3b}.}
\label{fig:phiDelta}
\end{figure}

In Fig.~\ref{fig:phiDelta} we have shown the azimuthal dependence of the elongation of trajectory ($\Delta$). Similar behaviour is expected from the azimuthal distribution of UHECRs, as can be seen in Fig.~\ref{fig:skymap}. In particular, one can see that along the ``equator'' the skymap presents azimuthal dependence similar to the one alluded to in Fig.~\ref{fig:phiDelta} and Eq.~\ref{eq:Delta_Rs}.

The relationship between $\Delta$ and $\varphi$ is shown in Fig.~\ref{fig:phiDelta} for $10^{20} \; \text{eV}$ protons travelling through a magnetic field of $B = 10^{-9} \; \text{G}$ and $\lambda = 10 \; \text{Mpc}$. While for the chosen magnetic field the values of $\varphi$ for the minima of $\Delta$ (hence the maxima of the energy) are independent of $\sigma$ (in our case being at $\varphi=90^\circ$ and $\varphi= -90^\circ$), the maxima of $\Delta$ (and hence the minima of the energy) are slightly shifted, thus resulting in the asymmetry seen in Fig.~\ref{fig:skymap}.

Note that while the simulations qualitatively behave as the analytical prediction from Eq.~\ref{eq:Delta_Rs}, the overall values of $\Delta$ are different. This is due to the fact that in the simulations, in contrast to the analytical predictions, we have included energy losses, which elongates the cosmic-ray trajectories, thus explaining why the curves for simulations are above the ones for the theoretical expectations in Fig.~\ref{fig:phiDelta}.

\subsection{The Effect of the Galactic Magnetic Field} \label{subsec:GMF}

We have considered the GMF model by Jansson \& Farrar~\cite{Jansson:2012pc,Jansson:2012rt}. It is comprised of multiple components, including a turbulent and a regular one. The effect of the former on UHECRs is a smearing around the source position. The latter systematically shifts the arrival directions. Deflections in the GMF are estimated to be $\sim 5.2^\circ$ for protons with energies $6 \times 10^{19} \; \text{eV}$; in about a quarter of the sky deflections are $\lesssim 2.2^\circ$ in this model. Recently it has been argued that the turbulent component of the Jansson-Farrar model might have been overestimated~\cite{Unger:2017kfh}. If this is true, then the large-scale anisotropy patterns we expect in the presence of helical magnetic fields may not be significantly affected by the GMF.

The coordinate system of the magnetic field defined by Eq.~\ref{helB} is adopted. We apply a rotation $\mathcal{R}$ to the arrival directions in order to mimic a rotation of the magnetic field. 

The rotations applied can be described as the product of three individual rotation matrices with angles $\alpha$, $\beta$, and $\gamma$:
\begin{equation} \label{rotmat}
  \mathcal{R} = \mathcal{R}_z(\alpha)\mathcal{R}_y(\beta)\mathcal{R}_x(\gamma),
\end{equation}
where $\mathcal{R}_i$ are the canonical rotation matrices around the corresponding axes. 

We consider four cases defined by combinations of the yaw, pitch, and roll angles, as defined in Table~\ref{tab:gmf}. These scenarios are introduced in order to move the patterns shown in Fig.~\ref{fig:skymap} with respect to the galactic coordinate system. The values of $\alpha$, $\beta$, and $\gamma$ are chosen to encompass the most extreme cases.
\begin{table}
  \centering
  \caption{Scenarios derived from the original simulations by applying rotations defined by combinations of the angles defined in Eq.~\ref{rotmat}.}.

  \begin{tabular}{c|ccc}
  \hline
  \hphantom{scenario} & $\alpha \; \text{[} ^\circ \text{]}$ & $\beta \; \text{[} ^\circ \text{]}$ & $\gamma \; \text{[} ^\circ \text{]}$ \\
  \hline
  scenario 0 &  0 &  0 &  0 \\
  scenario 1 & -90 &  0 &  0 \\
  scenario 2 &  0 &  0 & 90 \\
  scenario 3 &  150 & 0 & 60 \\
  \end{tabular}
  \label{tab:gmf}
\end{table}
\section{Constraining Helical Magnetic Fields} \label{sec:Constraints}

\subsection{The Sign of Magnetic Helicity}

In Sec.~\ref{sec:Simulations} we have argued that the sign of the helicity can be constrained from the azimuthal angle of the dipole ($\varphi_{\rm d}$). In this section we attempt to constrain this quantity. To this end, we use the sign of $\varphi_{\rm d}$. 

For a magnetic field ($\mathbf{B}$) conveniently defined in the $xy$-plane blue, the dipole moment points to $(\theta_{\rm d}, \varphi_{\rm d})$, defined in Eq.~\ref{eq:dirDip}. In this frame, the effects of helicity would be visible about the meridian $\varphi = 0$ of the skymap. The zenithal angle of the dipole should, in principle, be $\theta_{\rm d} \simeq 0$. If we rotate the magnetic field vector using an arbitrary global rotation matrix $\mathcal{R}$, the new magnetic field can be written as $\mathbf{B}^\prime = \mathcal{R} \mathbf{B}$, such that the new direction of the dipole would be $(\theta_{\rm d}^\prime, \varphi_{\rm d}^\prime)$. By applying the same rotation operator to all position vectors of the skymap, we carry out a passive transformation of the arrival directions and obtain a new skymap which merely represents a change of basis. Therefore, we can use our choice of the coordinate system without loss of generality, while being able to easily transform it, for example, to compare the results with data.

\begin{figure}[htb!]
  \includegraphics[width=0.495\columnwidth]{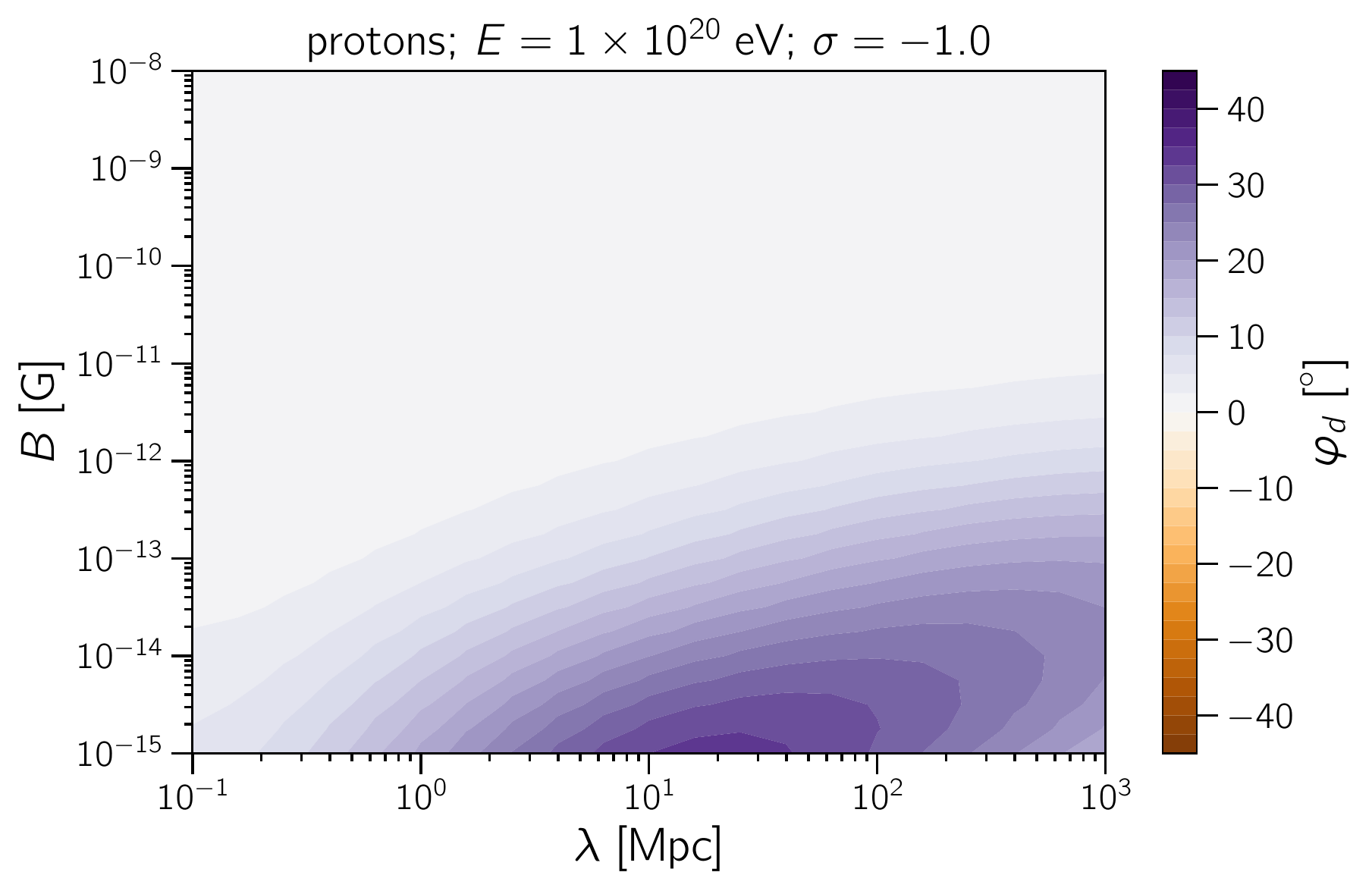}
  \includegraphics[width=0.495\columnwidth]{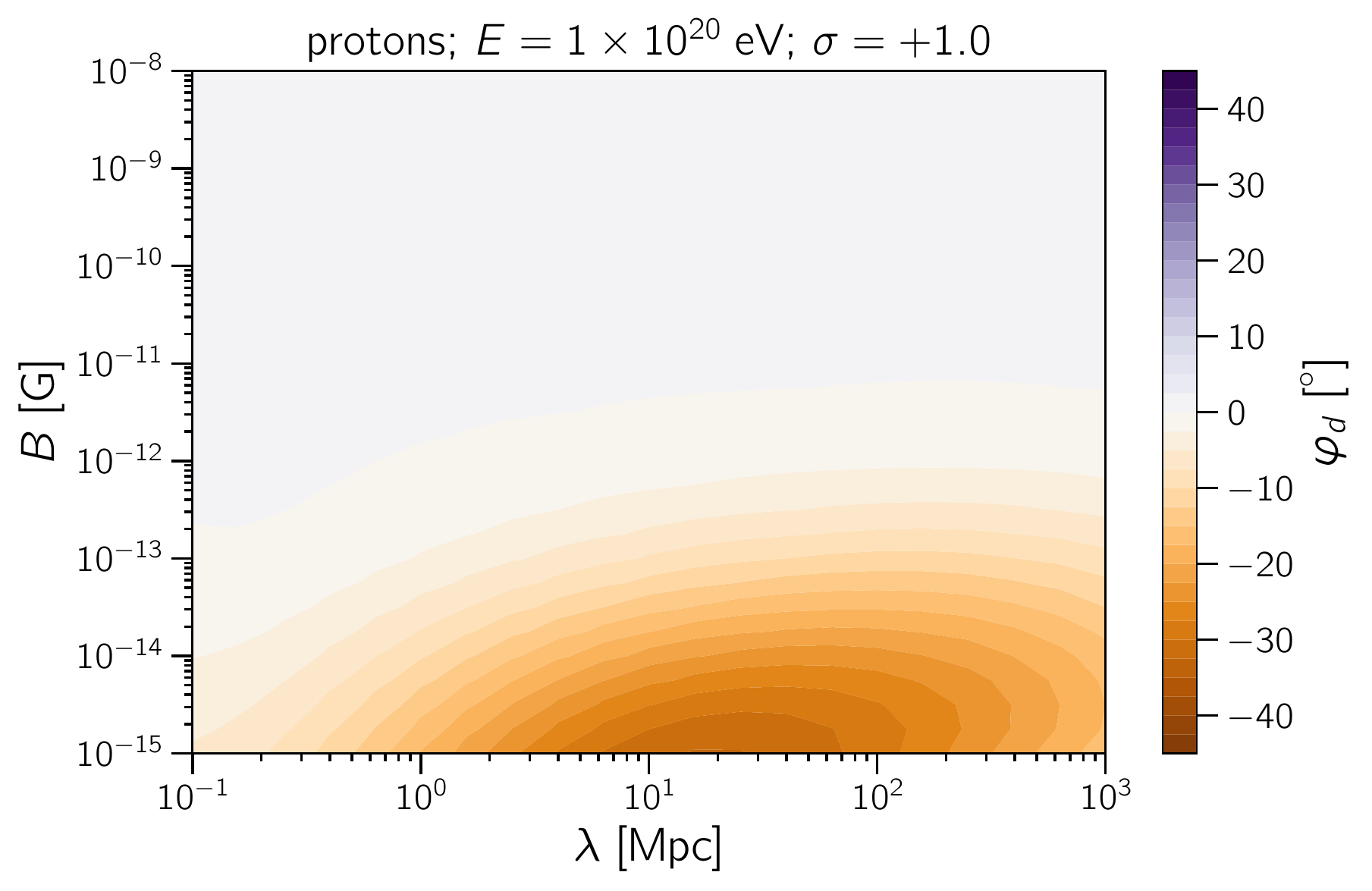}
  \includegraphics[width=0.495\columnwidth]{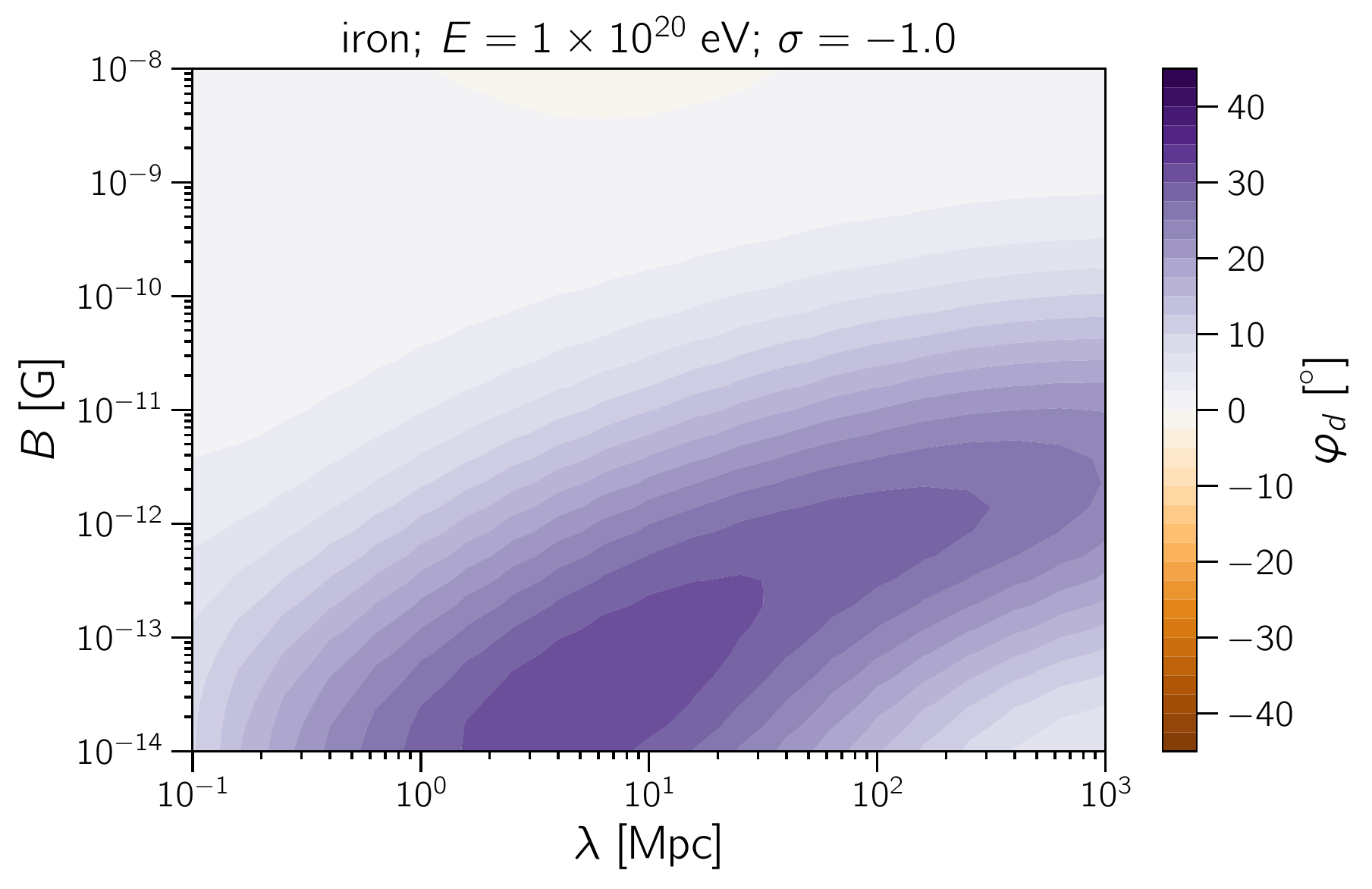}
  \includegraphics[width=0.495\columnwidth]{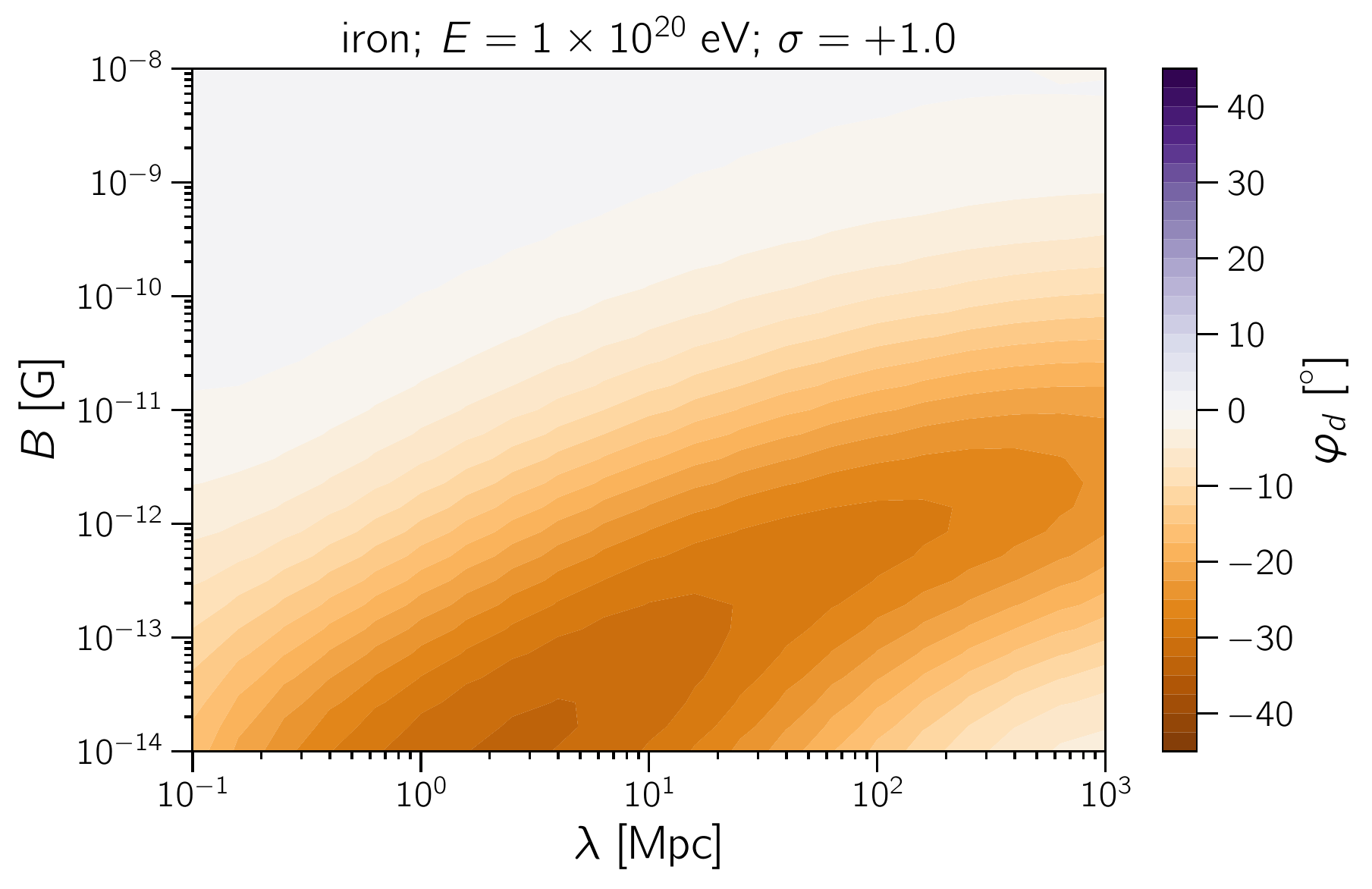}
  \includegraphics[width=0.495\columnwidth]{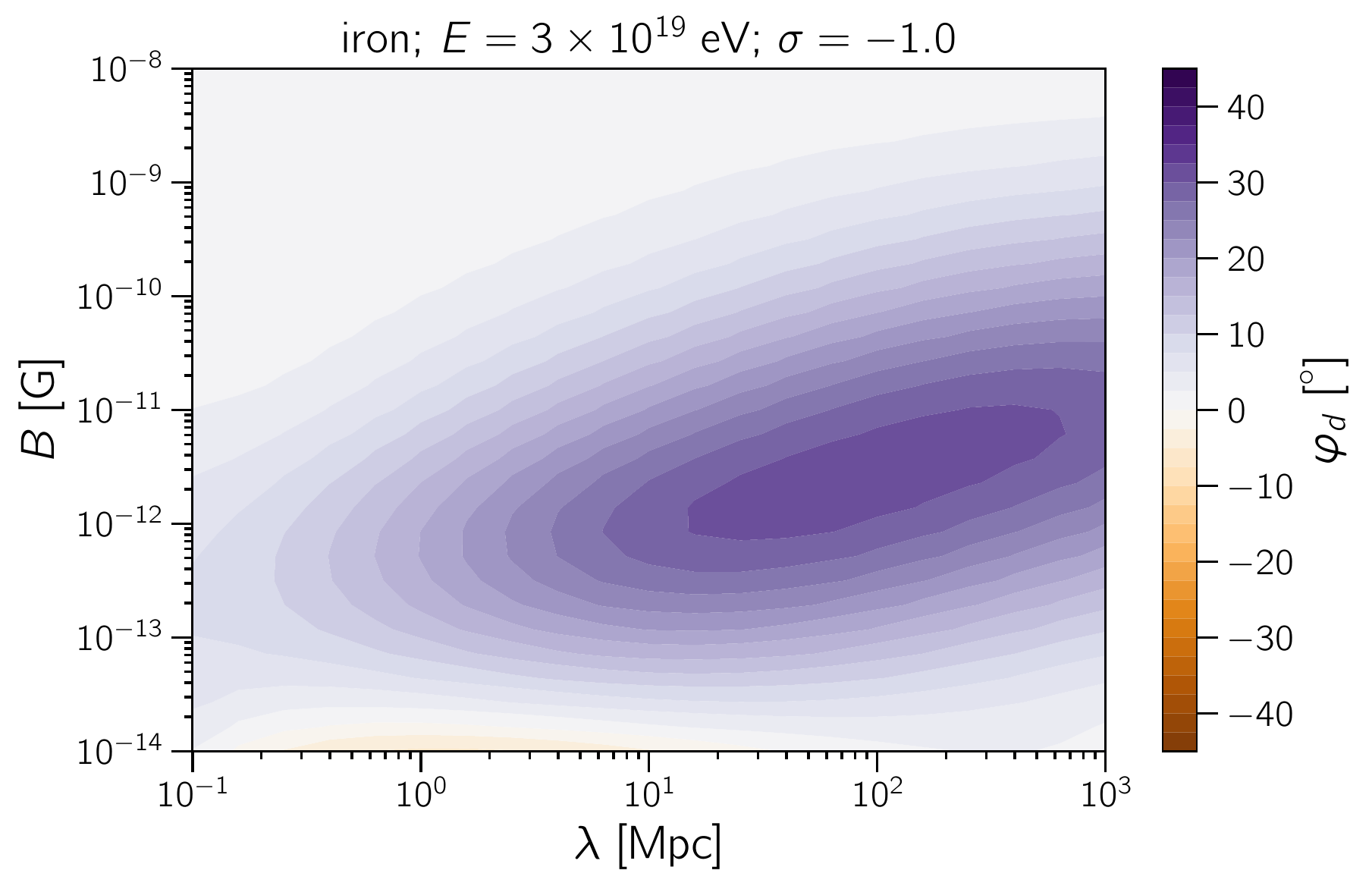}
  \includegraphics[width=0.495\columnwidth]{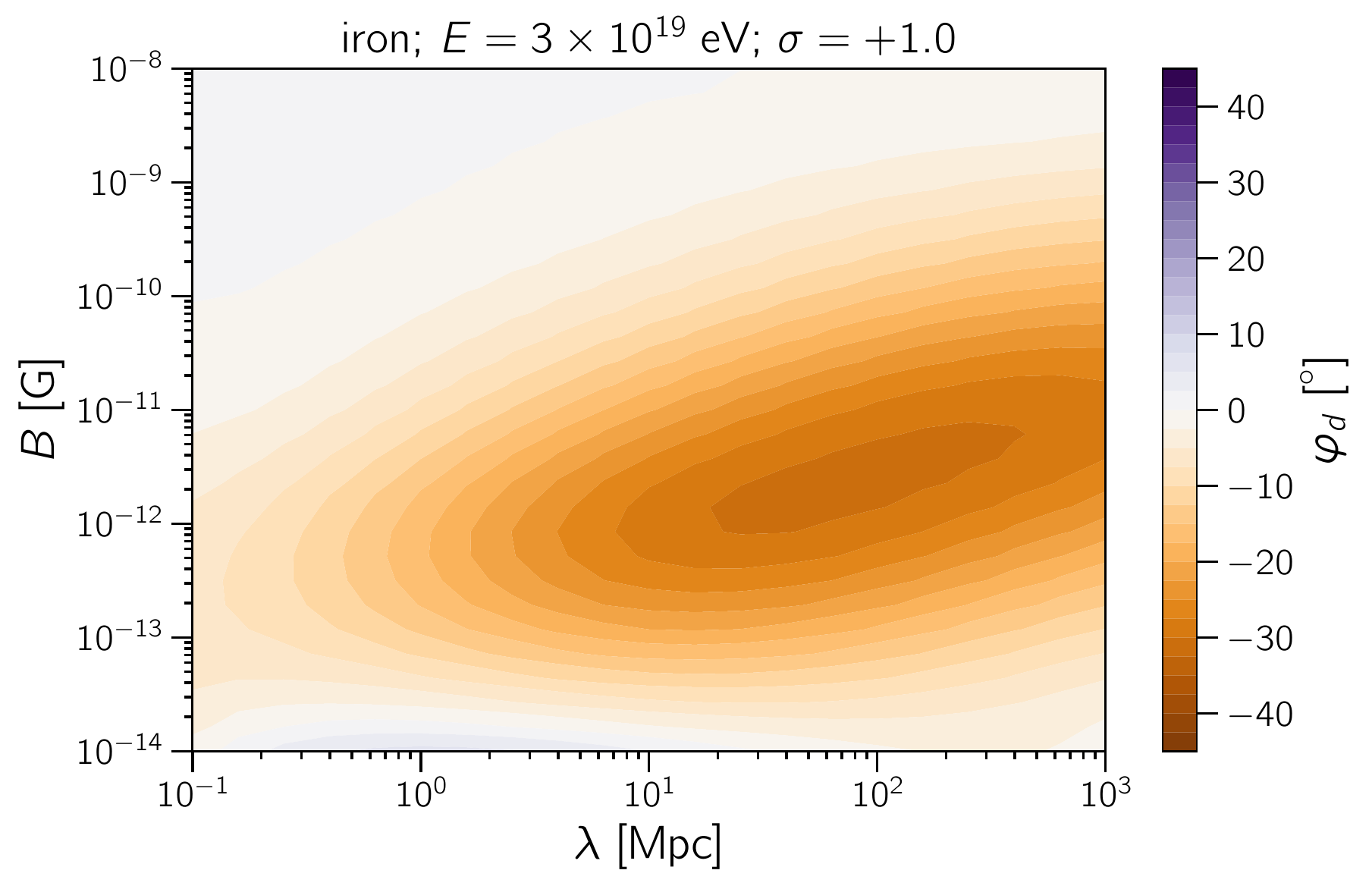}
  \caption{Azimuthal angle of the dipole ($\varphi_{\rm d}$) for different combinations of magnetic field strength ($B$) and coherence length ($\lambda$). Maximally negative helicities are shown in the left column, positive helicities on the right, and the case of null helicity s omitted because $\varphi_d \simeq 0$. The upper and central rows correspond to the case of proton and iron primaries, respectively, with $E = 10^{20} \; \text{eV}$, whereas the lower row corresponds to iron primaries with $E = 3 \times 10^{19} \; \text{eV}$.}
  \label{fig:constraint_ld}
\end{figure}

In Fig.~\ref{fig:constraint_ld} the dependence of the dipole azimuthal angle is presented for various combinations of magnetic field strength and coherence length. Note that in Fig.~\ref{fig:constraint_ld} the region in the parameter space that produces the largest effects on the dipole direction is, to first order, $\mathcal{B} = B \lambda \sim 10^{-13.5} \; \text{G} \, \text{Mpc}$ for $10^{20} \; \text{eV}$ protons and iron nuclei, and $\mathcal{B} \sim 10^{-10} \; \text{G} \, \text{Mpc}$ for iron with $E = 3 \times 10^{19} \; \text{eV}$.

The identification of the sign of helicity depends on the strength of the dipole and on a reliable estimate of $\varphi_{\rm d}$ whose uncertainty ($\Delta \varphi_{\rm d}$) can be relatively high at $E \gtrsim 10^{19} \; \text{eV}$ due to the low flux. Moreover, for typical UHECR observatories the angular resolution is $\sim 1^\circ$. Therefore, one expects $\Delta \varphi_{\rm d} \simeq 1 - 10^\circ$, which limits our ability to constrain the sign of the helicity for $\mathcal{B} \lesssim Z 10^{-16} \; \text{G} \, \text{Mpc}$ and $\mathcal{B} \gtrsim Z 10^{-12} \; \text{G} \, \text{Mpc}$ for $E \simeq 10^{20} \; \text{eV}$.

We have also studied the effect of the galactic magnetic field on $\varphi_\text{d}$, assuming the scenarios from Table~\ref{tab:gmf}. The results are qualitatively similar to those shown in Fig.~\ref{fig:constraint_ld}, and approximately hold regardless of the orientation of the helical extragalactic magnetic field with respect to the GMF. Thus we conclude that the GMF does not compromise the measurement of the sign of the helicity.

\subsection{The Absolute Value of Magnetic Helicity}

To constrain the absolute value of the helicity of a magnetic field configuration, we consider two quantities: the dipole ($w_{\rm d}$) and quadrupole ($w_{\rm q}$) amplitudes. We compute the relative differences between the amplitudes for the cases $\sigma = \pm 1$ and the case $\sigma = 0$. We find that the constraining power of the quadrupole moment alone is higher in a region of the $B-\lambda$ parameter space different from the region probed by the dipole amplitude. In particular, the relative difference between $w_{\rm q}(\sigma = \pm 1)$ and $w_{\rm q}(\sigma=0)$ is enhanced for $B \gtrsim 10^{-14} \; \text{G}$ and $\lambda \lesssim 10 \; \text{Mpc}$, whereas $w_{\rm d}(\sigma = \pm 1)$ and $w_{\rm d}(\sigma = 0)$ do not differ significantly in this region. Therefore, to expand the parameter space that can be constrained, we introduce a new quantity, the dipole-to-quadrupole ratio ($r$), given by:
\begin{equation}
  r = \dfrac{w_{\rm d}}{w_{\rm q}}.
\end{equation}

\begin{figure}
  \includegraphics[width=0.495\columnwidth]{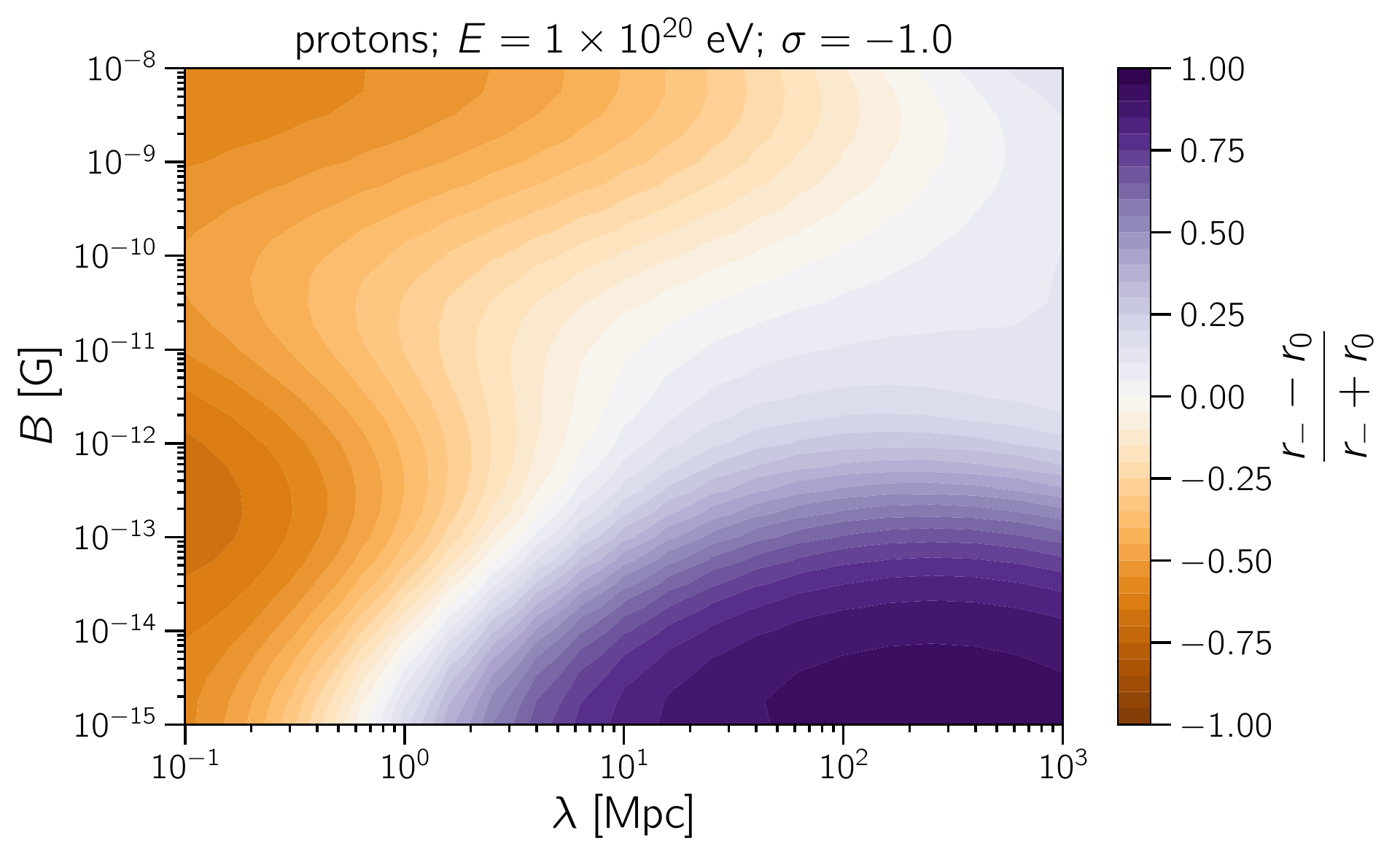}
  \includegraphics[width=0.495\columnwidth]{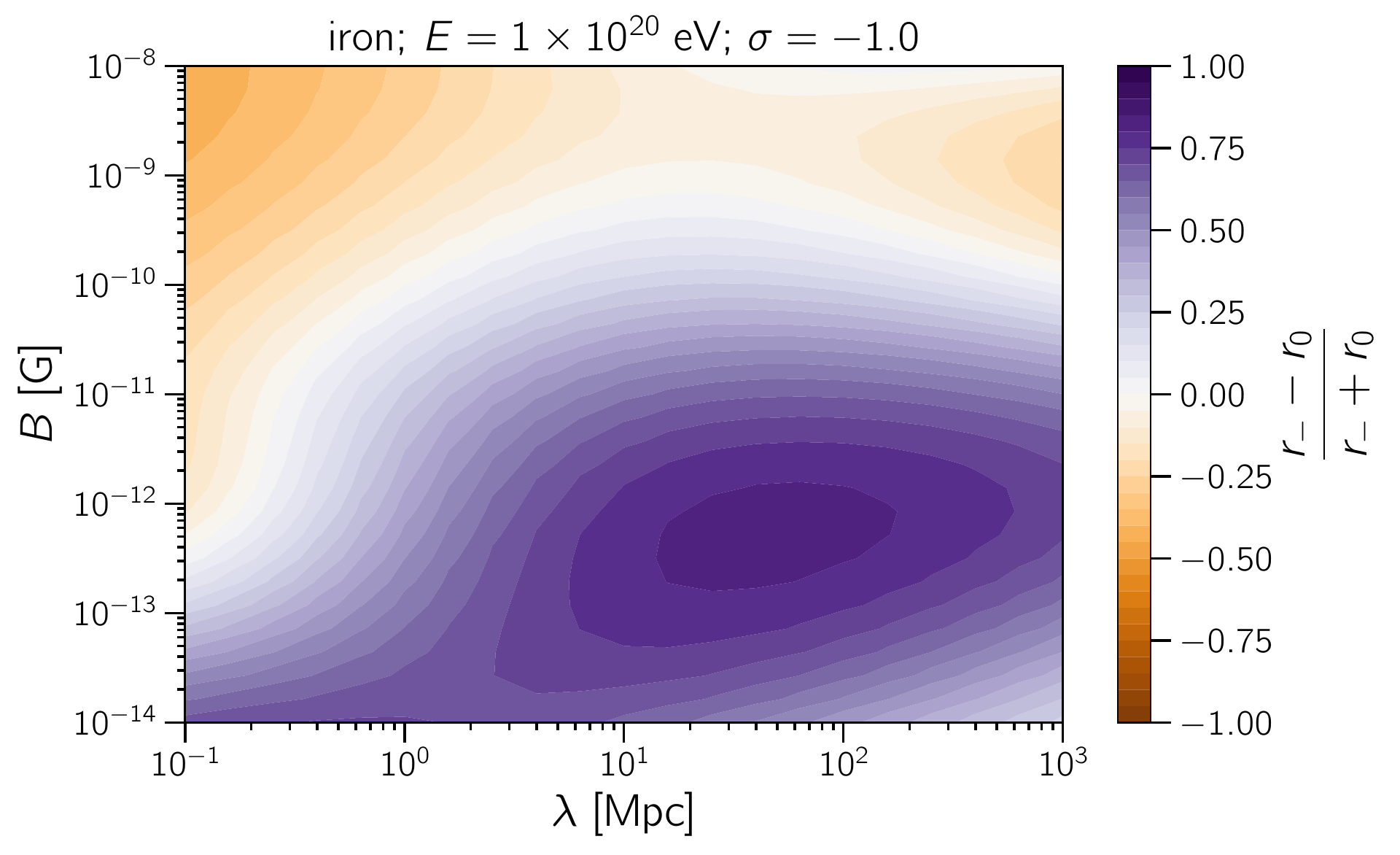}
  \includegraphics[width=0.495\columnwidth]{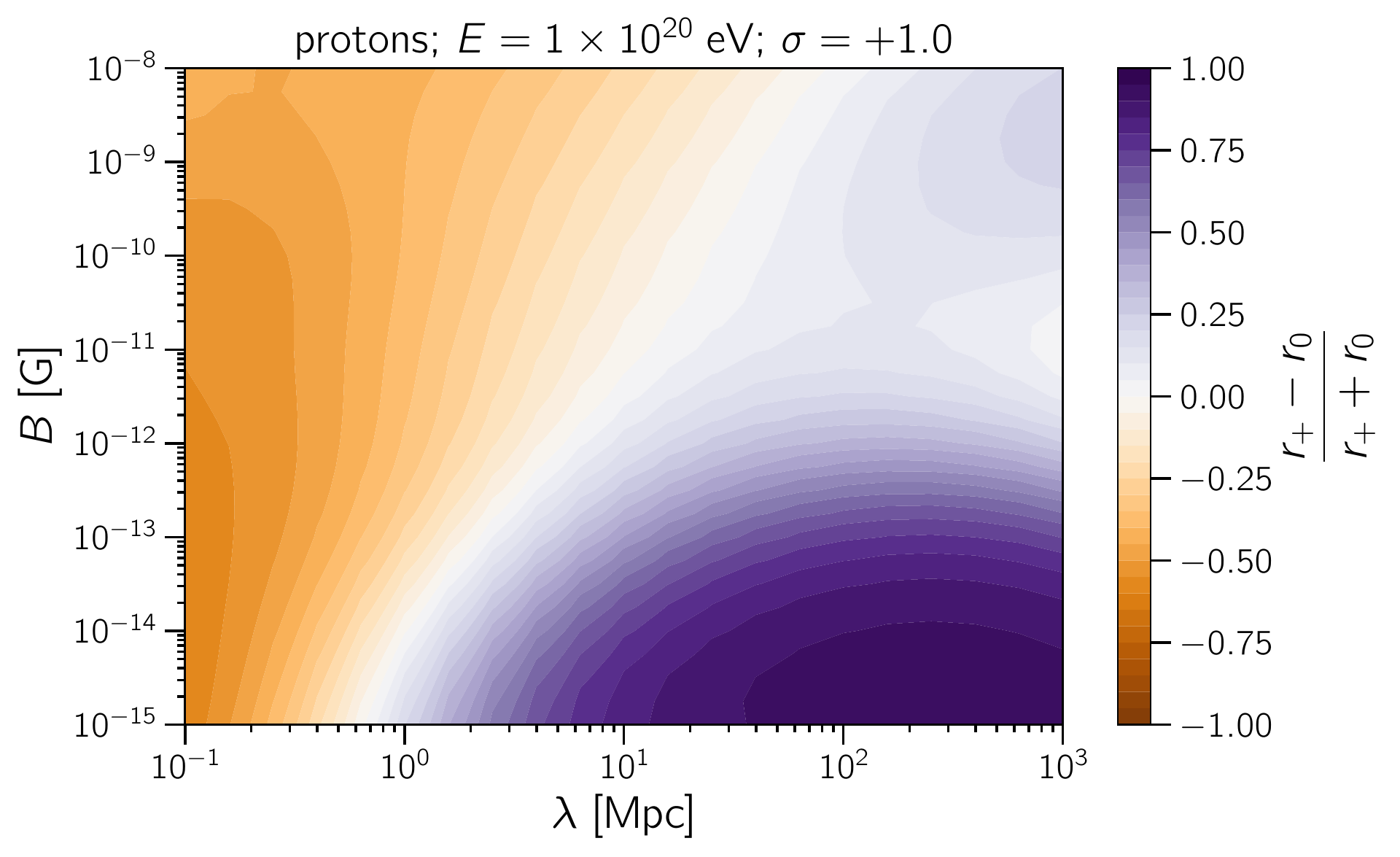}
  \includegraphics[width=0.495\columnwidth]{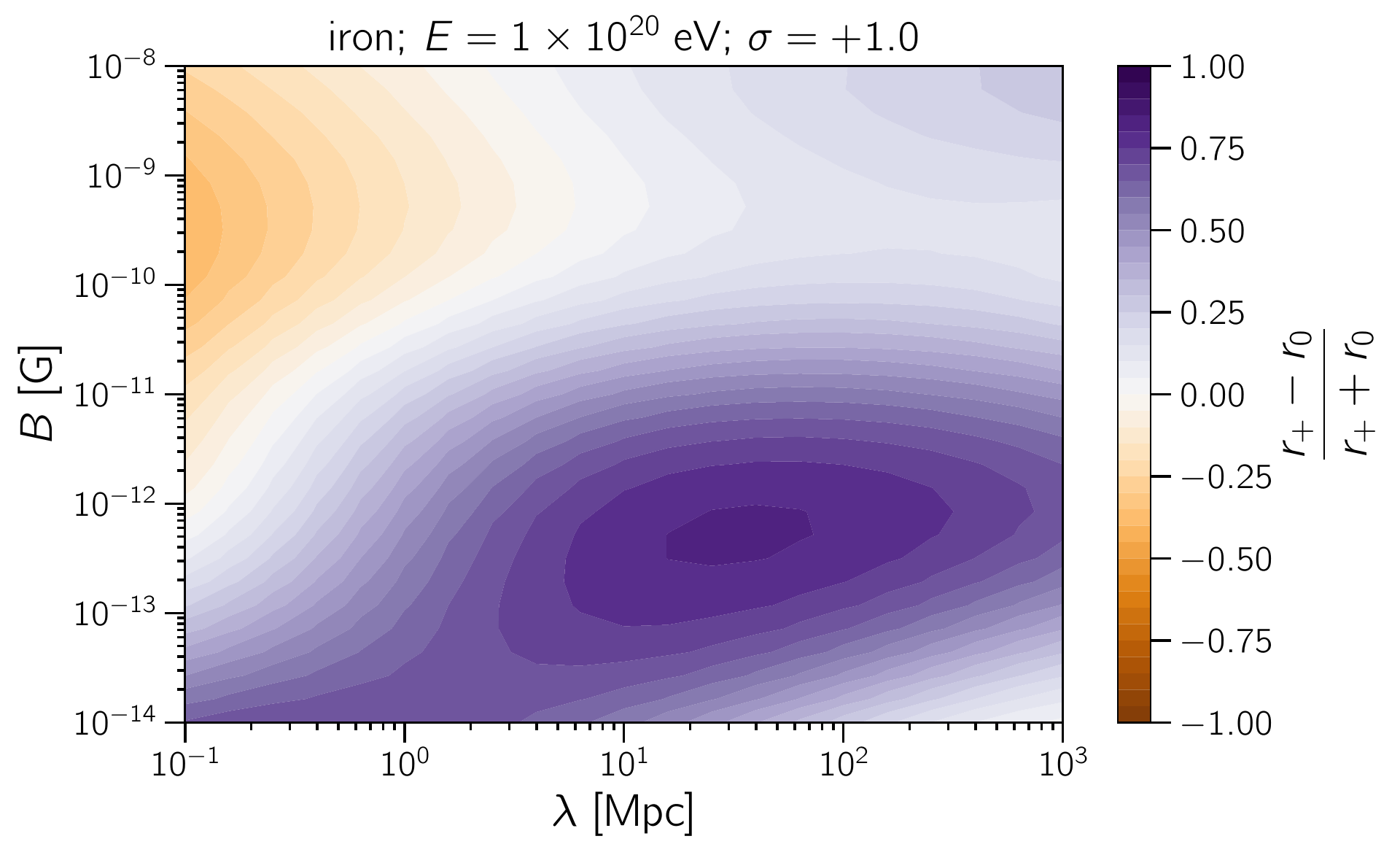}
  \caption{Dipole-to-quadrupole ratios ($r$) for different combinations of magnetic field strength ($B$) and coherence length ($\lambda$). The difference between corresponding scenarios for maximally positive helicities are shown on the upper panels, and on the  lower panels for maximally negative helicities. The upper row is for the case of proton primaries, whereas the lower row corresponds to iron primaries, for $10^{20} \; \text{eV}$ primary cosmic rays.}
  \label{fig:constraint}
\end{figure}

In Fig.~\ref{fig:constraint} we present the relative difference of the dipole-to-quadrupole ratio between $\sigma = \pm 1$ and $\sigma = 0$, for several combinations of $B$ and $\lambda$.

The interpretation of Fig.~\ref{fig:constraint} follows immediately from the discussion about the diffusive regime presented in Section~\ref{subsec:harmonicAnalysis}. We can distinguish two regimes for $E > E_c$. In the non-resonant diffusion case ($\lambda \ll D_\text{min}$), we see that the observable chosen ($\Delta r_- \equiv (r_- - r_0) / (r_- + r_0)$) enhances the contrast between maximally negative and null helicities. There is a region where $\Delta r_- > 0$, approaching 1 for high values of $\lambda$ and low $B$. 
Note that $\lambda$ is constrained by the Hubble horizon, i.e., $\lambda \lesssim 4000 \; \text{Mpc}$, thus implying $B \lesssim 10^{-12} \; \text{G}$ in the quasi-rectilinear regime.  Because the nearest sources are located at $D_\text{min} \sim 20 \; \text{Mpc}$, as described in Section~\ref{sec:Simulations}, for protons with $E \gtrsim 10^{19} \; \text{eV}$ we expect a transition from non-resonant to resonant diffusion around $\mathcal{B} \sim 10 \; \text{Mpc} \, \text{nG}$ for $\lambda \sim D_\text{min}$, wherein $D_\text{min}$ is the distance to the closest (or brightest) sources. Similar arguments apply to the iron case (right column). 

It is tempting to attempt to derive a relation for $\sigma$ as function of $r$, i.e., $\sigma(r)$. However, this relation is highly dependent upon the model of choice and very sensitive to the distribution of sources. 

For any non-maximal helicities, i.e., if $| \sigma | < 1$, we expect both the dipole and the quadrupole to be fainter with respect to the cases of $| \sigma | = 1$. Because the assumptions made in this first study are very simple, it is not instructive to study the scenarios with $|\sigma| < 1$. 

Note that the estimation of $|\sigma|$ depends on $r$, not on $w_{\rm d}$ and $w_{\rm q}$ alone. Therefore, it is not unreasonable to expect $r$ to provide satisfactory measurements of the absolute value of the helicity even for small $|\sigma|$.

\begin{figure}
  \includegraphics[width=0.495\columnwidth]{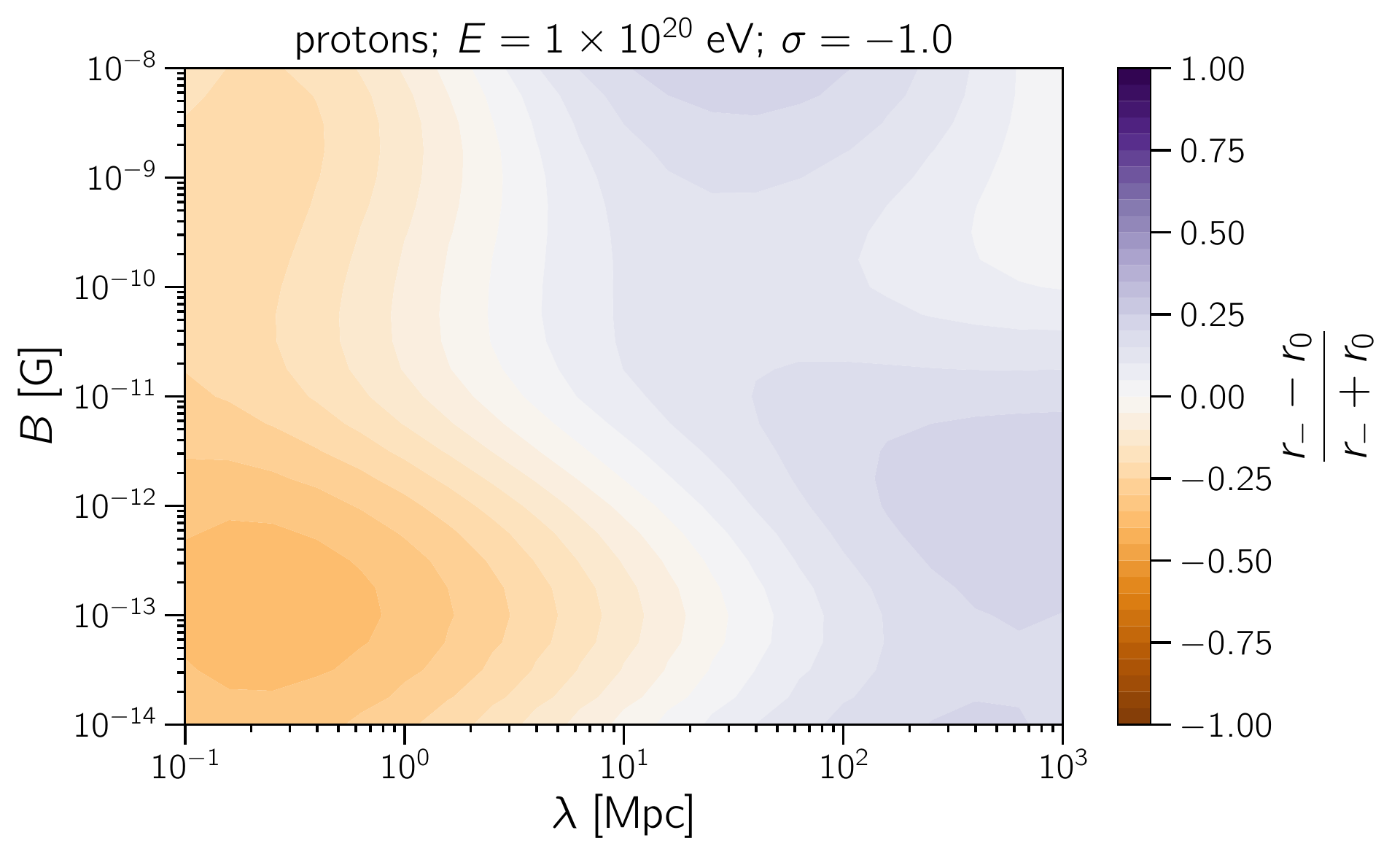}
  \includegraphics[width=0.495\columnwidth]{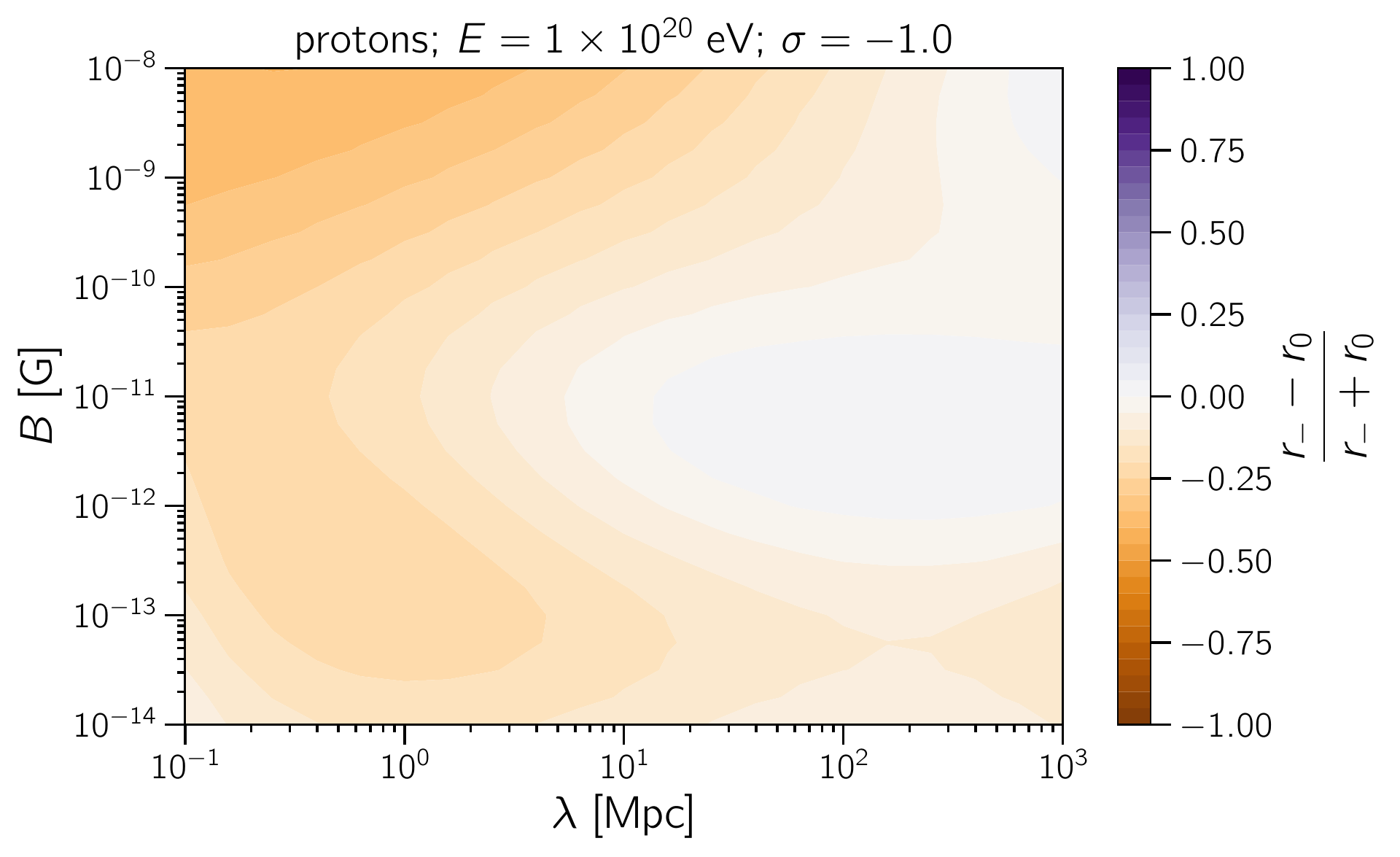}
  \includegraphics[width=0.495\columnwidth]{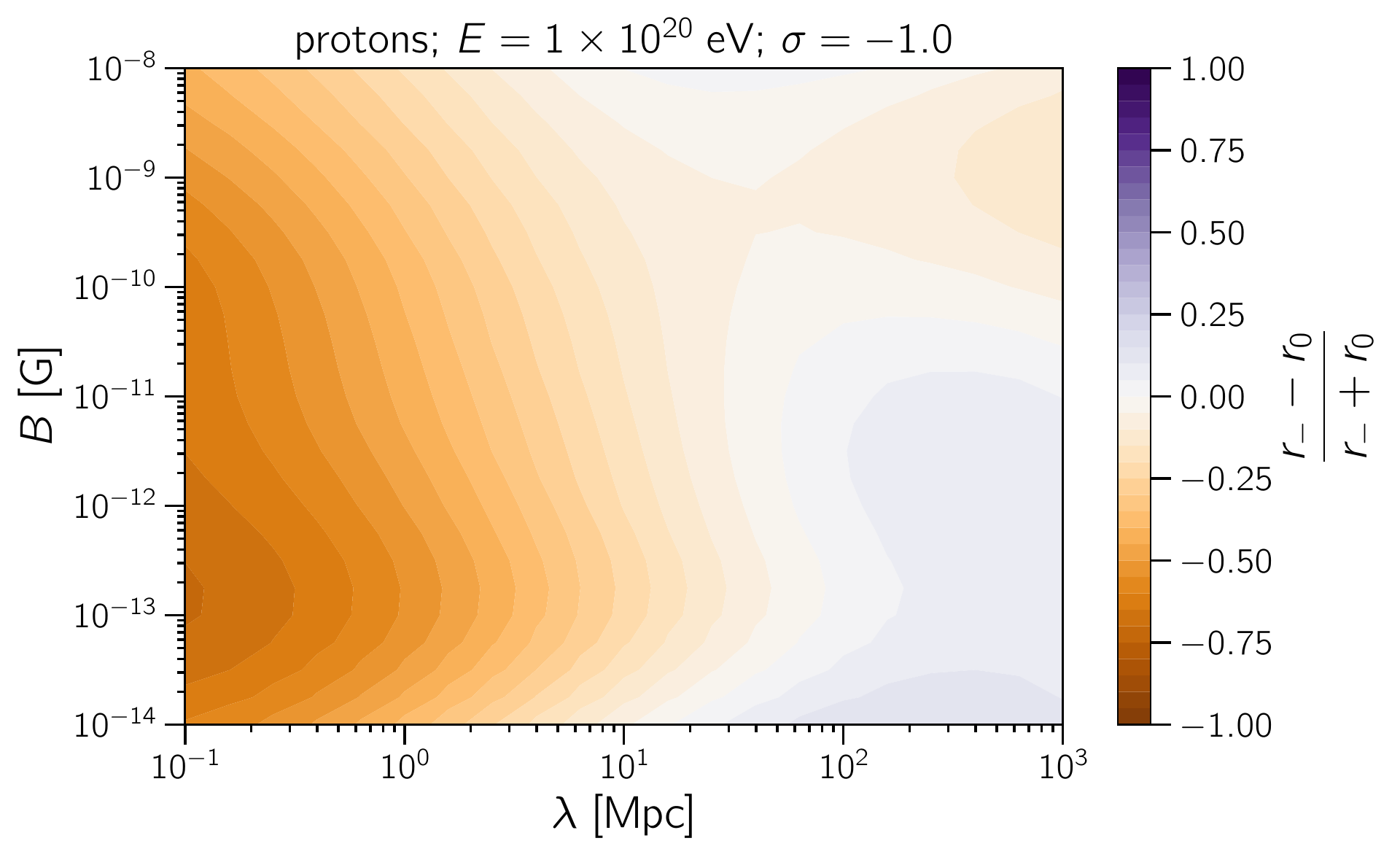}
  \includegraphics[width=0.495\columnwidth]{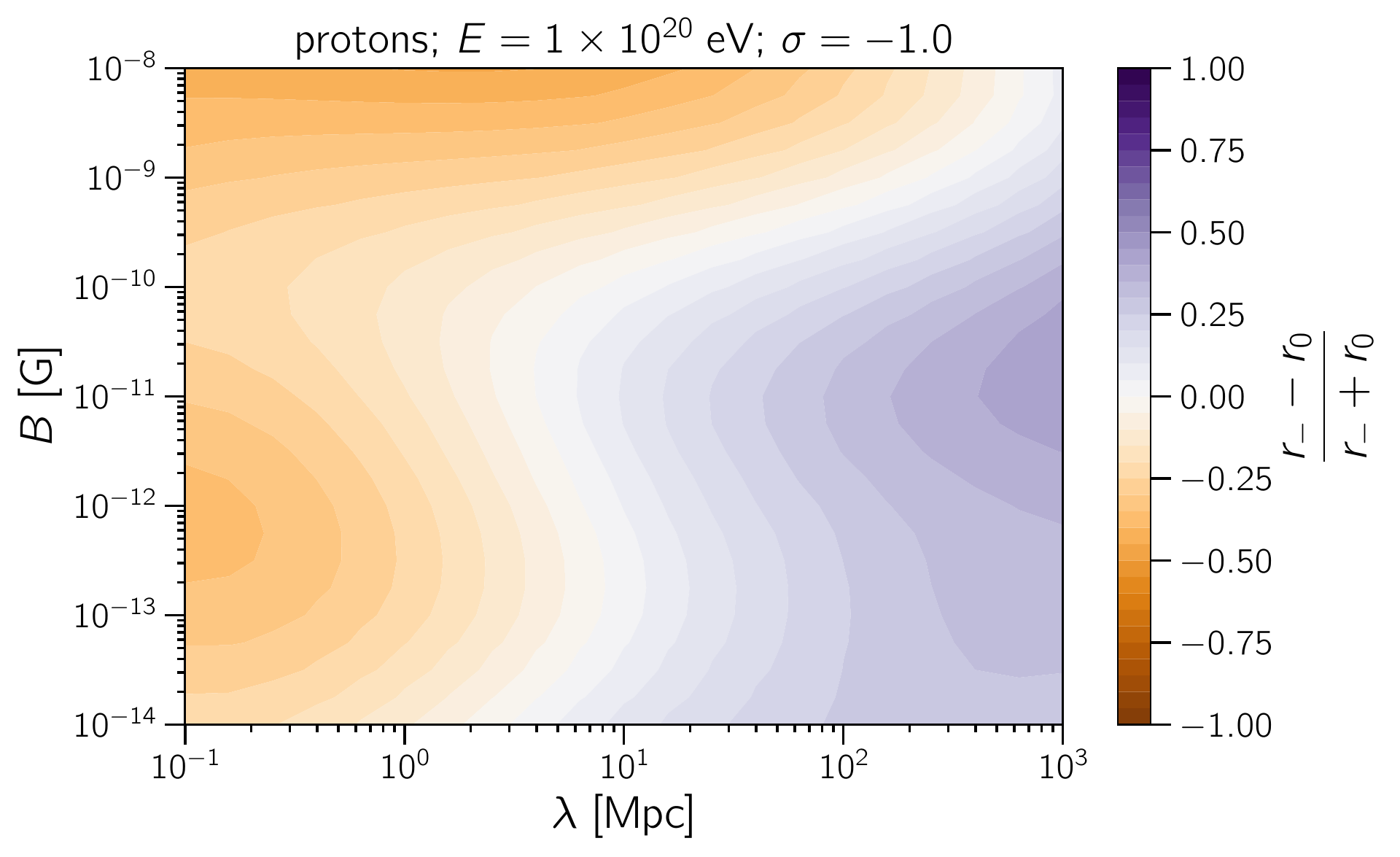}
  \caption{Dipole-to-quadrupole ratios ($r$) for different combinations of magnetic field strength ($B$) and coherence length ($\lambda$), considering the effect of the galactic magnetic field. The upper left panel corresponds to scenario 0, the upper right to scenario 1, and the lower left and lower right panels to scenarios 2 and 3, respectively, as defined in Table~\ref{tab:gmf}. These results are for maximally negative helicities, assuming proton primaries with $10^{20} \; \text{eV}$.}
  \label{fig:constraint_gmf}
\end{figure}

We have investigated the impact of the galactic magnetic field on dipole-to-quadrupole ratio. For that, we considered the scenarios shown in Table~\ref{tab:gmf}. We compute the values of $r(B,\lambda)$, as in Fig.~\ref{fig:constraint}. The results are shown in Fig.~\ref{fig:constraint_gmf}. One can see that this observable is sensitive to the effects of the GMF, which severely restricts the region of the parameter space that can be probed with this method. In some of the scenarios  $|\Delta r_\pm |$ is small, thus compromising the measurement of the absolute value of the helicity.

\section{Discussion}\label{sec:discussion}

In this work we have assumed simple models with homogeneous helical IGMFs. In reality, the distribution of magnetic fields is closer to turbulent than to uniform. However, in the limit of large coherence lengths, the local magnetic field effectively behaves as uniform for sources that are distant approximately less than one coherence length, i.e., $R_\text{s} \lesssim \lambda$. Thus it is reasonable to consider the case of a uniform magnetic field for a first study, since large coherence lengths, in particular, arise in models with non-zero helicity via inverse cascade, which enhances the transfer of power from small to large scales. It is worth stressing that the coherence scale of IGMFs is poorly constrained, lying between $10^{-12} \; \text{Mpc}$ and $10^{3} \; \text{Mpc}$~\cite{DuNe}. 

Our analysis was motivated by helical IGMFs. However, the exact same arguments can be applied to study helical magnetic fields in structures such as filaments and clusters, provided that the distribution of sources does not mask the signal. We have considered a distribution of sources extending up to a distance, $D_\text{max}$, which relates to the minimal energy of interest in the analysis. To probe IGMFs up to very large distances one could, for instance, decrease the minimal energy considered in the analysis. However, for $E \simeq 10^{18.7} \; \text{eV}$ there could be a contamination of the signal by a possible galactic component~\cite{Deligny:2014opa}, so that the search of signatures of helical IGMFs should be done above this threshold.

In Ref.~\cite{Kahniashvili:2005yp} the authors suggest that some specific source distributions could be used to constrain helicity.
In our work, we consider a more general case, showing that even if the sources are distributed isotropically, the cosmic-ray arrival directions can be anisotropic. In the context of our analysis, this follows from the minimum energy threshold we have imposed for the arriving particles.
Note, however, that our treatment is not generic and knowledge about the source positions is required. Thus, some sources located at similar distances would need to be identified before the helicity of IGMFs could be constrained.

If at least one single source of UHECRs were known, this could be enough to constrain the helicity of the intervening magnetic field, depending on the angle between the line of sight and the magnetic field. In this case, energy-ordered multiplets could be detected~\cite{Zimbres:2013zba,Aab:2014dha}. Nevertheless, past analyses by Auger have not found any indications of multiplets in their data~\cite{Abreu:2011md}.

At the highest energies ($E \sim 10^{20} \; \text{eV}$), the typical energy loss length for cosmic-ray protons is $\sim 100 \; \text{Mpc}$, and $\sim 300 \; \text{Mpc}$ for iron, being of the order of $\sim 1 - 100 \; \text{Mpc}$ for helium and intermediate-mass nuclei such as nitrogen, carbon, and oxygen~\cite{kotera2011a}. Therefore, at these energies UHECRs originate in our local Universe, being the distance of the closest sources possibly comparable to typical coherence lengths, within a factor of a few. This means that even our simple scenarios with uniform magnetic fields may be used to adequately constrain the helicity of intervening magnetic fields. Note that such composition is favoured by Auger measurements~\cite{Aab:2014aea}.

The ideas outlined here could, in principle, be compared with experimental data, as long as one bears in mind the simplified nature of our assumptions. For instance, magnetic fields are not homogeneous. Instead, in order to recreate a more realistic situation, one has to choose a magnetogenesis scenario and carry out numerical or (semi-)analytical simulations of the time evolution of the IGMFs from their creation to the present day~\cite{Fromang:2006aw,Tevzadze:2012kk,Kahniashvili:2012uj,Kahniashvili:2012vt,Saveliev:2012ea,Saveliev:2013uva,Brandenburg:2014mwa,Kahniashvili:2015msa,DokMath.95.1.68,AML.72.75}, resulting in complex stochastic field configurations. Furthermore, in order to constrain the helicity of intervening magnetic fields, besides improving the modelling of the magnetic field distribution, one would also need to take into account the non-uniform sky coverage of the observatory whose data is being analysed, which we have ignored. While this is relatively simple for dipolar patterns~\cite{Aublin:2005nv}, the analysis for quadrupoles is more intricate~\cite{Auger:2012an,Sommers:2000us,Billoir:2007kb}. 

The Pierre Auger Collaboration has recently reported~\cite{Aab:2017tyv} the existence of a dipole in the arrival directions of UHECRs with energies $E > 8 \times 10^{18} \; \text{eV}$. To which extent this result can be attributed solely to the distribution of UHECR sources or if magnetic fields dominate the anisotropy signal, is a matter of debate. While our model is too simple to be boldly compared with observations, we have shown that an enhancement in the dipole moment is possible for many configurations of helical magnetic fields, with respect to non-helical scenarios, as shown in Fig.~\ref{fig:dipoleAmplitude}.
\section{Conclusion and Outlook} \label{sec:CO}

In this paper we have shown, as a proof of principle, that it may be possible to constrain the helicity of magnetic fields using UHECRs. We have outlined a methodology to look for the imprints of helical fields in the arrival directions of UHECRs. By performing a harmonic analysis of simulated data sets, we have demonstrated that the direction to which the dipole points correlates with the sign of the helicity. We have also suggested that the ratio between the dipole and quadrupole amplitudes may be used to constrain the absolute value of the helicity. 

The galactic magnetic field does not compromise the measurement of the sign of the helicity. However, it may compromise the value of the absolute value of the helicity, depending on the orientation of the helical magnetic field with respect to the galactic plane.

In our analysis we have discussed in detail the case of IGMFs, but similar ideas can be applied to measure the helicity of magnetic fields in clusters or filaments, for example. Nevertheless, the conclusions that could be drawn would strongly depend upon the distribution of sources, which is unknown.

We have demonstrated the impact of magnetic helicity on UHECR propagation, and how this may affect UHECR arrival distributions. Given the elongation in the trajectory described by a cosmic ray in the presence of a helical magnetic field, it is not unreasonable to expect this to have an impact on observables other than the arrival directions, namely the spectrum and inferred composition. We will defer this investigation to future works.

If UHECR sources are ever found, the observation of energy-ordered multiplets could also be used to constrain the helicity of intervening magnetic fields, provided that enough events are detected. 

With this study we have laid the foundations for constraining magnetic helicity with UHECRs. In the future we intend to extend our analysis to more realistic cases of helical turbulent magnetic fields and source distributions. Then, by comparison with data collected by the two largest UHECR observatories, the Pierre Auger Observatory and the Telescope Array, it might be possible to constrain not only the sign, but also the absolute value of the magnetic helicity.

\acknowledgments

RAB is supported by grant \#2017/12828-4, S\~ao Paulo Research Foundation (FAPESP). The work of AS was supported by the Russian Science Foundation under grant no.~17-71-10040, carried out at the Immanuel Kant Baltic Federal University.
We thank G\"unter Sigl for valuable comments.

\appendix
\section{On the reliability of the predictions} \label{sec:AppIso}

We dedicate this appendix to address the seemingly counter-intuitive issue of how an isotropic distribution may lead to the anisotropies discussed in the previous sections. As described in Section~\ref{sec:Simulations}, the setup of the simulations is such that the sources lie on the surface of concentric spheres, aiming to mimic a homogeneous source distribution. 
Here we demonstrate that the anisotropy stems from the magnetic field configuration rather than from the source distribution.
To this end, we choose a scenario with $10^{20} \; \text{eV}$ protons, $B=10^{-9} \; \text{G}$, and $L_{\rm c} = 10 \; \text{Mpc}$. The arguments outlined in this section hold for all cases studied and this choice is for illustration purposes only.

\begin{figure}
  \includegraphics[width=0.325\columnwidth]{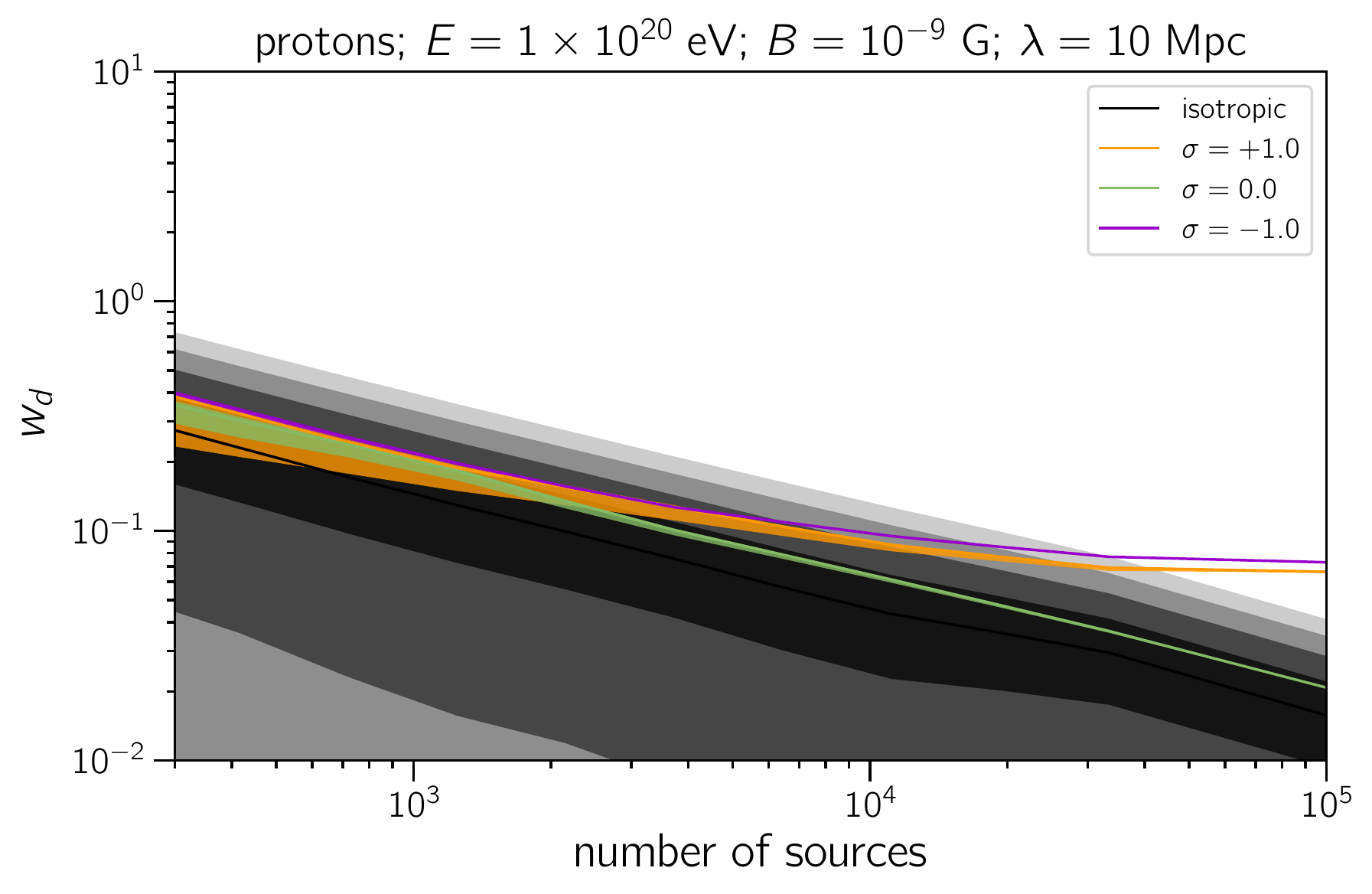}
  \includegraphics[width=0.325\columnwidth]{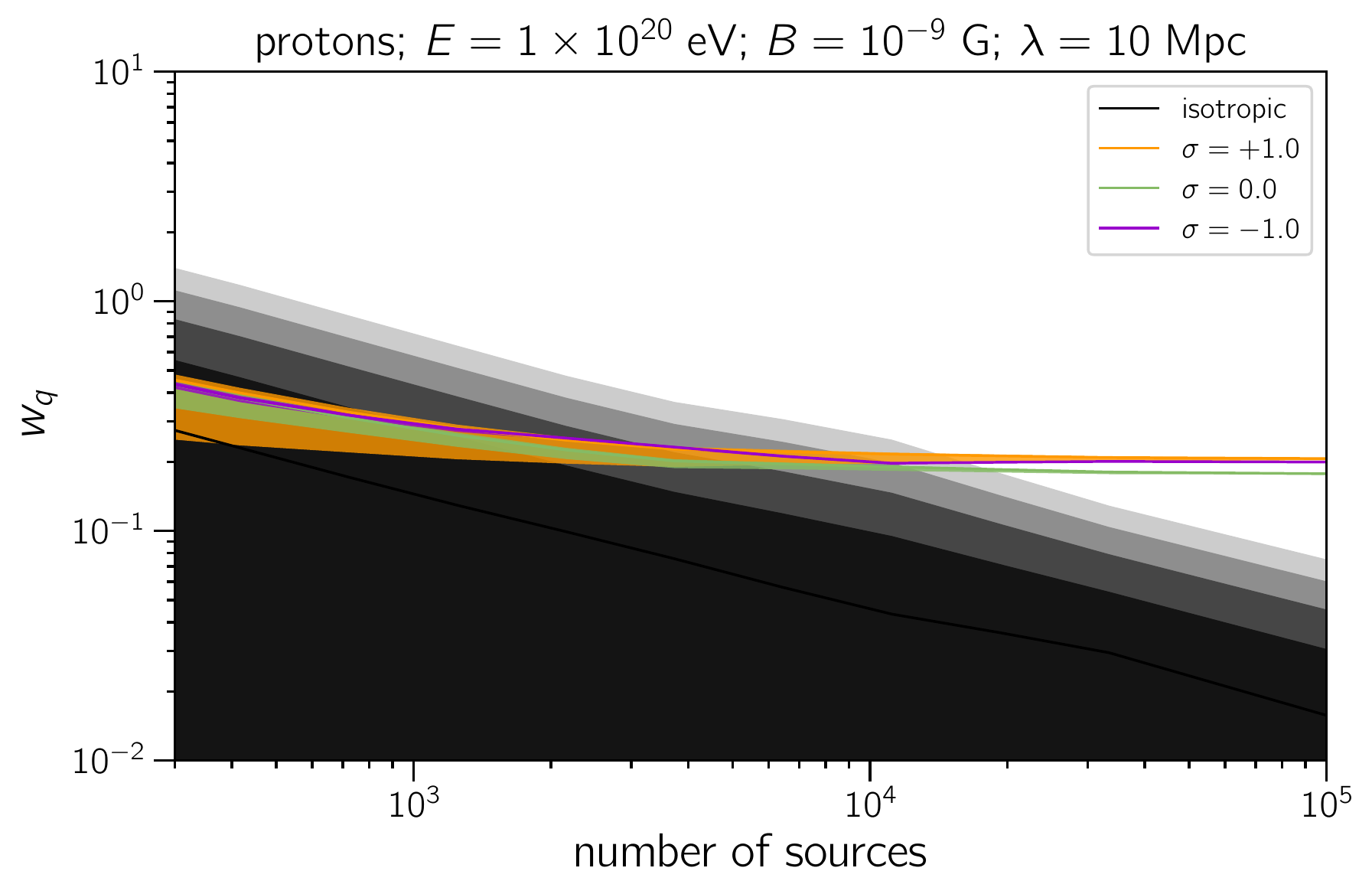}
  \includegraphics[width=0.325\columnwidth]{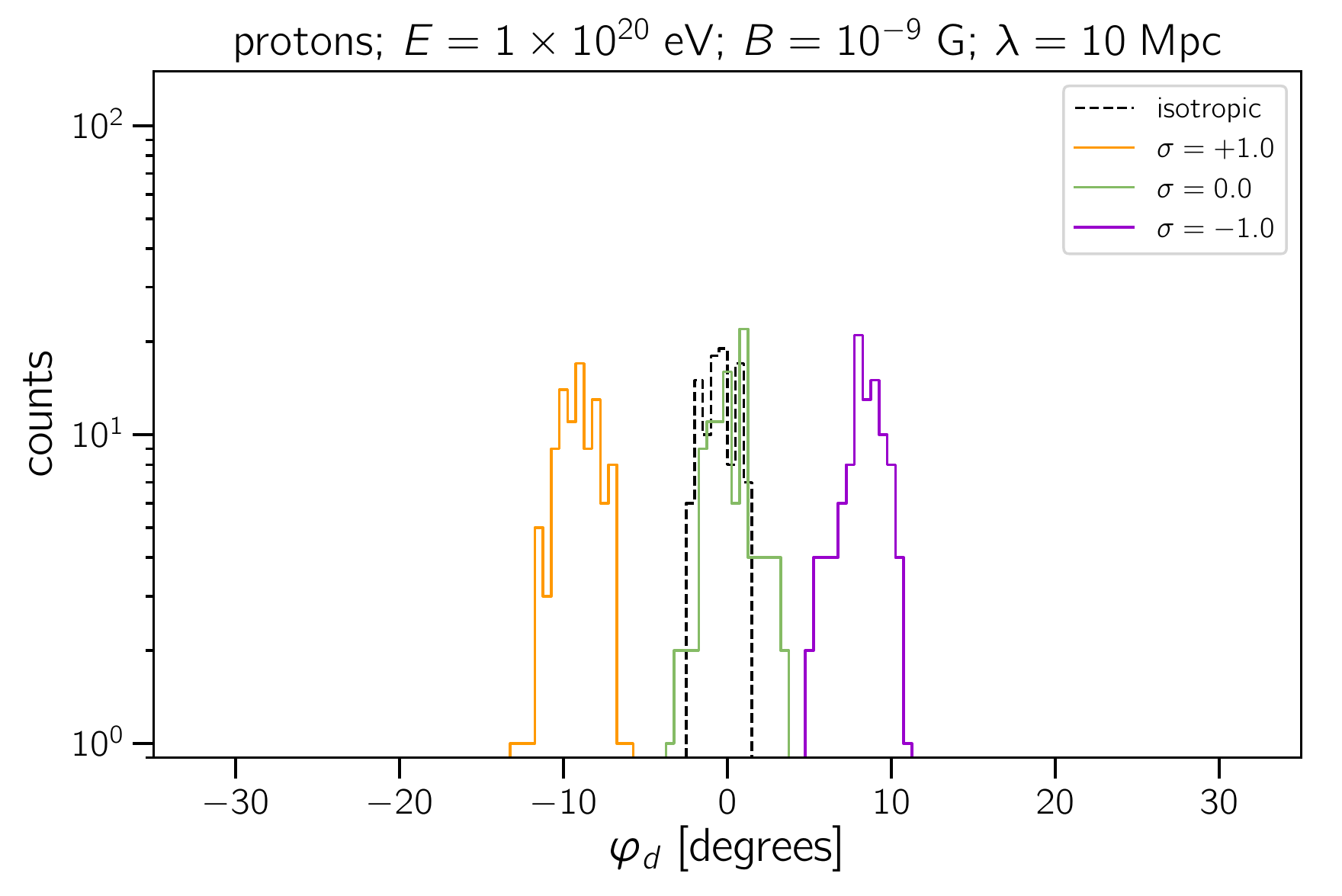}
  \caption{The dipole (left) and quadrupole (middle) are shown as a function of the number of sources contributing to the signal. The bands indicate the confidence intervals from one to four standard deviations, from darker to lighter shades of grey, respectively. The distribution of azimuthal angles ($\varphi_\text{d}$) of the dipole is shown for 100 realisations (right panel). Orange lines correspond to the $\sigma=+1$ case, green lines to $\sigma=0$, and purple lines to $\sigma=-1$.}
  \label{fig:spheres_combined}
\end{figure}

\begin{figure}
  \includegraphics[width=0.325\columnwidth]{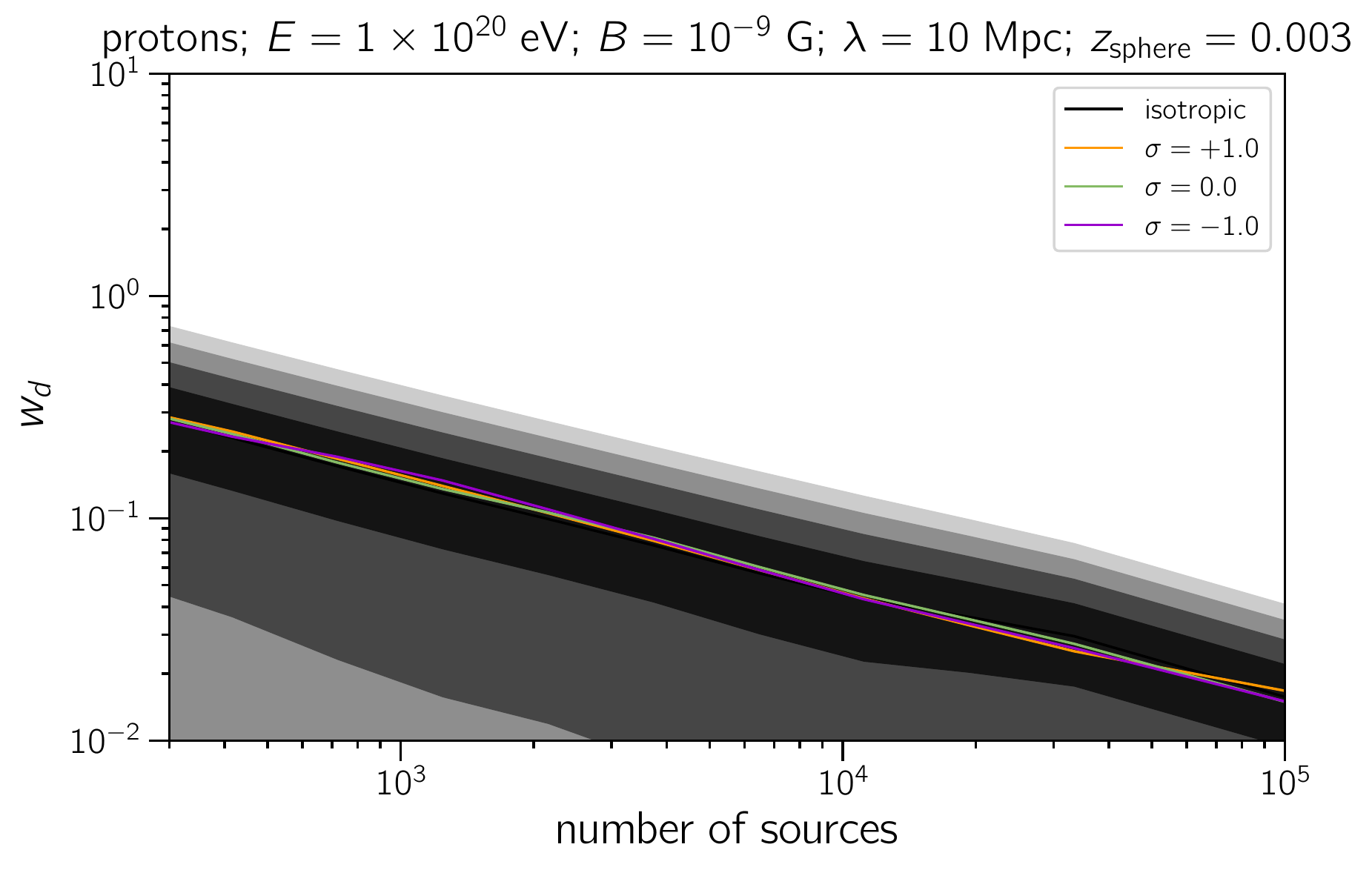}
  \includegraphics[width=0.325\columnwidth]{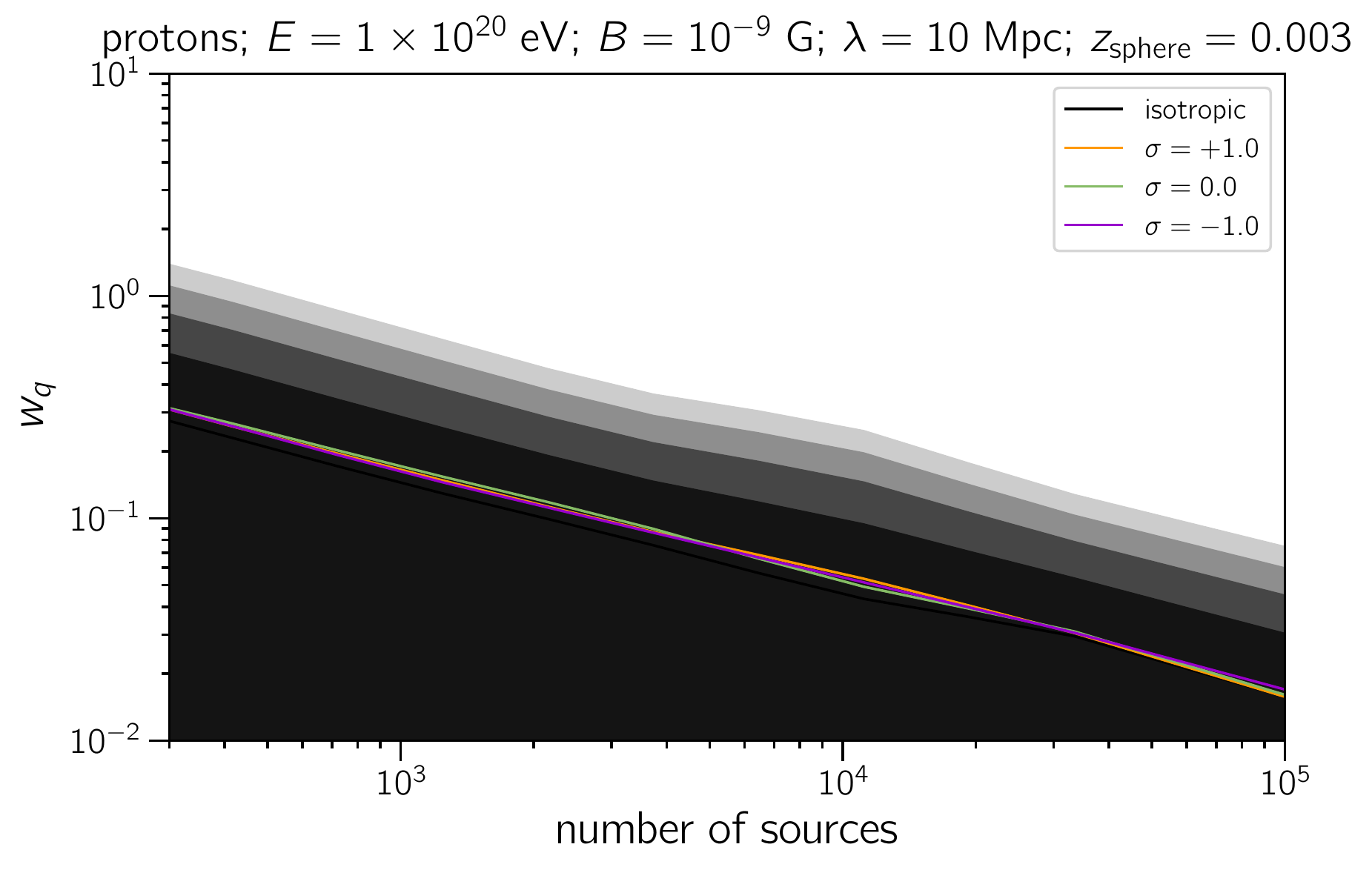}
  \includegraphics[width=0.325\columnwidth]{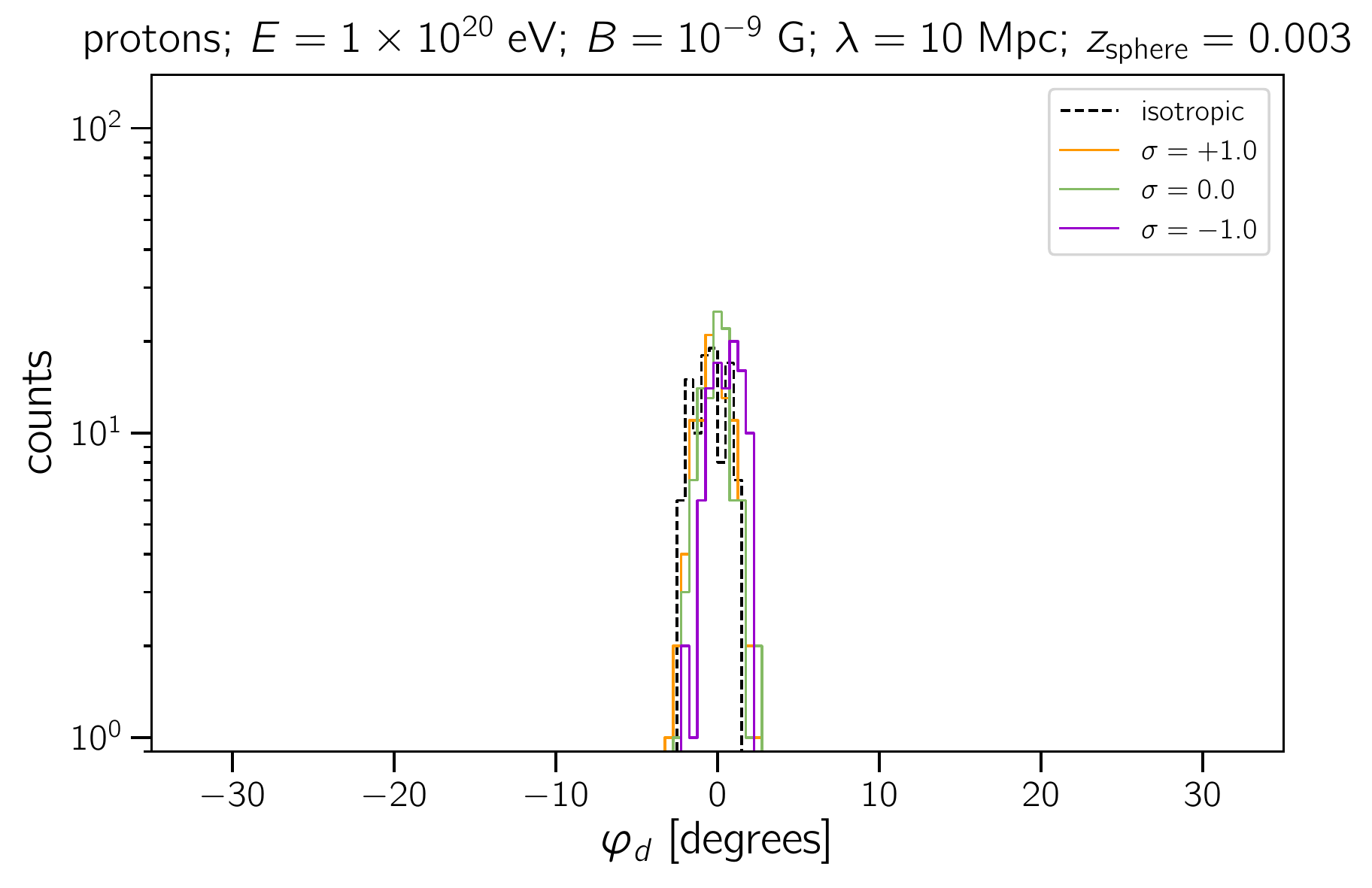}
  \includegraphics[width=0.325\columnwidth]{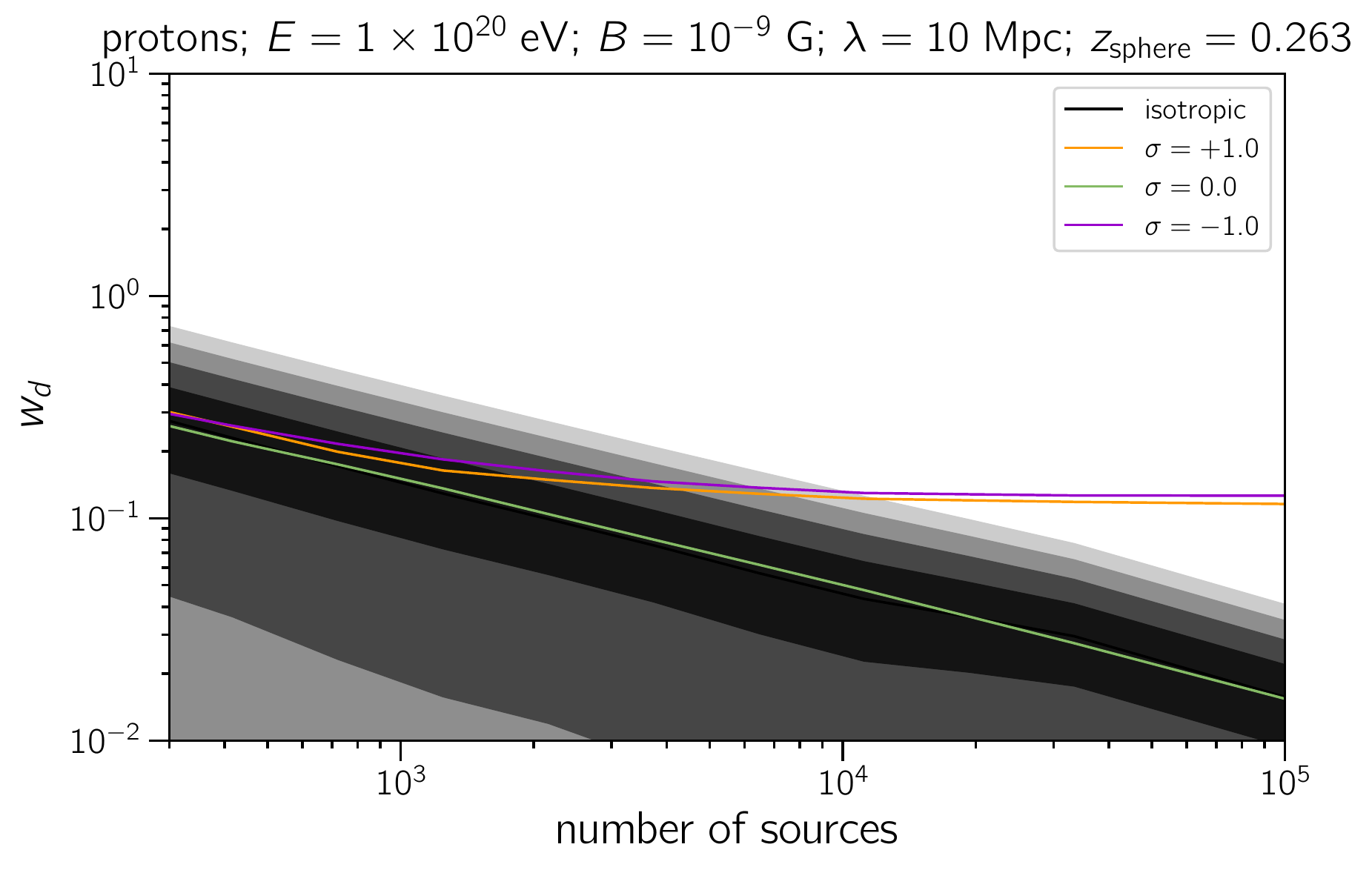}
  \includegraphics[width=0.325\columnwidth]{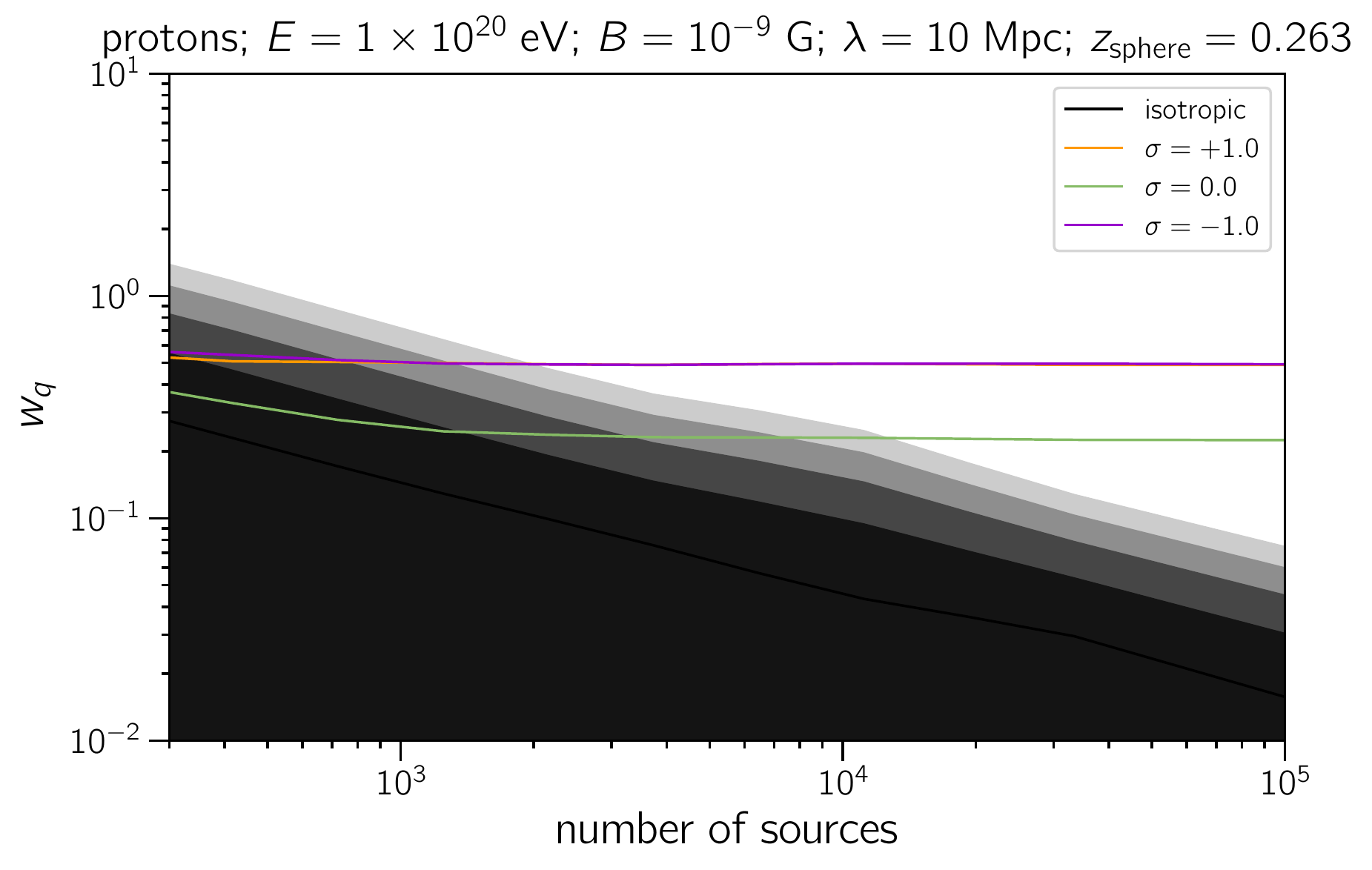}
  \includegraphics[width=0.325\columnwidth]{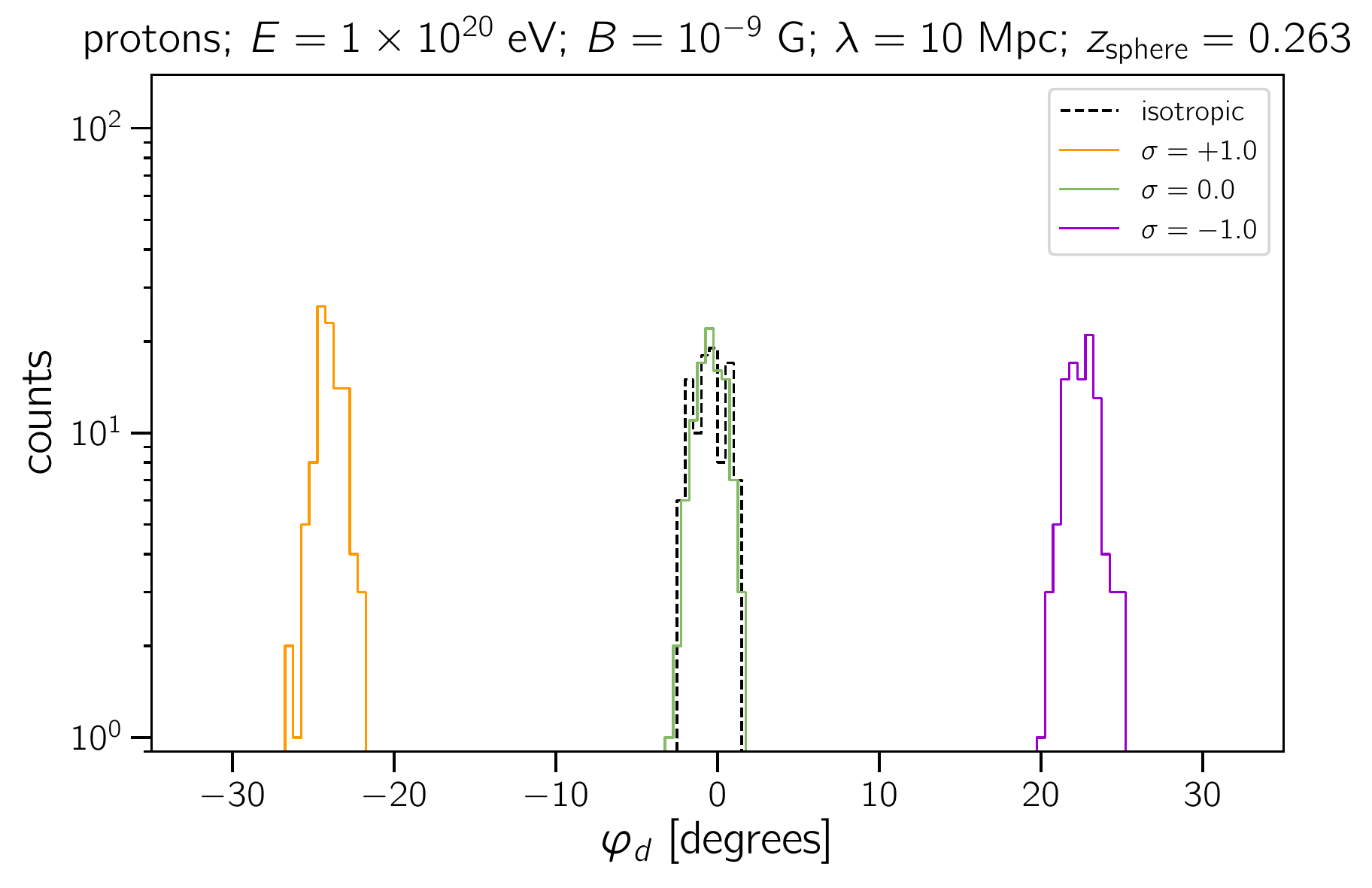}
  \caption{The dipole (left) and quadrupole (middle) are shown as a function of the number of sources contributing to the signal. The bands indicate the confidence intervals from one to four standard deviations, from darker to lighter shades of grey, respectively. The distribution of azimuthal angles ($\varphi_\text{d}$) of the dipole is shown for 100 realisations (right panel). Orange lines correspond to the $\sigma=+1$ case, green lines to $\sigma=0$, and purple lines to $\sigma=-1$. The upper panels are for a sphere located at $z_\text{sphere} \approx 0.003$, and the lower panels correspond to $z_\text{sphere} = 0.263$.}
  \label{fig:individualSpheres}
\end{figure}

We first compare the total source distribution, comprised of all spheres, with the case of $B=0$, which by construction is an isotropic distribution. In Fig.~\ref{fig:spheres_combined} (left), we analyse the behaviour of the dipole ($w_{\rm d}$) and quadrupole amplitudes ($w_{\rm q}$) as a function of the number of events. We also present the distribution of the azimuthal angles in the simulation frame ($\varphi_{\rm d}$), which is intrinsically connected to the sign of the magnetic helicity, for 100 realisations; this is shown in Fig.~\ref{fig:spheres_combined} (right).

From Fig.~\ref{fig:spheres_combined} one notices that this particular scenario can be distinguished from isotropy if more than $\gtrsim 10^{4} - 10^{5} $  sources are contributing to the signal; for equally luminous sources, this number is approximately the number of events required. Furthermore, the distribution of the dipole azimuthal angles ($\varphi_\text{d}$) suggests that this observable allows us to clearly distinguish the $\sigma=\pm 1$ from the $\sigma=0$ and isotropic case.

Fig.~\ref{fig:individualSpheres} is similar to Fig.~\ref{fig:spheres_combined}, but it corresponds to single spheres located at $D_\text{sphere} \approx 14.24 \; \text{Mpc}$ ($z_\text{sphere} \approx 0.0032$) (upper row) and $D_\text{sphere} \approx 1096 \; \text{Mpc}$ ($z_\text{sphere} \approx 0.263$) (lower row). Once again, the estimation of the absolute value of the helicity could be done if at least $\sim 10^{4}$ sources contribute to the signal, or $\sim 10^3$ events in the case of a uniform luminosity distribution. For other scenarios the required number of events for the analysis is also $\sim 10^4$, although for larger coherence lengths this may decrease to $\sim 10^{3}$. Hence, by using $10^{5}$ sources for each sphere in our simulations, we have ensured that the observed anisotropy signal is indeed a consequence of the (helical) magnetic field configuration and \textit{not} the result of insufficient statistics.

\begin{figure}
	\includegraphics[width=0.325\columnwidth]{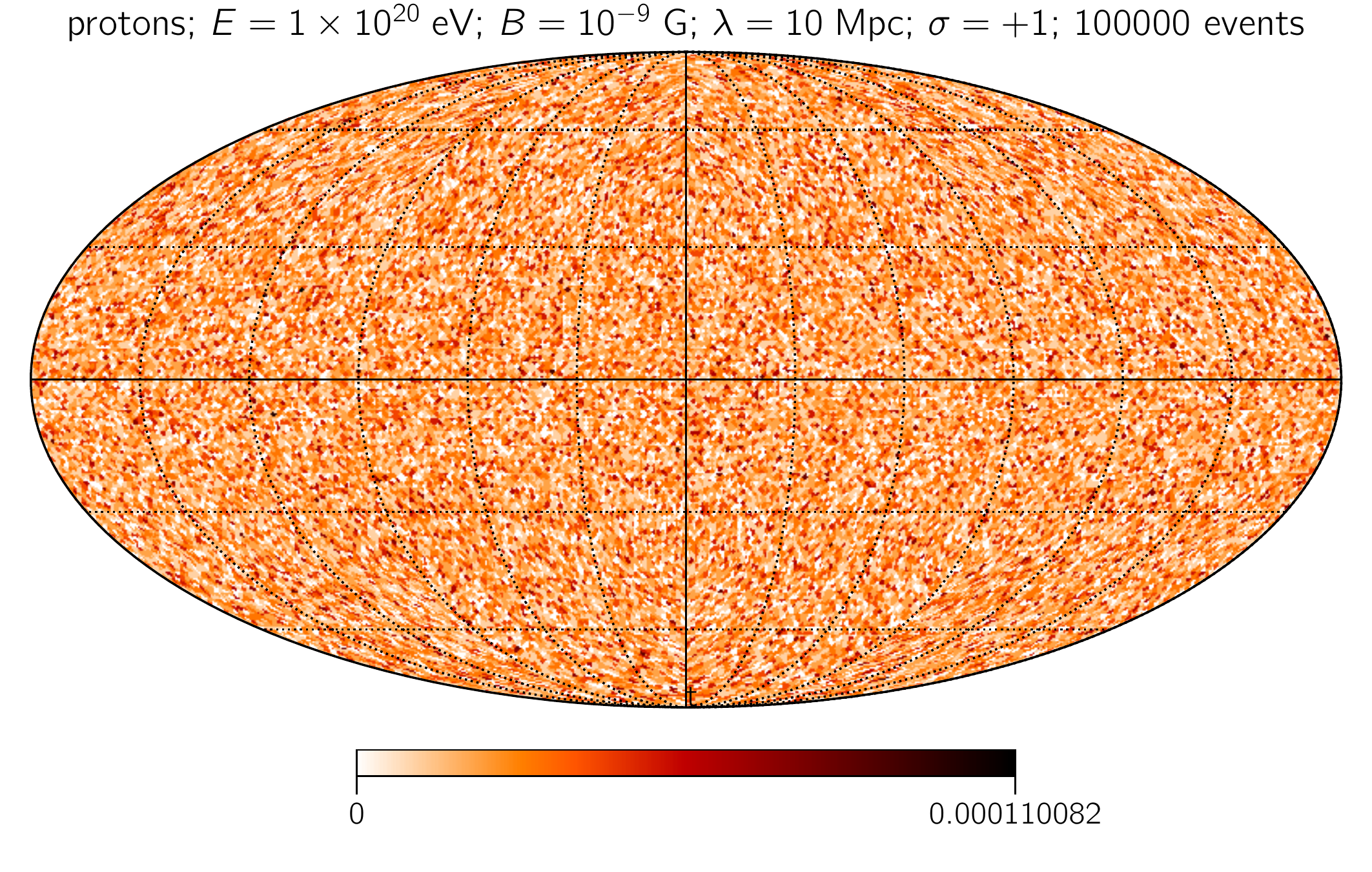}
	\includegraphics[width=0.325\columnwidth]{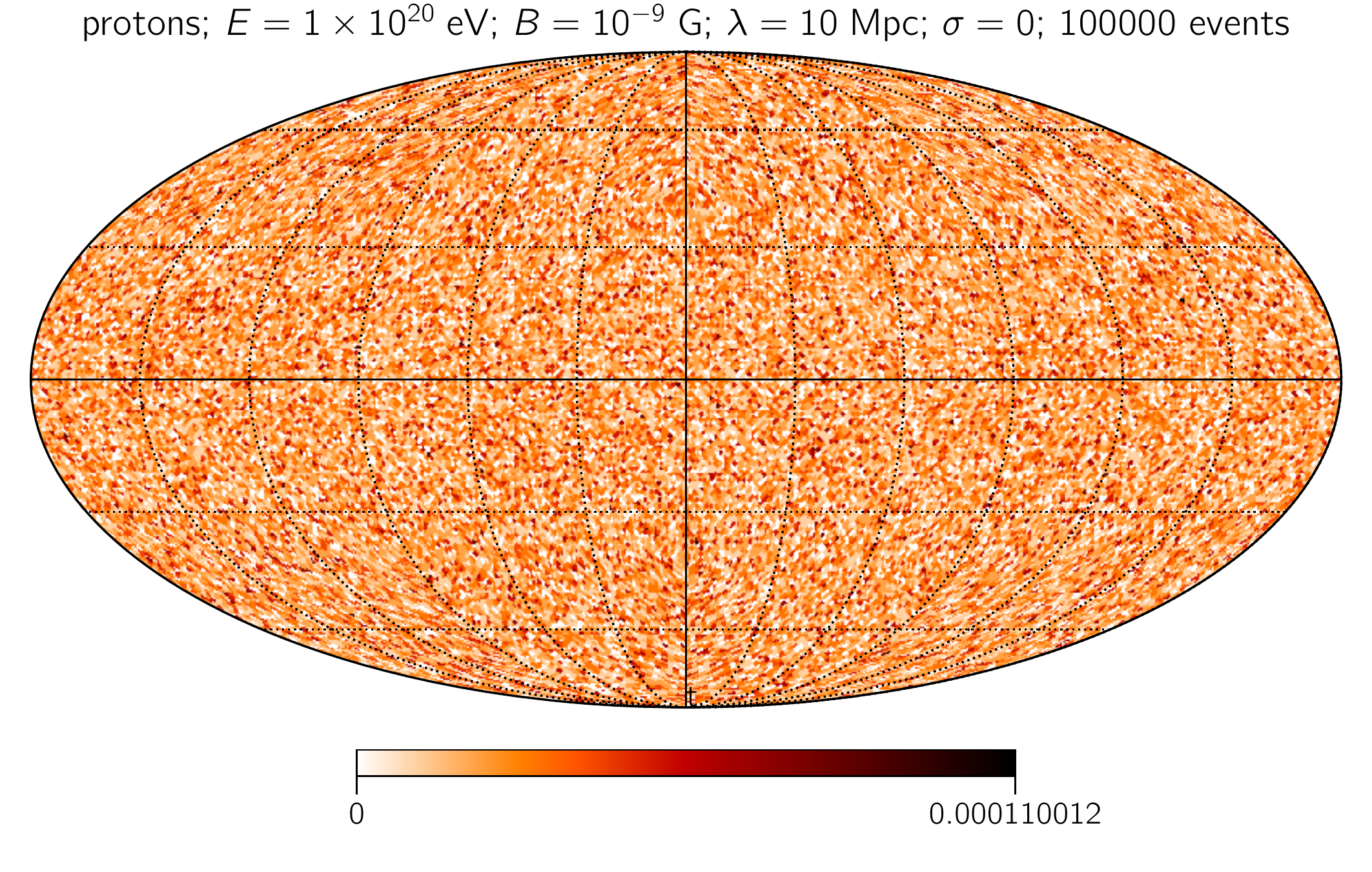}
	\includegraphics[width=0.325\columnwidth]{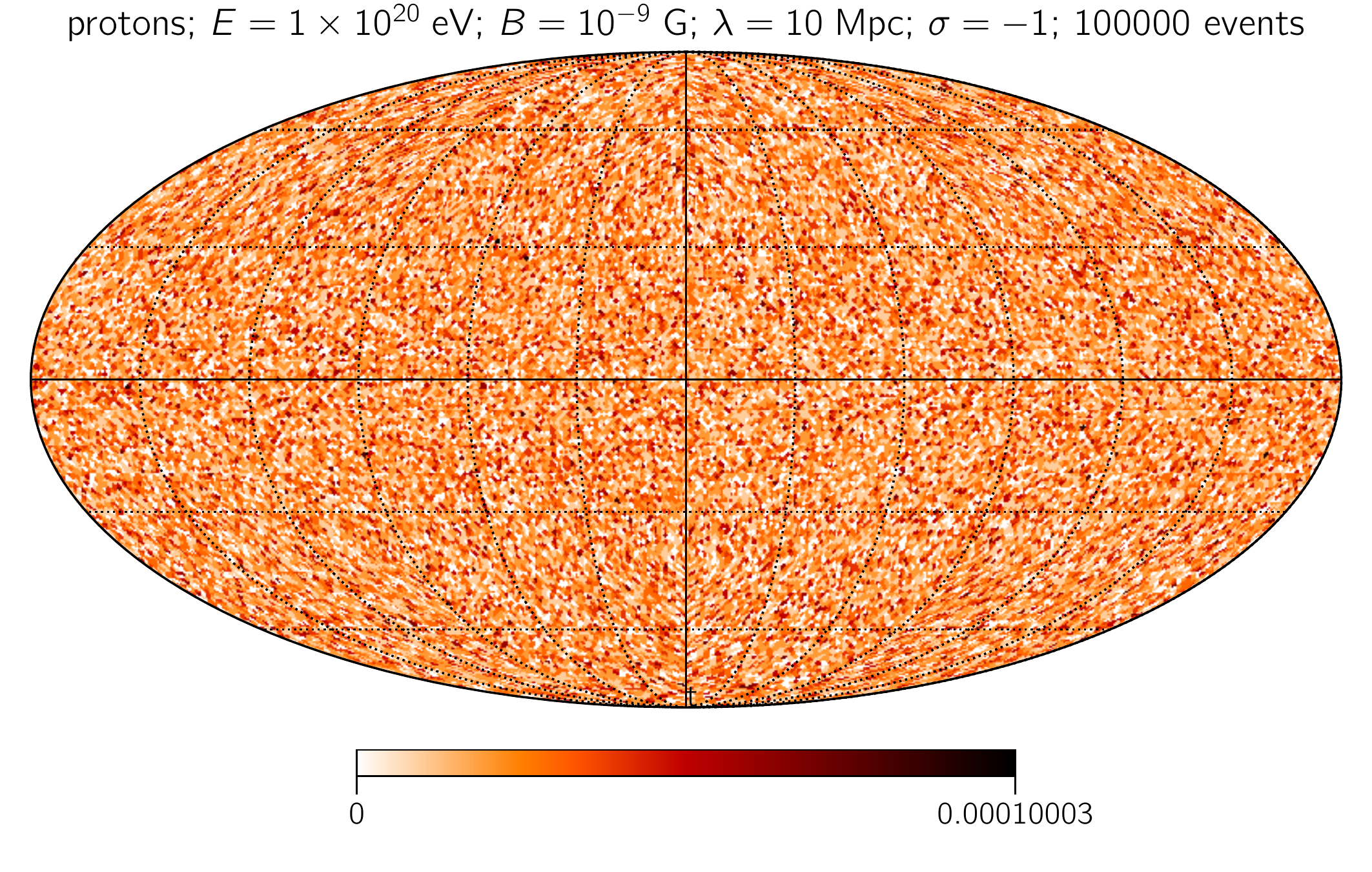}
	\includegraphics[width=0.325\columnwidth]{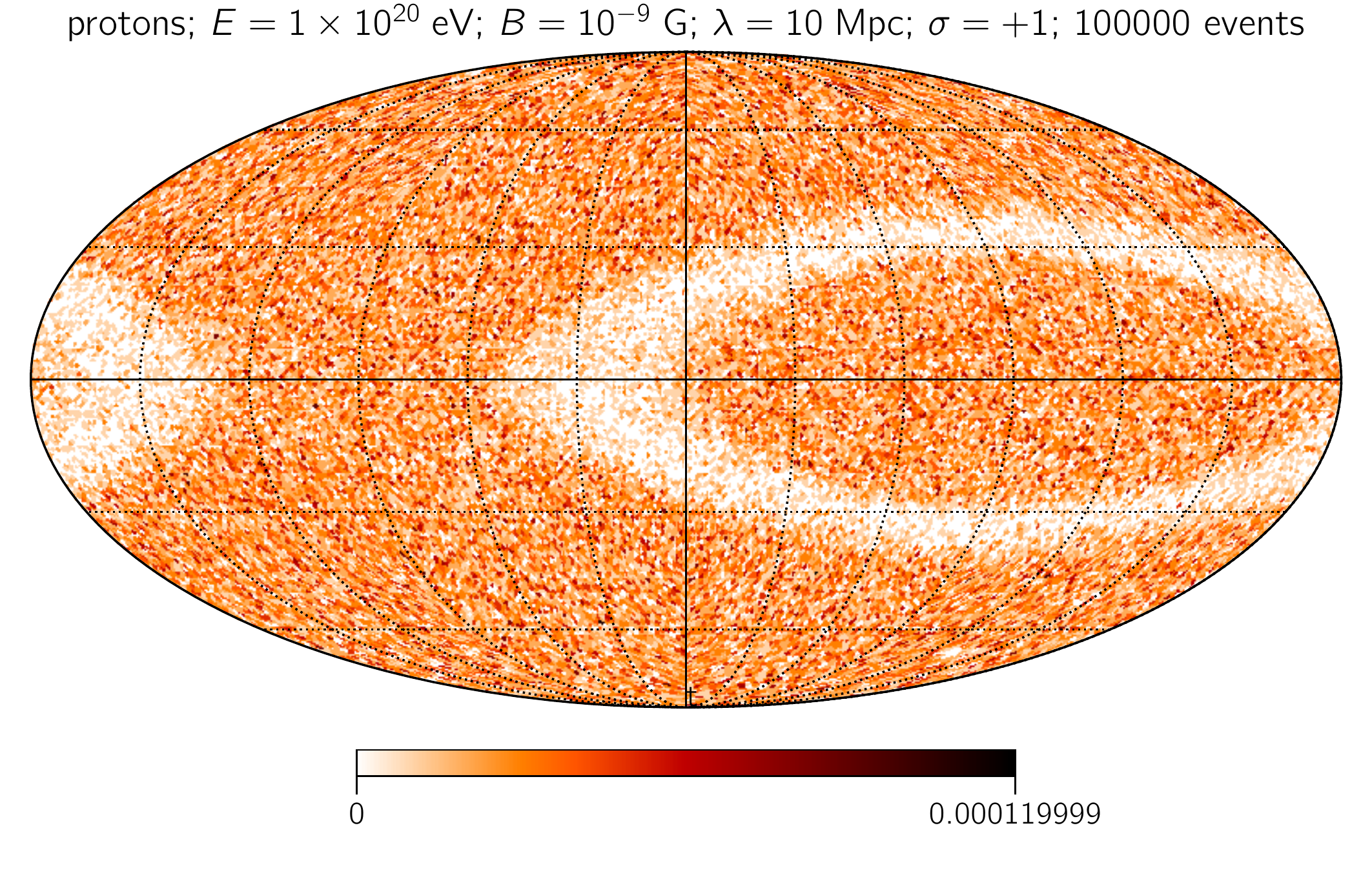}
	\includegraphics[width=0.325\columnwidth]{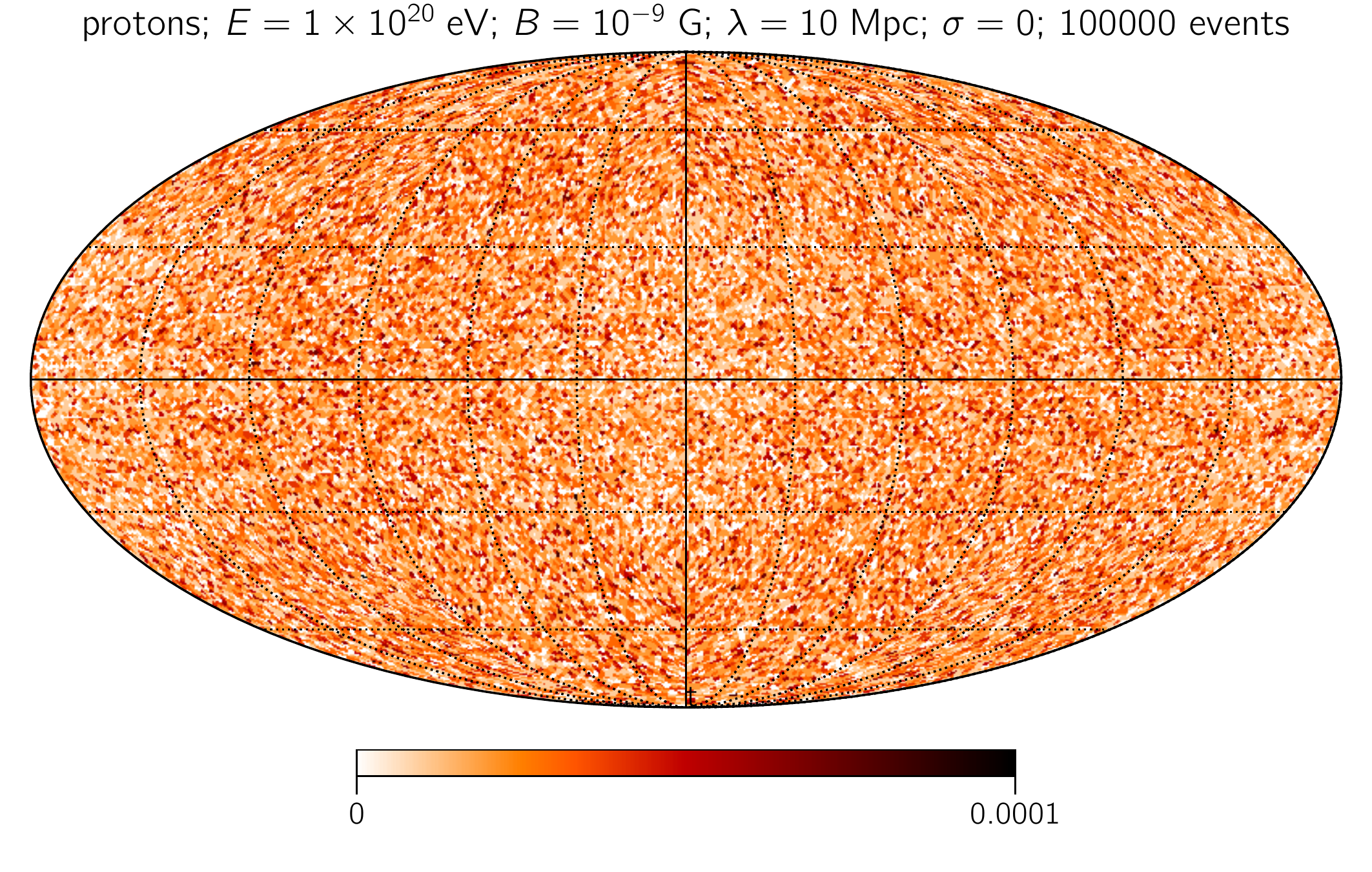}
	\includegraphics[width=0.325\columnwidth]{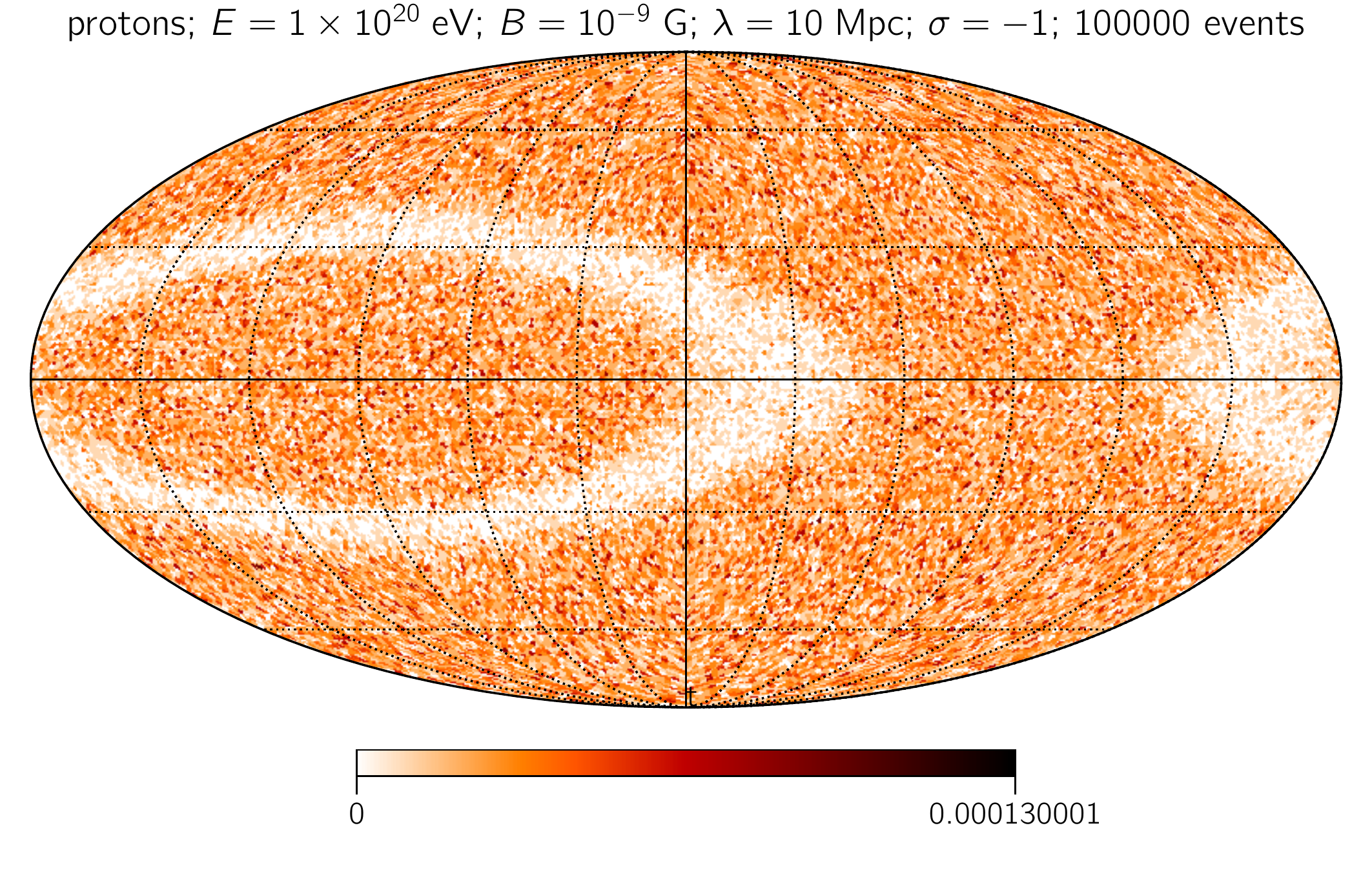}
	\caption{Skymaps containing the arrival directions of UHECRs for $\sigma=+1$ (left), $\sigma=0$ (centre), $\sigma=-1$ (right). These skymaps are for the case of $10^{20} \; \text{eV}$ protons, assuming $B = 10^{-9} \; \text{G}$ and $\lambda = 10 \; \text{Mpc}$. The upper row corresponds to the arrival directions of UHECRs coming from a sphere at $z_\text{sphere}=0.0032$, whereas the lower row is for $z_\text{sphere}=0.263$. The colour bar shows the normalised number of events per pixel.}
	\label{fig:spheres_skymaps}
\end{figure}

An analysis of Fig.~\ref{fig:individualSpheres} suggests that the outer spheres contribute more to the anisotropy signal than nearby ones. As discussed in Section~\ref{sec:Simulations}, our results depend on a delicate interplay of energy losses, coherence lengths, and source distances. Therefore, for sources that are farther away, interaction horizons will incur energy losses that would ultimately suppress the more elongated trajectories and hence cause the anisotropic signal. As a consequence, even for an isotropic distribution of sources one might expect an anisotropic UHECR distribution depending on the magnetic field configuration.
This is confirmed by comparing the skymaps shown in Fig.~\ref{fig:spheres_skymaps}.

Interestingly, the strength of the anisotropy signal would depend on the redshift evolution of the sources. If sources have positive evolution with redshift, i.e. $(1+z)^m$ with $m>0$, like gamma-ray bursts, AGNs, or star formation rate, then the detectability of helical magnetic fields would be favoured. On the other hand, if the source evolution is negative ($m<0$), as suggested by phenomenological fits of UHECR data~\cite{Taylor:2015rla,AlvesBatista:2018zui}, then it would be harder to constrain the helicity of IGMFs, as nearby sources would dominate over distant ones.

\bibliography{references}
\bibliographystyle{JHEP}

\end{document}